# Advances in Phonons: From Band Topology to Phonon Chirality

Tiantian Zhang*

*Institute of Theoretical Physics,*
*Chinese Academy of Sciences, Beijing 100190,*
*China*

Yizhou Liu†

*School of Physics Science and Engineering,*
*Tongji University,*
*Shanghai 200092,*
*China*

Hu Miao‡

*Materials Science and Technology Division,*
*Oak Ridge National Laboratory,*
*Oak Ridge, TN 37830,*
*USA*

Shuichi Murakami§

*Department of Applied Physics,*
*University of Tokyo,*
*7-3-1 Hongo, Bunkyo-ku,*
*Tokyo 113-8656,*
*Japan*
*International Institute for Sustainability with Knotted Chiral Meta Matter (WPI-SKCM$^2$),*
*Hiroshima University,*
*Hiroshima, 739-8526,*
*Japan*
*Center for Emerged Matter Science,*
*RIKEN, 2-1 Hirosawa,*
*Wako, Saitama, 351-0198,*
*Japan*

Phonons, the quantized collective vibrations of a crystal lattice, are among the most fundamental bosonic excitations in condensed matter systems. They govern thermal transport, mediate electron–phonon coupling, and drive symmetry-breaking orders such as charge density waves and conventional superconductivity. Long regarded as spin-0 bosons characterized only by their vibrational frequencies and linear or circular polarization, phonons have recently been revealed to host a much richer internal structure. Recent advances in topological band theory and quantum geometry have shown that phonon eigenstates, encoded in both their eigenvalues and eigenvectors, can exhibit nontrivial topological and geometric properties. These developments have established topological and circularly polarized phonons as two major frontiers in phonon physics, motivating this review of recent theoretical and experimental advances. We present a unified framework for classifying phonon modes in both reciprocal and real space, encompassing symmetry-protected topological phases, topological invariants, and phonon polarization. We then examine their coexistence in $\mathcal{PT}$-broken systems through Weyl phonons, highlighting the simultaneous emergence of topological and rotational chirality. Finally, we discuss outstanding challenges and future research directions, including the role of topology and quantum geometry in phonon-mediated interactions and the controlled manipulation of phonon angular momentum for prospective quantum technologies.

## CONTENTS

* ttzhang@itp.ac.cn
† yizhouliu@tongji.edu.cn
‡ miaoh@ornl.gov
§ murakami@ap.t.u-tokyo.ac.jp

In condensed matter physics, phonons are ubiquitous lattice excitations that play a key role in thermal properties of crystalline materials and critical for emergent electronic orders such superconductivity and charge density waves. In contrast to electrons, which offer a rich palette of intrinsic degrees of freedom such as charge, spin, and orbital character for manipulation (Fig. 1), phonons have traditionally been viewed as spin-0 bosons with limited intrinsic tunability. This fundamental disparity has historically confined phonons to a passive role, motivating the search for new strategies to control phononic properties and harness them for information processing and energy management.

A significant step in this direction came from the proposal of phonon-based devices such as transistors (Li *et al.*, 2006) and logic gates (Wang and Li, 2007) utilizing phonon nonlinearity (Chang *et al.*, 2006; Li *et al.*, 2012). These developments established the field of "phononics" (Li *et al.*, 2012; Maldovan, 2013)—the phononic analogue of electronics—demonstrating that phonons can play an active, rather than detrimental, role in solid-state device functionality.

More recently, inspired by topological insulators, superconductors, and semimetals in electronic systems (Ando and Fu, 2015; Armitage *et al.*, 2018; Hasan and Kane, 2010; Lv *et al.*, 2021; Moore, 2010; Qi and Zhang, 2010a, 2011; Sato and Ando, 2017; Yan and Felser, 2017), a parallel perspective has emerged from applying topological band theory to phonons, giving rise to "topological phononics" (Chen *et al.*, 2026b; Liu *et al.*, 2020b) and "chiral phononics" (Juraschek *et al.*, 2025; Wang *et al.*, 2024a). Topological phononics offers new possibilities in traditional phononics by avoiding fine-tuning parameters of lattice dynamics. The robust edge or surface modes are predicted useful in thermoelectric effects (Singh *et al.*, 2018; Xu *et al.*, 2014) and superconductivity (Di Miceli *et al.*, 2022). Chiral phononics on the other hand enables phonon-based magnetism (Cheng *et al.*, 2020c; Juraschek and Spaldin, 2019), chirality-controlled phonon diode effect (Chen *et al.*, 2022a), and phonon-mediated angular momentum transfer processes (Chen *et al.*, 2025; Holanda *et al.*, 2018; Kim *et al.*, 2023; Liu *et al.*, 2021f; Tauchert *et al.*, 2022; Wolf *et al.*, 2022). The topological characterization of excitations—via invariants such as the Berry phase and Chern number (Berry, 1984a; Thouless *et al.*, 1982a)—is not exclusive to fermions but extends naturally to bosonic systems satisfying Bloch's theorem. Key milestones include the extension of topological band theory to photonic crystals by Haldane and Raghu (Haldane and Raghu, 2008; Onoda *et al.*, 2004; Raghu and Haldane, 2008), followed by Prodan's demonstration that nontrivial phononic Chern numbers produce topologically protected edge modes (Prodan and Prodan, 2009a). Later theoretical work showed that nonzero phononic Chern numbers arise in systems with broken time-reversal symmetry (Sheng *et al.*, 2006; Zhang *et al.*, 2010) or broken spatial inversion symmetry (Zhang *et al.*, 2018a), leading to experimental realizations across acoustic metamaterials (Ma *et al.*, 2019a; Xue *et al.*, 2022), mechanical lattices (Süsstrunk and Huber, 2016), and crystalline solids (Li *et al.*, 2021a, 2023a; Liu *et al.*, 2022; Miao *et al.*, 2018; Zhang *et al.*, 2023b, 2018a, 2020b).

A notable distinction from electronic topology is that phonons originate from Newton's second law (second-order in time), whereas electrons obey the Schrödinger equation (first-order in time), making the formulation of phonon geometric phases less direct (Berry, 1984a) (detailed in Secs. I–II). Nevertheless, topological phonons share robust, defect-immune edge and surface channels with their electronic counterparts (Liu *et al.*, 2020b; Ma *et al.*, 2019b; Qin *et al.*, 2024;

Wang *et al.*, 2022; Xue *et al.*, 2022; Yang *et al.*, 2024b). Since phonons are the primary heat carriers in insulators and semiconductors, these states hold promise for thermal management and low-loss waveguiding applications (Xi *et al.*, 2025). One striking manifestation is the intrinsic phonon thermal Hall effect (Jin *et al.*, 2025; Liu *et al.*, 2020b, 2017b; Qin *et al.*, 2012; Sheng *et al.*, 2006; Shragai *et al.*, 2026; Strohm *et al.*, 2005; Zhang *et al.*, 2010, 2011), where a transverse heat current emerges under a magnetic field in charge-neutral insulators—a phenomenon inexplicable by the Lorentz force. Its topological origin, formulated in terms of the phonon Berry curvature (Zhang *et al.*, 2010) and refined through molecular Berry curvature (Berry, 1984b; Mead and Truhlar, 1979; Qin *et al.*, 2011, 2012; Saparov *et al.*, 2022; Zhang *et al.*, 2026b), has been recently demonstrated in $\alpha$-$RuCl_3$ (Shragai *et al.*, 2026), confirming that the Berry curvature or associated Hall viscosity (Barkeshli *et al.*, 2012; Flebus and MacDonald, 2023; Morales, 2026; Saito *et al.*, 2019) can rotate polarization vectors and deflect heat currents (Fig. 6).

While topological phonon focuses on momentum-space band geometry, a complementary perspective has emerged from the real-space rotational dynamics of the lattice (Kittel and McEuen, 2018; McLellan, 1988). In the continuum limit, phonons carry spin-1 angular momentum (Bliokh, 2022; Bliokh *et al.*, 2017; Bliokh and Nori, 2019; Bliokh *et al.*, 2022; Jones, 1973; Lei *et al.*, 2025; Levine, 1962; Longuet-Higgins, 1980; Ren, 2022; Shi *et al.*, 2019; Voronskii and Svirskii, 1961, 1962; Yang *et al.*, 2023); in crystals, however, the angular momentum is not quantized and its definition becomes context-dependent. In continuous media, the spin originates from the vector displacement field (Bliokh, 2022; Lei *et al.*, 2025; Ren, 2022), while in crystals the collective nature of lattice vibrations can produce collective interference phonon spin (Liu *et al.*, 2026).

Under broken inversion or time-reversal symmetry, atomic displacement eigenvectors acquire a complex phase, yielding circularly polarized vibrations that carry finite intrinsic angular momentum (Kittel and McEuen, 2018; McLellan, 1988; Zhang *et al.*, 2026a), as shown in Fig. 2(b). These modes (Bermudez *et al.*, 2008b; Kagan and Maksimov, 2008; Komiyama and Murakami, 2021; Lin *et al.*, 1985; McLellan, 1988; Rebane, 1983; Schaack, 1976; Zhang and Niu, 2015; Zhu *et al.*, 2018) have attracted intense interest for their ability to couple with magnetism, electron spins, and valley degrees of freedom (Bao *et al.*, 2026; Baydin *et al.*, 2022b; Che *et al.*, 2025a; Chen *et al.*, 2019; Cheng *et al.*, 2020a; Coh, 2023; Fransson, 2023; Grissonnanche *et al.*, 2020, 2019; Ishito *et al.*, 2023b; Juraschek *et al.*, 2025; Kusuno *et al.*, 2026; Li *et al.*, 2024b, 2020d; Mankovsky *et al.*, 2022; Minakova *et al.*, 2025; Nabei *et al.*, 2026; Ohe *et al.*, 2024; Tauchert *et al.*, 2022; Ueda *et al.*, 2023; Wang *et al.*, 2024b; Yang *et al.*, 2025a, 2026; Zhang *et al.*, 2025b; Zhang and Niu, 2014, 2015; Zhang *et al.*, 2026a,b, 2023b; Zhang and Murakami, 2022; Zhu *et al.*, 2018), driving phenomena including circularly polarized Raman scattering (Che *et al.*, 2025a; Ishito *et al.*, 2023a,b; Zhang *et al.*, 2026a, 2023b, 2025c), circular infrared absorption and emission (He *et al.*, 2020; Kong *et al.*, 2022; Li *et al.*, 2019; Liu *et al.*, 2019a; Yang *et al.*, 2025a; Zhu *et al.*, 2018), nonreciprocal thermal transport (Chen *et al.*, 2020, 2022b; Grissonnanche *et al.*, 2020, 2019; Kasahara *et al.*, 2018; Li *et al.*, 2020d, 2023b; Ohe *et al.*, 2024; Qin *et al.*, 2012; Zhang *et al.*, 2019c), and spin Seebeck effects (Fransson, 2023; Fu *et al.*, 2019; Kim *et al.*, 2023; Li *et al.*, 2024b; Yao and Murakami, 2025).

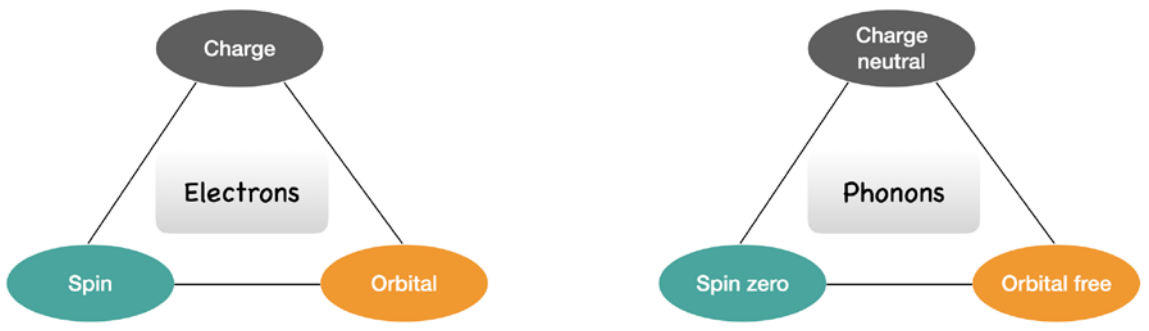

Figure 1 Comparison bewteen electrons and phonons. Electrons have multiple degrees of freedom, such as charge, spin, orbital, etc. Thus, electrons can be modulated by external fields (e.g., electric fields and magnetic fields) and thereby realizing relevant applications like spin devices. However, phonons are charge neutral, and are traditionally considered to be spin zero and orbital free. Therefore, new degrees of freedom are needed to be introduced to effectively modulate relevant physical properties.

Despite their distinct conceptual origins, i.e., momentum-space topology vs. real-space rotational motion shown in Fig. 2, these two paradigms are deeply connected. Weyl phonons (Zhang *et al.*, 2023b, 2025c) exemplify this unification: topological chirality (nonzero Chern number) and rotational chirality (finite phonon angular momentum) coexist. At present, several experimentally observed phenomena attributed to topological or circularly polarized phonons remain subjects of active debate, and careful re-evaluation is needed to ensure consistency with underlying physical mechanisms.

This review is organized as follows. Section I provides an overview of theoretical and experimental foundations of phonons in crystalline solids. Section II introduces the classification of topological phonon bands based on symmetries. Section III focuses on topological phononic models in both 2D and 3D systems, while Section IV discusses topological phonons in realistic materials. Section V discuss two notions of chirality in detail, i.e., topological chirality and rotational chirality, together with representative material examples. Parallel experimental progress on topological and circularly polarized phonons is reviewed in Sec. VI. We close with perspectives and open questions in Sec. VII.

## I. PHONONS: THEORETICAL BASIS AND EXPERIMENTAL PROBES

In this section, we describe non-interacting phonons within the framework of harmonic approximation. We focus on physical observable that is related to experimental measurements.

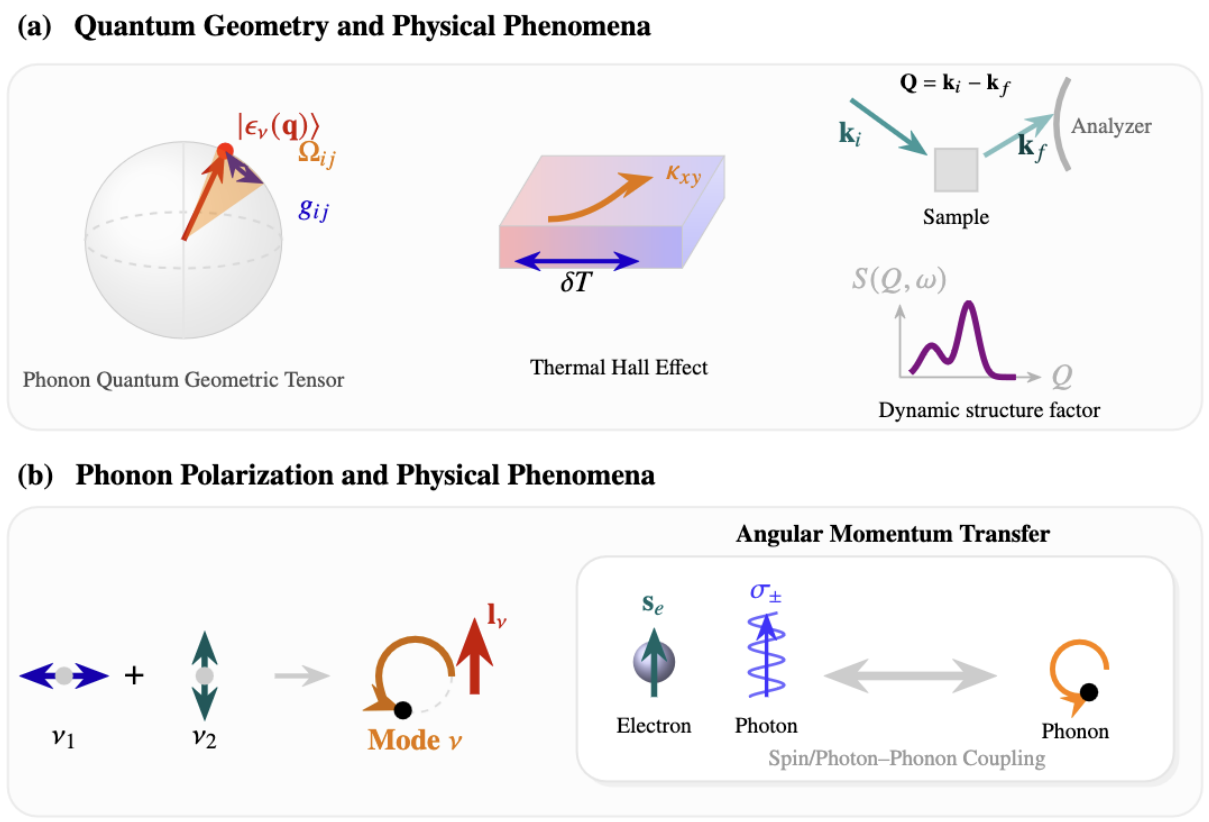

Figure 2 Integrated framework of phonon wave function properties. (a) The quantum geometric tensor (QGT) and Hilbert space. The phonon wave function (eigenvector) $|\epsilon_\nu(\mathbf{q})\rangle$ at a given momentum $\mathbf{q}$ is represented as a point on the Bloch sphere. The Berry curvature ($\Omega_{ij}$) contributes directly to the thermal Hall effect. Remarkably, both the quantum metric ($g_{ij}$) and the Berry curvature are, in principle, accessible through measurements of the dynamic structure factor. (b) Phonon modes in real space. Two orthogonal linear lattice vibrations, mode $\nu_1$ and $\nu_2$, both characterized by zero net angular momentum ($\mathbf{l}_{\nu_{1,2}} = 0$), can be superposed with a 90° phase difference to generate a circularly polarized mode $\nu$. This finite phonon angular momentum $\mathbf{l}_\nu$ can be transferred to other quasiparticles, for example through spin-phonon coupling.

## A. A Brief Overview of Lattice Dynamics

Phonons are ubiquitous quasiparticles describing the collective motions of the underlying lattice. The ordinary lattice dynamics can be well described by the following lattice Lagrangian (Kittel and McEuen, 2018):

$$L = \frac{1}{2}\sum_{\ell i\alpha} M_i \dot{u}^\alpha_{\ell i}\dot{u}^\alpha_{\ell i} - V(\{\boldsymbol{u}\}), \tag{1}$$

where $u^\alpha_{\ell i}$ represents $\alpha$ th component of atomic displacement of $i$ th atom belonging to $\ell$ th unit cell. Suppose we consider a three-dimensional lattice with $r$ atoms within one unit cell, then we have $\alpha = x, y, z$, $i = 1, 2, \cdots, r$ and $\ell = 1, 2, \cdots, N$ ($N$ is the total number of unit cells). $M_i$ refers to the atomic mass, and $V(\{\boldsymbol{u}\})$ is the potential energy of lattice as a function of the collection of atomic displacements $\{\boldsymbol{u}\}$. Near the equilibrium positions, the lattice potential energy can be approximated as:

$$V(\{\boldsymbol{u}\}) = V_0 - \sum_{\ell i\alpha} F^\alpha_{\ell i} u^\alpha_{\ell i} + \frac{1}{2}\sum_{\ell i\alpha}\sum_{\ell' j\beta} \Phi^{\alpha,\beta}_{\ell i,\ell' j} u^\alpha_{\ell i} u^\beta_{\ell' j} + O(|\boldsymbol{u}|^3), \tag{2}$$

where $V_0$ is the potential energy without atomic displacements. Since we assume the lattice is in equilibrium condition, the atomic force acting on $i$-th atom should be zero, *i.e.*, $F^\alpha_{\ell i} = -\frac{\partial V}{\partial u^\alpha_{\ell i}}\big|_{\boldsymbol{u}=0} = 0$. Under the harmonic approximation, only the second-order force constants $\Phi^{\alpha,\beta}_{\ell i,\ell' j} = \frac{\partial^2 V}{\partial u^\alpha_{\ell i}\partial u^\beta_{\ell' j}}\big|_{\boldsymbol{u}=0}$ are considered. In the presence of transnational symmetry, $\Phi^{\alpha,\beta}_{\ell i,\ell' j} = \Phi^{\alpha\beta}_{ij}(\boldsymbol{R}_\ell - \boldsymbol{R}_{\ell'}) = \Phi^{\alpha\beta}_{ij}(\boldsymbol{R}_{\ell''})$. Here $\boldsymbol{R}_{\ell''} = \boldsymbol{R}_\ell - \boldsymbol{R}_{\ell'}$ is the relative distance between the $\ell$ th and $\ell'$ th unit cells.

The lattice dynamics governed by Eqs. (1) and (2) can be derived by solving the Euler-Lagrangian equation $\frac{\mathrm{d}}{\mathrm{d}t}\frac{\partial L_0}{\partial \dot{u}^\alpha_{\ell i}} - \frac{\partial L_0}{\partial u^\alpha_{\ell i}} = 0$ as

$$M\ddot{u}^\alpha_{\ell i} = -\sum_{\ell' j\beta} \Phi^{\alpha,\beta}_{\ell i,\ell' j} u^\beta_{\ell' j}. \tag{3}$$

Consider the plane-wave ansartz:

$$u^\alpha_{\ell i} = \frac{1}{\sqrt{M_i}}\varepsilon^{i\alpha}_{n\boldsymbol{q}} \exp[i(\boldsymbol{q}\cdot\boldsymbol{R}_\ell - \omega_{n\boldsymbol{q}}t)], \tag{4}$$

Eq. (3) can be rewritten as:

$$\mathsf{D}_{\boldsymbol{q}}|\varepsilon_{n\boldsymbol{q}}\rangle = \omega^2_{n\boldsymbol{q}}|\varepsilon_{n\boldsymbol{q}}\rangle. \tag{5}$$

$\mathsf{D}_{\boldsymbol{q}}$ is the dynamical matrix whose matrix elements are given by

$$\begin{aligned}(\mathsf{D}_{\boldsymbol{q}})_{i\alpha,j\beta} =& \sum_{\ell'} \frac{\Phi^{\alpha,\beta}_{\ell i,\ell' j}}{\sqrt{M_i M_j}} \exp[-i\boldsymbol{q}\cdot(\boldsymbol{R}_\ell - \boldsymbol{R}_{\ell'})] \\ =& \sum_{\ell''} \frac{\Phi^{\alpha\beta}_{ij}(\boldsymbol{R}_{\ell''})}{\sqrt{M_i M_j}} \exp(-i\boldsymbol{q}\cdot\boldsymbol{R}_{\ell''}).\end{aligned} \tag{6}$$

$|\varepsilon_{n\boldsymbol{q}}\rangle = (\varepsilon^{1x}_{n\boldsymbol{q}}, \varepsilon^{1y}_{n\boldsymbol{q}}, \varepsilon^{1z}_{n\boldsymbol{q}}, \cdots, \varepsilon^{rx}_{n\boldsymbol{q}}, \varepsilon^{ry}_{n\boldsymbol{q}}, \varepsilon^{rz}_{n\boldsymbol{q}})^\top$ ($\top$ stands for matrix transpose) is the eigenvector of $n$ th branch at wave vector $\boldsymbol{q}$. Equation (5) gives rise to $3r$ eigenvalues $\omega_{n\boldsymbol{q}}$ and corresponding orthogonal eigenvectors $|\varepsilon_{n\boldsymbol{q}}\rangle$ ($n = 1, 2, \cdots, 3r$). The acoustic sum rule "$\sum_j \Phi^{\alpha\beta}_{ij} = 0$" requires that there must be three zero modes ($\omega_{n\boldsymbol{q}} = 0$) at $\Gamma$ ($\boldsymbol{q} = 0$), which are called acoustic modes. The other $3r - 3$ nonzero-frequency branches are optical phonon branches.

When going beyond single-particle picture, it is usually very useful to express the atomic displacement $u^\alpha_{\ell i}$ in the second-quantization form as:

$$\begin{aligned}u^\alpha_{\ell i} =& \sum_{n\boldsymbol{q}} \sqrt{\frac{\hbar}{2NM_i\omega_{n\boldsymbol{q}}}}\varepsilon^{i\alpha}_{n\boldsymbol{q}} \exp(i\boldsymbol{q}\cdot\boldsymbol{R}_\ell) \\ &\times \left(\hat{a}_{n\boldsymbol{q}} e^{-i\omega_{n\boldsymbol{q}}t} + \hat{a}^\dagger_{n-\boldsymbol{q}} e^{i\omega_{n\boldsymbol{q}}t}\right).\end{aligned} \tag{7}$$

Based on Eq. (7), it is easily checked that $u^\alpha_{\ell i}$ satisfies the quantum commutation relations

$$\begin{aligned}&[u^\alpha_{\ell i}, u^\beta_{\ell' j}] = [\dot{u}^\alpha_{\ell i}, \dot{u}^\beta_{\ell' j}] = 0, \\ &[u^\alpha_{\ell i}, \dot{u}^\beta_{\ell' j}] = (i\hbar/M_i)\delta_{\ell\ell'}\delta_{ij}\delta_{\alpha\beta},\end{aligned} \tag{8}$$

as long as the operators $\hat{a}_{n\boldsymbol{q}}$ satisfies the Bosonic commutation relations

$$\begin{aligned}&[\hat{a}_{n\boldsymbol{q}}, \hat{a}_{n'\boldsymbol{q}'}] = [\hat{a}^\dagger_{n\boldsymbol{q}}, \hat{a}^\dagger_{n'\boldsymbol{q}'}] = 0, \\ &[\hat{a}_{n\boldsymbol{q}}, \hat{a}^\dagger_{n'\boldsymbol{q}'}] = \delta_{nn'}\delta_{\boldsymbol{q}\boldsymbol{q}'},\end{aligned} \tag{9}$$

where $\delta$ refers to the Kronecker symbol. Thus $\hat{a}_{n\boldsymbol{q}}$ and $\hat{a}^{\dagger}_{n\boldsymbol{q}}$ have the physical meaning of phonon annihilation and creation operators, respectively. By making use of Eq. (7), the lattice vibration Hamiltonian can be rewritten as:

$$
\begin{aligned}
H_0 =& \sum_{\ell i \alpha} \frac{1}{2} M_i \dot{u}^{\alpha}_{\ell i} \dot{u}^{\alpha}_{\ell i} + \sum_{\ell i \alpha} \sum_{\ell' j \beta} \frac{1}{2} \Phi^{\alpha,\beta}_{\ell i, \ell' j} u^{\alpha}_{\ell i} u^{\beta}_{\ell' j} \\
=& \sum_{n\boldsymbol{q}} \hbar \omega_{n\boldsymbol{q}} \left( \hat{a}^{\dagger}_{n\boldsymbol{q}} \hat{a}_{n\boldsymbol{q}} + \frac{1}{2} \right).
\end{aligned}
\tag{10}
$$

It is worthy noting that when solving the lattice dynamics equation, different gauge choices for the plane-wave ansartz can be used:

$$
\begin{aligned}
u^{\alpha}_{\ell i} =& \underbrace{\frac{1}{\sqrt{M_i}} \varepsilon^{i\alpha}_{n\boldsymbol{q}} \exp[i(\boldsymbol{q} \cdot \boldsymbol{R}_{\ell} - \omega_{n\boldsymbol{q}} t)]}_{\boldsymbol{R}\text{ gauge}} \\
=& \underbrace{\frac{1}{\sqrt{M_i}} \tilde{\varepsilon}^{i\alpha}_{n\boldsymbol{q}} \exp[i(\boldsymbol{q} \cdot \boldsymbol{r}_{\ell i} - \omega_{n\boldsymbol{q}} t)]}_{\boldsymbol{r}\text{ gauge}},
\end{aligned}
\tag{11}
$$

where $\boldsymbol{r}_{\ell i} = \boldsymbol{R}_{\ell} + \boldsymbol{d}_i$ is the coordinate of $i$ th atom of $\ell$ th unit cell with $\boldsymbol{d}_i$ being the intracell coordinate of $i$ th atom. Both $\boldsymbol{R}$-gauge and $\boldsymbol{r}$-gauge satisfy Bloch's theorem. The comparison between $\boldsymbol{R}$-gauge and $\boldsymbol{r}$-gauge is shown in Tab. I.

When considering anharmonic effects in a crystal lattice, the traditional harmonic approximation for phonons is no longer sufficient, and corrections must be introduced to account for the interactions between phonons. Anharmonicity arises due to the deviation of atomic potentials from perfect quadratic forms, leading to phonon-phonon scattering, thermal expansion, and temperature-dependent phonon frequencies. These effects can significantly alter the phonon dispersion relations, lifetimes, and thermal properties of materials.

To incorporate anharmonic corrections, one common approach is to use perturbation theory, where the anharmonic terms in the potential energy are treated as perturbations to the harmonic Hamiltonian. This leads to shifts in phonon frequencies (self-energy corrections) and finite phonon lifetimes due to decay processes such as three-phonon and four-phonon interactions. Additionally, anharmonic effects lead to temperature-dependent behavior in phonon properties. For example, the phonon frequency softens with increasing temperature due to lattice expansion and enhanced phonon-phonon interactions. These corrections are crucial for accurately predicting thermal conductivity, specific heat, and other thermodynamic properties of materials at finite temperatures. Advanced computational methods, such as density functional theory (DFT) combined with anharmonic lattice dynamics or molecular dynamics simulations, are often employed to capture these effects quantitatively. In the following, we will ignore the anharmonic effects since they usually do not influence the topological properties of phonons.

### B. Phonon Topological Chirality and Rotational Chirality

The phononic single-particle "wavefunction" $|\boldsymbol{\varepsilon}_{n\boldsymbol{q}}\rangle$ represents a mixed real-space and $\boldsymbol{k}$-space description: On one hand, the phononic eigenvector indicates the real-space polarization vector $\boldsymbol{\varepsilon}^{i}_{n\boldsymbol{q}} = (\varepsilon^{x}_{n\boldsymbol{q}}, \varepsilon^{y}_{n\boldsymbol{q}}, \varepsilon^{z}_{n\boldsymbol{q}})$ $(i = 1, 2, \cdots, r)$ for each atom; on the other hand, the wavefunction represent a Bloch lattice wave in momentum-space with wave vector $\boldsymbol{q}$. In this sense, the chirality of phononic wavefunction $|\boldsymbol{\varepsilon}_{n\boldsymbol{q}}\rangle$ should be two-sided: it can be defined either in momentum space or in real space, as shown in Fig. 3.

In momentum space, the topological chirality of phononic wavefunction is represented by the phononic Berry curvature

$$
\Omega^{\alpha}_{n\boldsymbol{q}} = -\epsilon_{\alpha\beta\gamma} \mathrm{Im}[\partial_{q_\beta} (\boldsymbol{\varepsilon}^{i}_{n\boldsymbol{q}})^* \cdot \partial_{q_\gamma} \boldsymbol{\varepsilon}^{i}_{n\boldsymbol{q}}], \tag{12}
$$

where Einstein's convention of summation over repeated index is assumed. $(.)^*$ means complex-conjugate quantities. The topological chirality characterizes the phase winding of wavefunction in $\boldsymbol{q}$ space (Berry, 1984a). It is the imaginary part of a more general quantity called quantum geometric tensor (QGT) (Provost and Vallee, 1980) as shown in Fig. 2(a). The Berry curvature is responsible for the phononic thermal Hall effect (Qin *et al.*, 2011; Shragai *et al.*, 2026; Zhang *et al.*, 2010) and various topological phononic states including phononic Chern insulators (Liu *et al.*, 2017b; Wang *et al.*, 2025b; Zhang *et al.*, 2010) and Weyl phonons (Zhang *et al.*, 2018a). This part will be illustrated in detail in Secs. II to IV.

In real space, the rotational chirality is characterized by the phonon angular momentum (McLellan, 1988; Zhang and Niu, 2014, 2015)

$$
l^{\alpha}_{n\boldsymbol{q}} = -\epsilon_{\alpha\beta\gamma} \mathrm{Im}[(\varepsilon^{i\beta}_{n\boldsymbol{q}})^* \varepsilon^{i\gamma}_{n\boldsymbol{q}}]. \tag{13}
$$

The rotational chirality characterizes the real-space circular atomic motion as shown in Fig. 2(b). It can couple with other quasiparticles (e.g. electrons, photons, magnons) resulting in angular-momentum-transfer processes including phonon-magnon (Holanda *et al.*, 2018; Liu *et al.*, 2021f), electron-phonon (Chen *et al.*, 2025; Wang *et al.*, 2025a; Zhang *et al.*, 2026b) couplings, etc. This part will be illustrated in detail in Sec. V.

While the formulation of topological invariants in terms of the phonon Bloch wavefunction elucidates the global topological property of phonon bands, it does not fully capture the geometric information encoded in these states. A more complete description is provided by the QGT, which unifies topology and geometry within a single framework. In this formulation, the Berry curvature, given by the imaginary part of the QGT, characterizes the global topological properties of the band structure, whereas the real part defines the quantum metric, quantifying the distance between neighboring phonon states in momentum space, as illustrated in Fig. 3. Together, the Berry curvature and quantum metric furnish a comprehensive geometric description of phonon bands. While the study of topological phonons remains at an early stage, investigations of their quantum geometry are only just beginning, with relatively few works to date. In the following, we will review the definition of the QGT and discuss recent proposals for its experimental detection in phonon systems.

Table I  Comparison between $\boldsymbol{R}$-gauge and $\boldsymbol{r}$-gauge of lattice dynamics.

| | $\boldsymbol{R}$-gauge | $\boldsymbol{r}$-gauge |
|---|---|---|
| Plane-wave ansartz | $u^{\alpha}_{\ell i} = \frac{1}{\sqrt{M_i}} \varepsilon^{i\alpha}_{n\boldsymbol{q}} e^{i\boldsymbol{q}\cdot\boldsymbol{R}_\ell}$ | $\boldsymbol{u}^{\alpha}_{\ell i} = \frac{1}{\sqrt{M_i}} \tilde{\varepsilon}^{\alpha}_{n\boldsymbol{q}} e^{i\boldsymbol{q}\cdot\boldsymbol{r}_{\ell i}}$ |
| Eigenvalue equation | $\mathsf{D}_{\boldsymbol{q}}\|\varepsilon_{n\boldsymbol{q}}\rangle = \omega^2_{n\boldsymbol{q}}\|\varepsilon_{n\boldsymbol{q}}\rangle$ | $\tilde{\mathsf{D}}_{\boldsymbol{q}}\|\tilde{\varepsilon}_{n\boldsymbol{q}}\rangle = \omega^2_{n\boldsymbol{q}}\|\tilde{\varepsilon}_{n\boldsymbol{q}}\rangle$ |
| Dynamical matrix | $(\mathsf{D}_{\boldsymbol{q}})_{i\alpha,j\beta} = \sum_{\ell'} \frac{\Phi^{\alpha,\beta}_{\ell i,\ell' j}}{\sqrt{M_i M_j}} e^{-i\boldsymbol{q}\cdot(\boldsymbol{R}_\ell - \boldsymbol{R}_{\ell'})}$ | $\tilde{\mathsf{D}}_{\boldsymbol{q}} = \sum_{\ell'} \frac{\Phi^{\alpha,\beta}_{\ell i,\ell' j}}{\sqrt{M_i M_j}} e^{-i\boldsymbol{q}\cdot(\boldsymbol{r}_{\ell i} - \boldsymbol{r}_{\ell' j})}$ |
| Unit-cell dependence | $\mathsf{D}_{\boldsymbol{q}}$ and $\|\varepsilon_{n\boldsymbol{q}}\rangle$ rely on the choice of unit cell | $\tilde{\mathsf{D}}_{\boldsymbol{q}}$ and $\|\tilde{\varepsilon}_{n\boldsymbol{q}}\rangle$ do NOT rely on the choice of unit cell |
| Periodic condition | $\mathsf{D}_{\boldsymbol{q}+\boldsymbol{G}} = \mathsf{D}_{\boldsymbol{q}}$, $\|\varepsilon_{n\boldsymbol{q}+\boldsymbol{G}}\rangle = \|\varepsilon_{n\boldsymbol{q}}\rangle$ | $\tilde{\mathsf{D}}_{\boldsymbol{q}+\boldsymbol{G}} = e^{-i\boldsymbol{G}\cdot\mathbf{d}}\tilde{\mathsf{D}}_{\boldsymbol{q}} e^{i\boldsymbol{G}\cdot\mathbf{d}}$, $\|\tilde{\varepsilon}_{n\boldsymbol{q}+\boldsymbol{G}}\rangle = e^{-i\boldsymbol{G}\cdot\mathbf{d}}\|\tilde{\varepsilon}_{n\boldsymbol{q}}\rangle$ |

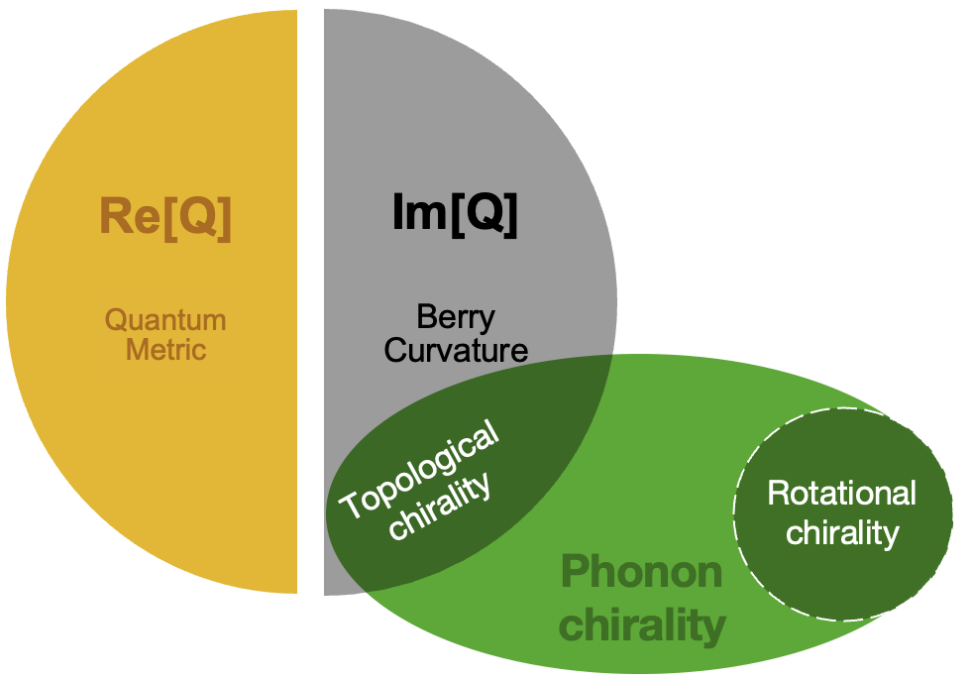

Figure 3  The phonon quantum geometric tensor ($Q$) decomposes into an imaginary part, the Berry curvature, and a real part, the quantum metric. Phonon chirality admits multiple definitions; in this review, we focus on topological chirality and rotational chirality, characterized by the sign of the topological invariant and the phonon angular momentum, respectively.

The QGT for the mode $|\varepsilon_{n\boldsymbol{q}}\rangle$ is defined as (Berry, 1989, 1984b; Bleu *et al.*, 2018; Cheng, 2010; Gianfrate *et al.*, 2020; Kang *et al.*, 2025; Ouyang *et al.*, 2025; Provost and Vallee, 1980; Sun *et al.*, 2024; Yu *et al.*, 2025, 2020)

$$Q^{\alpha\beta}_{n\boldsymbol{q}} = \left\langle \partial_{q_\alpha}\varepsilon_{n\boldsymbol{q}} \middle| (1 - |\varepsilon_{n\boldsymbol{q}}\rangle\langle\varepsilon_{n\boldsymbol{q}}|) \middle| \partial_{q_\beta}\varepsilon_{n\boldsymbol{q}} \right\rangle . \tag{14}$$

The Berry curvature introduced in Eq. (12) is related to above QGT as

$$Q^{\alpha\beta}_{n\boldsymbol{q}} = g^{\alpha\beta}_{n\boldsymbol{q}} - \frac{i}{2}\epsilon_{\alpha\beta\gamma}\Omega^{\gamma}_{n\boldsymbol{q}} \tag{15}$$

i.e. it is the imaginary part of QGT. The real part $g^{\alpha\beta}_{n\boldsymbol{q}} = \mathrm{Re}\, Q^{\alpha\beta}_{n\boldsymbol{q}}$ is called quantum metric tensor. The relation between phonon QGT and phonon chirality is shown in Fig. 3.

To elucidate the role of quantum geometry in phononic systems, direct experimental access to the full QGT is essential. Huang and Fang 2026 and Wu *et al.* 2026 developed a general framework that enables such access for collective excitations via the dynamical structure factor (DSF), encompassing both the Berry curvature and the quantum metric. As a universal probe of density-fluctuation-driven quasiparticles, the DSF is broadly applicable to phonons, magnons, excitons, and other bosonic modes.

As illustrated in Fig. 4, by formulating both the DSF and the QGT within a unified pseudospin framework (Jin *et al.*, 2022), a direct correspondence is established between measurable scattering intensities and the underlying QGT. This approach provides a practical protocol, implementable using inelastic x-ray or neutron scattering, to resolve all components of the bosonic QGT throughout the Brillouin zone.

### C. Phonon Dynamical Structure Factor

Experimentally, the direct physical measurable quantity is the phonon DSF, $S(\boldsymbol{q},\omega)$, which is determined by the charge density correlation function (Hansen and Klein, 1976):

$$S(\boldsymbol{q},\omega) = \frac{1}{2\pi\hbar}\int_{-\infty}^{\infty} e^{-i\omega t}\langle\rho(\boldsymbol{q},0)\rho^{\dagger}(\boldsymbol{q},t)\rangle\,\mathrm{d}t, \tag{16}$$

where $\rho(\boldsymbol{q},t)$ is the $\boldsymbol{q}$-space atomic charge density

$$\rho(\boldsymbol{q},t) = \sum_{\ell i} f_i(\boldsymbol{q}) e^{-i\boldsymbol{q}\cdot\boldsymbol{x}_{\ell i}(t)}, \tag{17}$$

$f_i(\boldsymbol{q})$ and $\boldsymbol{x}_{\ell i}(t) = \boldsymbol{r}_{\ell i} + \boldsymbol{u}_{\ell i}(t)$ are the atomic form factor and time-dependent position of $i$ th atom in $\ell$ th unit cell, respectively. Substitute Eq. (17) into (16), the dynamical structure factor can then be written as:

$$S(\boldsymbol{q},\omega) = \frac{N}{2\pi\hbar}\sum_{\ell} e^{i\boldsymbol{q}\cdot\boldsymbol{R}_\ell}\sum_{ij} f_i(\boldsymbol{q}) f^*_j(\boldsymbol{q}) e^{-i\boldsymbol{q}\cdot(\boldsymbol{d}_i - \boldsymbol{d}_j)} \int_{-\infty}^{\infty} e^{-i\omega t}\langle e^{-i\boldsymbol{q}\cdot\boldsymbol{u}_{0i}(0)} e^{i\boldsymbol{q}\cdot\boldsymbol{u}_{\ell j}(t)}\rangle\,\mathrm{d}t. \tag{18}$$

Under harmonic approximation, the Baker-Hausdorff theorem gives:

$$\langle e^{-i\boldsymbol{q}\cdot\boldsymbol{u}_{\ell' i}(0)} e^{i\boldsymbol{q}\cdot\boldsymbol{u}_{\ell j}(t)}\rangle = e^{-\frac{1}{2}\langle|\boldsymbol{q}\cdot\boldsymbol{u}_{l' i}|^2\rangle} e^{-\frac{1}{2}\langle|\boldsymbol{q}\cdot\boldsymbol{u}_{\ell j}|^2\rangle} \cdot e^{\langle(\boldsymbol{q}\cdot\boldsymbol{u}_{\ell' i}(0))(\boldsymbol{q}\cdot\boldsymbol{u}_{\ell j}(t))\rangle}. \tag{19}$$

We introduce the Debye-Waller factor, $W_\ell \equiv \frac{1}{2}\langle|\boldsymbol{q}\cdot\boldsymbol{u}_{\ell j}|^2\rangle \approx \frac{3\hbar^2|\boldsymbol{q}|^2}{2M\beta(k_B\Theta_D)^2}$, where $\Theta_D$ is the Debye temperature and $\beta = 1/k_B T$, and define $U \equiv (\boldsymbol{q}\cdot\boldsymbol{u}_{\ell' i}(0))$, $V \equiv (\boldsymbol{q}\cdot\boldsymbol{u}_{\ell j}(t))$. We then obtain:

$$\langle e^{-i\boldsymbol{q}\cdot\boldsymbol{u}_{\ell' i}(0)} e^{i\boldsymbol{q}\cdot\boldsymbol{u}_{\ell j}(t)}\rangle = e^{-W_\ell} e^{-W_{\ell'}} (1 + \langle UV\rangle + \langle UV\rangle^2 + \ldots). \tag{20}$$

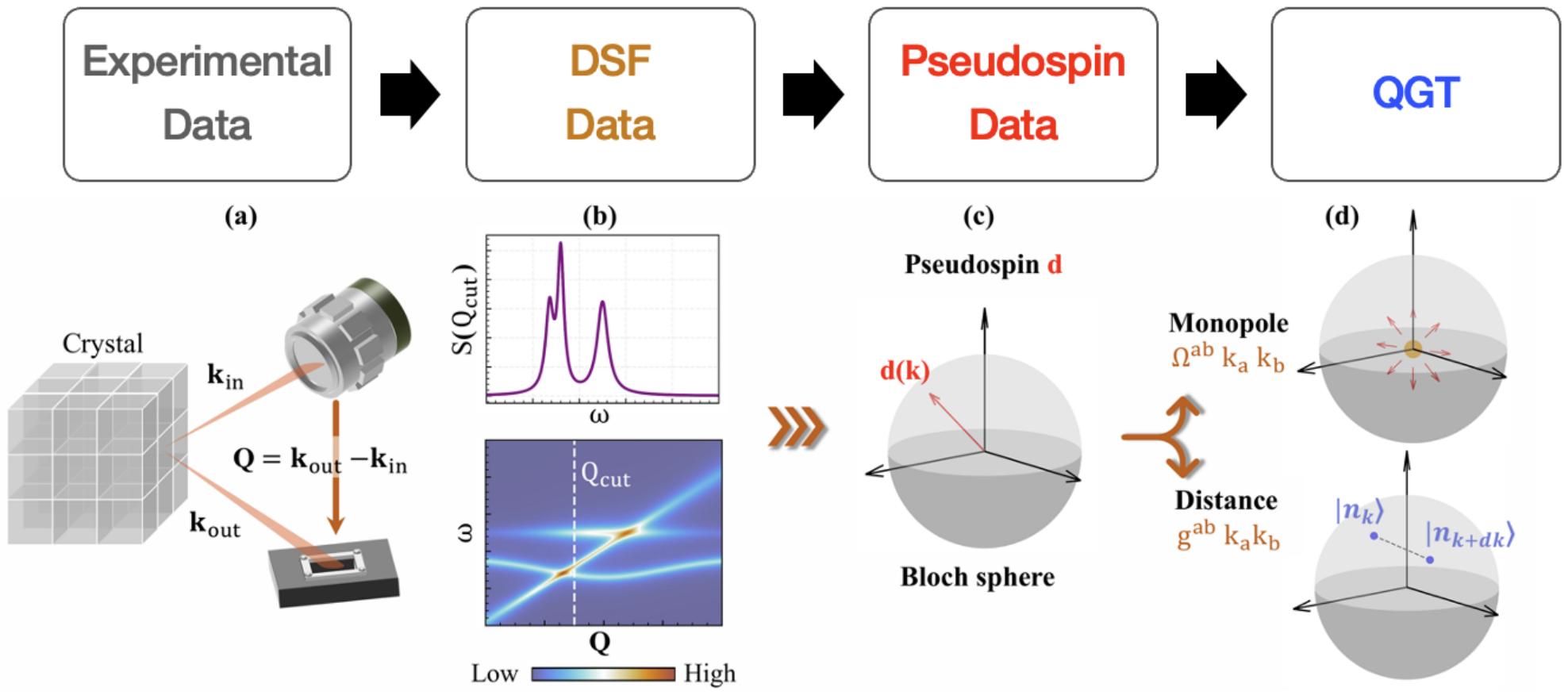

Figure 4 Illustration of the proposed direct measurement of the full quantum geometric tensor via the dynamical structure factor $S(\boldsymbol{Q}, \omega)$. (a) Schematic of the experimental setup. (b) Measured dynamical structure factor $S(\boldsymbol{Q}, \omega)$. (c) Pseudospin texture reconstructed from $S(\mathbf{Q}, \omega)$. (d) The whole quantum geometric tensor extracted from the reconstructed pseudospin texture. This figure is adapted from (Wu *et al.*, 2026).

We note that $\langle UV \rangle^n$ is corresponding to the $n^{\text{th}}$ order of phonon occupation number $\langle n_{\boldsymbol{q}\sigma} \rangle$. The phonon dynamical structure factor can thus be expanded as:

$$S(\boldsymbol{q}, \omega) = S(\boldsymbol{q}, \omega)_{0\text{p}} + S(\boldsymbol{q}, \omega)_{1\text{p}} + S(\boldsymbol{q}, \omega)_{2\text{p}} + ... \quad (21)$$

The elastic zero-phonon and inelastic single-phonon terms can respectively be derived as:

$$S(\boldsymbol{q}, \omega)_{0\text{p}} = N^2 \left| \sum_i f_i(\boldsymbol{q}) e^{-W_d} e^{i\boldsymbol{Q}\cdot\boldsymbol{d}_i} \right|^2 \delta_{\boldsymbol{q},\boldsymbol{G}} \delta(\hbar\omega), \quad (22)$$

$$S(\boldsymbol{q}, \omega)_{1\text{p}} = N \sum_{n\boldsymbol{q}'} \frac{1}{\omega_{n\boldsymbol{q}}} \left| \sum_i \frac{f_s(\boldsymbol{q})}{\sqrt{2M_i}} e^{-W_d} \boldsymbol{q} \cdot \boldsymbol{\varepsilon}^i_{n\boldsymbol{q}'} e^{i\boldsymbol{q}\cdot\boldsymbol{d}_i} \right|^2$$
$$\delta_{\boldsymbol{q}-\boldsymbol{q}',\boldsymbol{G}} \left[ \langle n_{n\boldsymbol{q}'} \rangle \delta(\omega - \omega_{n\boldsymbol{q}'}) + \langle n_{n\boldsymbol{q}'} + 1 \rangle \delta(\omega + \omega_{\boldsymbol{q}\sigma}) \right], \quad (23)$$

where $\boldsymbol{G}$ is the reciprocal lattice vector and $\boldsymbol{q}'$ is the reduced wave vector in the first Brillouin zone. Using Eq. (23), $S(\boldsymbol{q}, \omega)$ can be calculated from first-principles calculations to compare directly with various experimental probes described below. By measuring $S(\boldsymbol{q}, \omega)$, one can extract not only the eigenvalues of the dynamical matrix, which correspond to the phonon frequencies, but also the eigenvectors, which are directly related to the scattering intensity.

### D. Phonon Experimental Probes

Phonons can be probed, directly or indirectly, by absorption and scattering techniques. In this section, we provide a concise overview of spectroscopic techniques. For readers seeking in-depth information on a specific experimental probe, we refer to specialized technical review articles.

#### 1. Light-matter interactions

For light-matter interactions, the scattering and absorption cross-sections are related to the transition, $W$, between initial state, $|i\rangle$, and final state, $|f\rangle$, that is give by Fermi's golden rule:

$$W = \frac{1}{\mathcal{Z}} \sum_{i,f} e^{-\beta E_i} |M_{fi}|^2 \delta(E_f - E_i - \hbar\omega) \quad (24)$$

$\mathcal{Z}$ is the partition function. $M_{fi} = \langle f | H_I | i \rangle$ is the scattering matrix element, where $H_I$ is the scattering operator describing the interactions between the photon field and charged particles:

$$H_\text{I} = \frac{e\boldsymbol{A} \cdot \boldsymbol{P}}{m} + \frac{e^2 A^2}{2m}. \quad (25)$$

The vector potential $\boldsymbol{A}$ can be written in photon operators, $b_{u\boldsymbol{k}}$ and $b^\dagger_{u\boldsymbol{k}}$:

$$\boldsymbol{A}(\boldsymbol{r}, t) = \sum_{u\boldsymbol{k}} \hat{e}_u \sqrt{\frac{\hbar}{2\epsilon_0 V \omega_{u\boldsymbol{k}}}}$$
$$[b_{u\boldsymbol{k}} e^{i(\boldsymbol{k}\cdot\boldsymbol{r} - \omega_{u\boldsymbol{k}} t)} + b^\dagger_{u\boldsymbol{k}} e^{-i(\boldsymbol{k}\cdot\boldsymbol{r} - \omega_{u\boldsymbol{k}} t)}], \quad (26)$$

where $\hat{e}_u$ is the photon polarization vector. A general expression for the differential light scattering cross section is:

$$\frac{\mathrm{d}^2\sigma}{\mathrm{d}\Omega \mathrm{d}\omega} = \hbar r_0^2 \frac{\omega_f}{\omega_i} W, \quad (27)$$

where $r_0 = \frac{e^2}{mc^2}$ ~2.81 femtometer (fm) is the Thompson radius of the electron. The $\omega_i$ and $\omega_f$ represent the energy of incident and scattered photons with momentum $\boldsymbol{k}_i$ and $\boldsymbol{k}_f$, respectively.

## 2. Raman and Infrared Spectroscopy

Raman and infrared (IR) spectroscopy are well-known methods to probe lattice dynamics with nearly zero momentum transfer (Basov and Timusk, 2005; Devereaux and Hackl, 2007). In the presence of inversion symmetry, these techniques have opposite and complementary selection rules, and must be used in combination to measure all phonon modes near the Brillouin zone center.

The physical principle underlying IR spectroscopy is shown in Fig. 5(a). This technique involves measuring the absorption or reflection of IR radiation. Strong IR absorption occurs when the charge oscillations induced by the optical field resonate with lattice vibrations. In this process, the initial and final states contains one and zero photon, respectively. The cross-section is therefore determined by the parity-odd dipole term in Eq. (25), $\boldsymbol{A}\cdot\boldsymbol{P}$. Since the zero-phonon initial state is parity-even, the final states must be parity-odd phonon modes.

In contrast, Raman scattering is a "photon-in-photon-out" scattering process, as depicted in Fig. 5(b). It measures the change in wavelength of light as a phonon is either created or annihilated (Devereaux and Hackl, 2007). In this process, it measures the polarizability of the crystals under the photon field, where both linear and quadratic terms in Eq. (25) contribute to the cross-section. To understanding the selection rule, we can consider the light-induced dipole moment:

$$\boldsymbol{D} = \tilde{\alpha}\cdot\boldsymbol{E}_i\cos(\boldsymbol{k}_0\cdot\boldsymbol{r}-\omega_0 t), \tag{28}$$

where $\tilde{\alpha}$, a second-rank symmetric tensor, is the Raman polarizability tensor. Since the inelastic scattering of the incident light $\boldsymbol{E}_i$ can excite atomic vibrations, the polarizability tensor has frequency-dependent contributions at the vibration frequencies:

$$\tilde{\alpha} = \tilde{\alpha}_0 + \sum_{n\boldsymbol{q}}\tilde{\alpha}_{n\boldsymbol{q}}\cos(\boldsymbol{q}\cdot\boldsymbol{r}-\omega_{n\boldsymbol{q}}t). \tag{29}$$

So that

$$\begin{aligned}\boldsymbol{D} &= \tilde{\alpha}_0\cdot\boldsymbol{E}_i\cos(\boldsymbol{k}_0\cdot\boldsymbol{r}-\omega_0 t)\\ &+\sum_{n\boldsymbol{q}}\tilde{\alpha}_{n\boldsymbol{q}}\cdot\boldsymbol{E}_i\cos[(\boldsymbol{k}_0\pm\boldsymbol{q})\cdot\boldsymbol{r}-(\omega\pm\omega_{n\boldsymbol{q}})t].\end{aligned} \tag{30}$$

In Eq. (30), the $+\omega$ and $-\omega$ photon frequency shifts are known as Stokes and anti-Stokes process in the Raman scattering, respectively. For centrosymmetric materials, since both $\boldsymbol{D}$ and $\boldsymbol{E}$ are parity-odd, $\tilde{\alpha}_\sigma$ must be parity-even to ensure that the Raman-active modes are symmetry-allowed.

Beyond this simple selection rule, since Raman scattering involves virtual intermediate states, its scattering matrix element, $M_{fi}$, display strong photon polarization dependence (Devereaux and Hackl, 2007). By controlling the photon polarization, $\hat{\epsilon}_u$, Raman can selectively probe phonon modes with different chirality (Ishito *et al.*, 2023b; Zhang *et al.*, 2023b).

## 3. Inelastic x-ray and Neutron Scattering

Inelastic x-ray scattering (IXS) and inelastic neutron scattering (INS) are powerful techniques to directly measure the dynamic structure factor $S(\boldsymbol{Q},\omega)$ of bulk phonons across a broad energy and momentum range (Feigin *et al.*, 1987; Higgins and Maconnachie, 1986; Kotani and Shin, 2001).

For (non-resonant) IXS, the scattering cross-section is determined by the $A^2$ term in Eq. (25). The Thomson differential cross-section is given by:

$$\frac{\mathrm{d}^2\sigma}{\mathrm{d}\Omega\mathrm{d}\omega} = \frac{k_f}{k_i}r_0^2|\hat{\varepsilon}_i\cdot\hat{\varepsilon}_f|^2 S(\boldsymbol{q},\omega). \tag{31}$$

For a typical IXS experiment, the energy transfer, $\omega = \omega_f - \omega_i$, is about 5∼6 orders magnitude smaller than the incident photon energy. Therefore, $\frac{k_f}{k_i}\sim 1$, and $\frac{\mathrm{d}^2\sigma}{\mathrm{d}\Omega\mathrm{d}\omega}\propto S(\boldsymbol{Q},\omega)$.

Different from x-ray scattering, INS arises from the energy and momentum transfer between incident neutrons and the atomic nuclei. These interactions are short-range nuclear forces, which operate within the femtometer length scale. This length-scale is much smaller than the wavelength of neutrons. For this reason, the scattering center can be viewed as a delta function, corresponding to momentum independent form factor that is different from photon probes. The coherent scattering contribution to the INS cross-section is given by:

$$\frac{\mathrm{d}^2\sigma}{\mathrm{d}\Omega\mathrm{d}\omega} = \frac{\sigma_{\mathrm{coh}}}{4\pi}\frac{k_f}{k_i}NS^{\mathrm{N}}(\boldsymbol{q},\omega), \tag{32}$$

where $S^{\mathrm{N}}(\boldsymbol{q},\omega)$ is similar to $S(\boldsymbol{q},\omega)$ except $f_i(\boldsymbol{q}) = 1$ for the neutron scattering. The $\sigma_{\mathrm{coh}}$ depends on the nuclear isotope, spin relative to the neutron, and nuclear eigenstate.

While both IXS and INS are powerful techniques for probing phonon dynamics, they are often complementary in many aspects. For instance, the scattering intensity of IXS approximately scales with $Z^2$, where $Z$ is the atomic number, whereas INS displays complex isotope effect. As a result, the choice between the two techniques may depend on the specific chemical elements present in the sample.

In addition, IXS offers higher momentum resolution and the ability to study small-size samples ($\sim 10\ \mu m\times 30\ \mu m$) due to the high brilliance of modern synchrony light source. On the other hand, INS can achieve higher energy resolution and allow simultaneous collection of data across a large momentum space using time-of-flight spectrometer.

## 4. Resonant Inelastic x-ray Scattering

Resonant inelastic x-ray scattering (RIXS) is a photon scattering technique that has seen remarkable advancements over the past decade (Ament *et al.*, 2011; Mitrano *et al.*, 2024). RIXS is a two-step process involving photoelectric absorption and emission. As shown in Fig. 5(c), a core electron absorbs a photon and is excited into the valence band. This intermediate excited state then decays by emitting an x-ray photon.

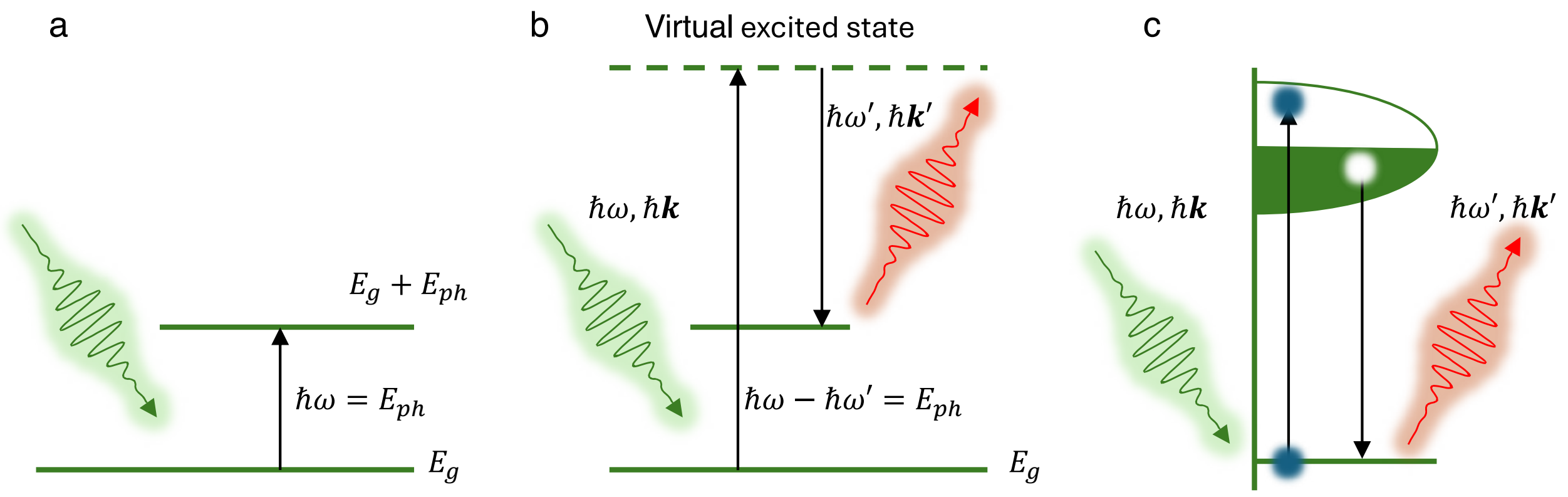

Figure 5 Schematics of photon scattering processes. (a) depicts the IR spectroscopy. When the incident photon energy, $\hbar\omega$, matches the phonon excitation energy, $E_{ph}$, the incident light will be absorbed, resulting reduced reflection or transmission light intensity. (b) shows the Raman scattering process. The incident photon drives the initial state to a virtual excited state. The system then relax to a final state and emit a photon. Following the energy and momentum conservation, $E_{\rm ph} = \hbar\omega - \hbar\omega'$ and $\hbar\boldsymbol{Q}_{\rm ph} = \hbar\boldsymbol{k} - \hbar\boldsymbol{k}'$, where $\hbar\boldsymbol{k}$ and $\hbar\boldsymbol{k}'$ are incident and scattered photon momentum. (c) illustrates the direct RIXS process. Similar to the Raman scattering, RIXS is also a scattering process. The incident photon energy is tuned to a element specific resonant energy. The core electron is excited to an intermediate state. The core hole is then filled by an electron from the conduction band.

Due to its resonant nature, RIXS is described by the Kramers-Heisenberg equations (Ament *et al.*, 2011; Healy, 1977). The RIXS scattering matrix element is given by:

$$M_{fi} = \sum_n \frac{\langle f|\mathcal{D}^\dagger_{\boldsymbol{k}',\hat{e}_f}|n\rangle\langle n|\mathcal{D}_{\boldsymbol{k},\hat{e}_i}|i\rangle}{E_n - E_i - \hbar\omega_{\boldsymbol{k}} + \mathrm{i}\Gamma_n/2}, \tag{33}$$

where $\Gamma_n/2$ is the inverse core-hole lifetime in units of energy, $E_n$ is the energy of the intermediate state. $\mathcal{D}^\dagger_{\mathbf{k}',\hat{e}_f}$ and $\mathcal{D}_{\mathbf{k},\hat{e}_i}$ are the emission and absorption operators that are given by the dipole term in Eq. (25).

Compared to IXS and INS, RIXS is sensitive to a wide range of quantum excitations, including spin, charge, orbital, lattice, and fractional quasiparticles (Ament *et al.*, 2011; Mitrano *et al.*, 2024). The intensity distribution in RIXS spectra also provides valuable insights into the strength of quasiparticle interactions. On the other hand, RIXS cross-section is not simply proportional to the phonon dynamical structure factor. Additionally, despite significant progress, the state-of-the-art energy resolution of RIXS remains on the order of 20 meV, making it challenging to quantitatively analyze the phonon dynamical structure factor.

## 5. Momentum Resolved Electron Energy Loss Spectroscopy

IXS and INS are generally bulk sensitive probes due to the deep penetration depth of hard x-rays and neutrons. For the interests in 2D materials and topological surface state, IXS and INS are limited by the small total scattering volume. To address this, high-resolution momentum resolved electron energy loss spectroscopy (HR-MREELS) has been developed as a powerful tool for investigating charge and phonon correlation functions on the surface (Brydson, 2020; Egerton, 2011; Ibach and Mills, 2013; Vig *et al.*, 2017). Unlike the transmission EELS, HR-MEELS experiments are typically conducted in reflection geometry with incident electron energy, $E_i$, less than 200 eV. It has been shown that by setting $E_i \leq$10 eV, HR-MEELS can achieve energy and momentum resolutions better than 1 meV and 0.002Å$^{-1}$, making HR-MEELS a highly effective techniques for probing phonon and charge excitations in 2D materials. For electron scattering with small momentum transfer, $\boldsymbol{q}$, the HR-MEELS cross section can be formulated as:

$$\frac{\mathrm{d}^2\sigma}{\mathrm{d}\Omega\mathrm{d}\omega} = \sigma_0 V^2_{\rm eff}(\boldsymbol{q}) \int_{-\infty}^{0} dz_1 dz_2 e^{-|\boldsymbol{q}||z_1+z_2|} \cdot S(\boldsymbol{q}, z_1, z_2, \omega), \tag{34}$$

where

$$V_{\rm eff}(\boldsymbol{q}) = \frac{4\pi e^2}{q^2 + (k_i^z + k_s^z)^2}. \tag{35}$$

$k^s_{i,s}$ represent the out-of-plane momentum of the incident and scattered electron. As shown in Eq. (34), the HR-MEELS cross-section is closely related to the surface $S(\boldsymbol{q}, \omega)$.

## 6. Other resonant and stimulating Techniques

Phonons can also be detected by other resonant and stimulating techniques. For instance, resonant ultrasound spectroscopy probes quasi-elastic phonon excitations in the MHz range (Migliori, 2016). This technique complements experimental probes described in previous sections that typically focus on phonons in THz-range. Additionally, pump-probe and time-domain spectroscopy (Fischer *et al.*, 2016; Fushitani, 2008; Hangyo *et al.*, 2005; Koch *et al.*, 2023), utilizing tabletop laser setup or advanced free-electron lasers, provide access to coherent optical and acoustic phonons, as well as their interactions with spin and charge excitations. These methods offer versatile tools for exploring the dynamic behavior of phonons and their coupling to other quantum excitations in materials.

Besides the spectroscopy probes mentioned above, the geometric or topological properties of phonon manifest themselves in transport signatures like phonon thermal Hall and acoustic Faraday effects as shown in Fig. 6. When time-reversal symmetry is broken (e.g. by an external magnetic field), the lattice Hamiltonian should be modified as (Mead and Truhlar, 1979; Saito *et al.*, 2019):

$$H = \sum_{\ell i} \frac{(\boldsymbol{p}_{\ell i} - Q_i \boldsymbol{A}(\boldsymbol{r}_{\ell i}) - \boldsymbol{\mathcal{A}}_{\ell i})^2}{2M_i} + \sum_{\ell i\alpha}\sum_{\ell' j\beta} \frac{1}{2}\Phi^{\alpha,\beta}_{\ell i,\ell' j} u^{\alpha}_{\ell i} u^{\beta}_{\ell' j}, \tag{36}$$

where $Q_i$ is the effective charge of $i$ th sublattice, $\boldsymbol{A}(\boldsymbol{r})$ is the magnetic vector potential, and $\boldsymbol{\mathcal{A}}_{\ell i}$ is the molecular Berry connection of the ground-state many-body electronic wavefunction (Mead and Truhlar, 1979; Qin *et al.*, 2011; Saparov *et al.*, 2022; Zhang *et al.*, 2026b)

$$\boldsymbol{\mathcal{A}}_{\ell i} = i\hbar\langle\Psi_0|\boldsymbol{\nabla}_{\ell i}\Psi_0\rangle = -\hbar \mathrm{Im}\Big\langle \Psi_0 \Big| \frac{\partial \Psi_0}{\partial \boldsymbol{u}_{\ell i}}\Big\rangle. \tag{37}$$

The lattice dynamics determined by Eq. (36) is

$$M_i \ddot{u}^{\alpha}_{\ell i} = \underbrace{-\sum_{\ell' j\beta} \Phi^{\alpha,\beta}_{\ell i,\ell' j} u^{\beta}_{\ell i}}_{\text{Hook's force}} + \underbrace{\sum_{\beta} Q_i F^{\alpha\beta} \dot{u}^{\beta}_{\ell i}}_{\text{Lorentz force}} + \underbrace{\sum_{\ell' j\beta} \mathcal{F}^{\alpha,\beta}_{\ell i,\ell' j} \dot{u}^{\beta}_{\ell' j}}_{\text{MBC force}}, \tag{38}$$

where $F^{\alpha\beta} = \partial^{\alpha} A^{\beta} - \partial^{\beta} A^{\alpha} = \sum_{\gamma} \epsilon^{\alpha\beta\gamma} B^{\gamma}$ ($\epsilon^{\alpha\beta\gamma}$ is the Levi-Civita tensor) gives rise to the magnetic field $\boldsymbol{B}$, and $\mathcal{F}^{\alpha,\beta}_{\ell i,\ell' j}$ is the gauge-invariant molecular Berry curvature (MBC) whose expression is given by

$$\mathcal{F}^{\alpha,\beta}_{\ell i,\ell' j} = -2\hbar \mathrm{Im}\Big\langle \frac{\partial \Psi_0}{\partial u^{\alpha}_{\ell i}} \Big| \frac{\partial \Psi_0}{\partial u^{\beta}_{\ell' j}} \Big\rangle. \tag{39}$$

The last term in Eq. (38) is the MBC force that is proportional to the atomic velocities $\dot{u}^{\beta}_{\ell' j}$, indicating that the MBC force performs no work. The MBC force is similar to Lorentz force where $\mathcal{F}^{\alpha,\beta}_{\ell i,\ell;j}$ takes the role of magnetic field $F^{\alpha\beta}$. Just like the Lorentz force can deflect the electric charge flow, the MBC force can deflect heat current flow thus resulting thermal Hall effect (Qin *et al.*, 2011; Saito *et al.*, 2019).

The phonon mode equation determined by the lattice equation (38) and plane-wave ansartz (11) is

$$\mathsf{D}_{\boldsymbol{q}} |\boldsymbol{\varepsilon}_{n\boldsymbol{q}}\rangle + i\omega_{n\boldsymbol{q}} \mathsf{F}_{\boldsymbol{q}} |\boldsymbol{\varepsilon}_{n\boldsymbol{q}}\rangle = \omega^2_{n\boldsymbol{q}} |\boldsymbol{\varepsilon}_{n\boldsymbol{q}}\rangle, \tag{40}$$

where $\mathsf{D}_{\boldsymbol{q}}$ is the dynamical matrix, and $\mathsf{F}_{\boldsymbol{q}}$ is an ansisymmetric matrix whose matrix elements are

$$\begin{aligned}(\mathsf{F}_{\boldsymbol{q}})^{\alpha\beta}_{ij} &= \frac{Q_i}{M_i} F^{\alpha\beta}\delta_{ij} + \underbrace{\sum_{\ell'} \frac{\mathcal{F}^{\alpha,\beta}_{\ell i,\ell' j}}{\sqrt{M_i M_j}} \exp[-i\boldsymbol{q}\cdot(\boldsymbol{r}_{\ell i} - \boldsymbol{r}_{\ell' j})]}_{\boldsymbol{r}\ \text{gauge}} \\ &= \frac{Q_i}{M_i} F^{\alpha\beta}\delta_{ij} + \underbrace{\sum_{\ell'} \frac{\mathcal{F}^{\alpha,\beta}_{\ell i,\ell' j}}{\sqrt{M_i M_j}} \exp[-i\boldsymbol{q}\cdot(\boldsymbol{R}_{\ell} - \boldsymbol{R}_{\ell'})]}_{\boldsymbol{R}\ \text{gauge}}.\end{aligned} \tag{41}$$

Equation (40) is no longer an eigenvalue equation, but it can be transformed into standard form of Hermitian eigenvalue equation (Huber, 2016a; Liu *et al.*, 2020b, 2018, 2017b):

$$\mathsf{H}_{\boldsymbol{q}} |\psi_{n\boldsymbol{q}}\rangle = \omega_{n\boldsymbol{q}} |\psi_{n\boldsymbol{q}}\rangle, \tag{42}$$

$$\mathsf{H}_{\boldsymbol{q}} = \begin{pmatrix} 0 & i\mathsf{D}^{1/2}_{\boldsymbol{q}} \\ -i\mathsf{D}^{1/2}_{\boldsymbol{q}} & i\mathsf{F}_{\boldsymbol{q}} \end{pmatrix}, \quad |\psi_{n\boldsymbol{q}}\rangle = \begin{pmatrix} \mathsf{D}^{1/2}_{\boldsymbol{q}} |\boldsymbol{\varepsilon}_{n\boldsymbol{q}}\rangle \\ -i\omega_{n\boldsymbol{q}} |\boldsymbol{\varepsilon}_{n\boldsymbol{q}}\rangle \end{pmatrix}. \tag{43}$$

Equation (42) is a Schrödinger-like equation for phonons with the phononic "wavefunction" $|\psi_{n\boldsymbol{q}}\rangle$ determined not only by the displacements $|\boldsymbol{\varepsilon}_{n\boldsymbol{q}}\rangle$, but also by the velocities $-i\omega_{n\mathbf{q}}|\varepsilon_{n\mathbf{q}}\rangle \sim \frac{\partial}{\partial t}|\varepsilon_{n\mathbf{q}}\rangle$. The velocities enters into the phononic eigenmode equation because both the Lorentz force and MBC force are proportional to the atomic velocities. In the presence of time-reversal breaking effects (e.g. Lorentz force and MBC force), the phononic topological quantities like Berry phase, Berry connection, Berry curvature, and Chern number should be defined based on $|\psi_{n\boldsymbol{q}}\rangle$ (Liu *et al.*, 2020b, 2018, 2017b).

In the long-wave length limit, the lattice dynamics equation is reduced to the elastic wave equation of motion. In a continuous elastic medium, the relationship between stress tensor $\sigma_{\alpha\beta}$ and strain tensor $\varepsilon_{\alpha\beta}$ is generally given by:

$$\sigma_{\alpha\beta} = \underbrace{\sum_{\gamma\xi} c_{\alpha\beta\gamma\xi}\varepsilon_{\gamma\xi}}_{\text{Hook's force}} + \underbrace{\sum_{\gamma\xi} \eta_{\alpha\beta\gamma\xi}\dot{\varepsilon}_{\gamma\xi}}_{\text{viscosity force}}. \tag{44}$$

$c_{\alpha\beta\gamma\xi}$ is the ordinary elastic tensor and $\eta_{\alpha\beta\gamma\xi}$ is referred to the phonon viscosity tensor (Barkeshli *et al.*, 2012; Flebus and MacDonald, 2023). The first term reflects the Hook's law, while the second term usually describes the phenomena of viscosity with energy dissipation. However, when time-reversal symmetry is broken, the viscosity tensor $\eta_{\alpha\beta\gamma\xi}$ acquires an antisymmetric component under the exchange between $(\alpha, \beta)$ and $(\gamma, \xi)$, i.e. $\eta_{\alpha\beta\gamma\xi} = \eta^{\mathrm{S}}_{\alpha\beta\gamma\xi} + \eta^{\mathrm{AS}}_{\alpha\beta\gamma\xi}$ with $\eta^{\mathrm{AS}}_{\alpha\beta\gamma\xi} = -\eta^{\mathrm{AS}}_{\gamma\xi\alpha\beta}$. The antisymmetric viscosity force becomes dissipationless with zero elastic power density i.e. $\sum_{\alpha\beta\gamma\xi} \dot{\varepsilon}_{\alpha\beta}\eta^{\mathrm{AS}}_{\alpha\beta\gamma\xi}\dot{\varepsilon}_{\gamma\xi} = 0$. Such an antisymmetric viscosity tensor is called phonon Hall viscosity tensor and it can lead to thermal Hall effect and acoustic Faraday effect as shown in Fig. 6. Recently, (Shragai *et al.*, 2026) managed to experimentally measure the value of phonon Hall viscosity tensor of $\alpha$-$RuCl_3$ based on acoustic Faraday effect.

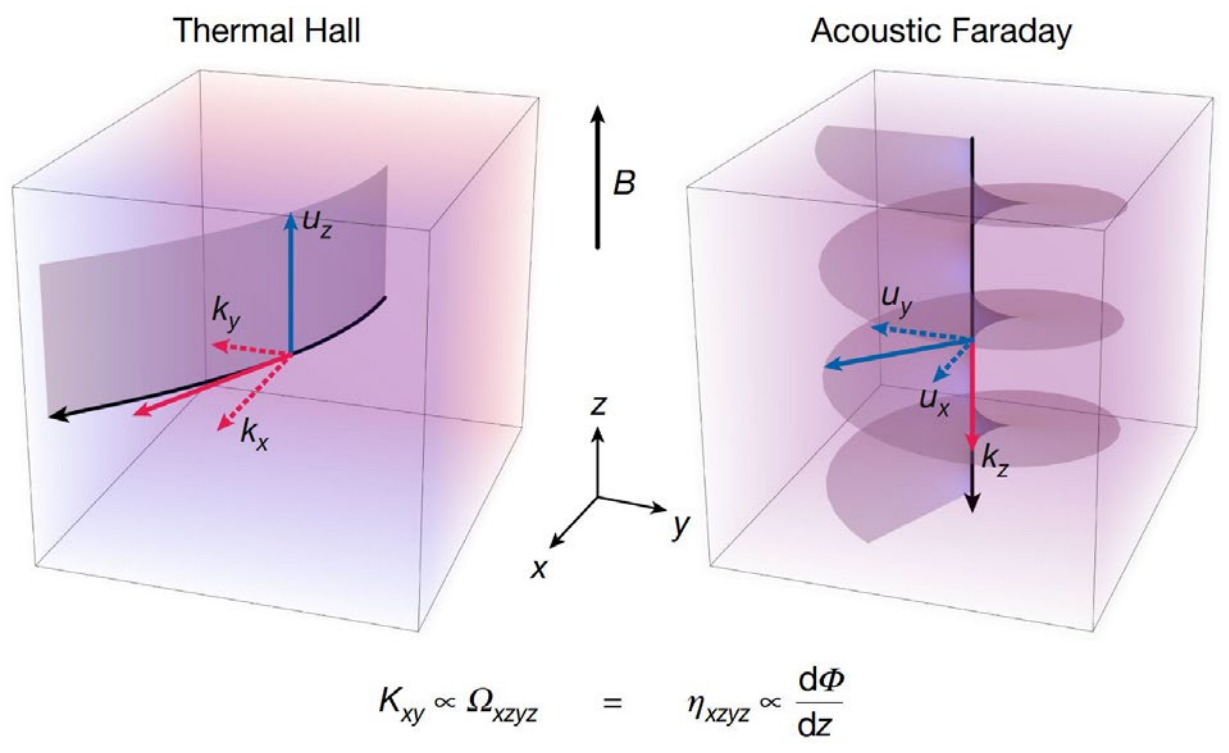

Figure 6 Schematic phonon thermal Hall and acoustic Faraday effects. An externally applied magnetic field breaks the time-reversal symmetry and allows for both effects. Adapted from (Shragai *et al.*, 2026).

## II. CLASSIFICATION OF TOPOLOGICAL PHONON BANDS BASED ON SYMMETRIES

Topological band theory is an extension of traditional band theory that incorporates the principles of topology. It has been widely applied to various condensed matter platforms, including electrons (Bradlyn *et al.*, 2017; Burkov, 2016; Fu *et al.*, 2007; Hasan and Kane, 2010; Lv *et al.*, 2021; Murakami, 2007; Po *et al.*, 2017; Qi and Zhang, 2011; Tang *et al.*, 2019a; Tokura *et al.*, 2019; Vergniory *et al.*, 2019b; Zhang *et al.*, 2019a), phonons (Li *et al.*, 2021b; Liu *et al.*, 2017a, 2018, 2017b; Sheng *et al.*, 2006; Xu *et al.*, 2024c; Zhang *et al.*, 2010, 2018a, 2019b), magnons (McClarty, 2022; Owerre, 2016; Zhang *et al.*, 2013), photons (Lu *et al.*, 2016a, 2014, 2016b; Ozawa *et al.*, 2019), and artificial systems (Huber, 2016a; Imhof *et al.*, 2018; Kane and Lubensky, 2014; Khanikaev and Shvets, 2017; Lu *et al.*, 2015; Luo *et al.*, 2018; Ma *et al.*, 2019b; Nash *et al.*, 2015; Ningyuan *et al.*, 2015; Süsstrunk and Huber, 2015; Wang *et al.*, 2015a; Yang *et al.*, 2015; Zhang *et al.*, 2019d). In this section, we first compare topological phonons with other topological quasiparticle excitations. We then introduce the topological classification of gapped phonon spectra and subsequently extend the framework of topological band theory to gapless phonon excitations, including topological states protected by different symmetries and characterized by distinct topological invariants. Since phonon systems generically preserve time-reversal symmetry ($\mathcal{T}$), our discussion focuses primarily on time-reversal-invariant systems.

### A. Topological Phonons vs. Other Topological Quasiparticles

Topological phonons, grounded in the principles of band theory, exhibit parallels with other topological quasiparticles, particularly those within spinless electronic band structures. They share characteristics such as various types of Weyl points, Dirac points, and nodal lines/rings, along with the emergence of topological surface states due to the "bulk-surface correspondence". In electronic systems without spin-orbit coupling, the spin degree of freedom is decoupled from other degrees of freedom. Therefore, the spin system has the full SU(2) symmetry. After decoupling the spin degree of freedom, we apply the topological band theory for spinless electronic systems. Phonons, being spin-0 bosons, can be treated within the same topological framework as spinless electronic systems. Topological states of phonons can be defined both in the presence and absence of time-reversal symmetry ($\mathcal{T}$). However, we note that breaking $\mathcal{T}$-symmetry in phonon spectra is uncommon and typically requires specific conditions—such as external magnetic fields (Baydin *et al.*, 2022a; Chaudhary *et al.*, 2024a; Cheng *et al.*, 2020b; Juraschek *et al.*, 2022a; Lujan *et al.*, 2024a; Luo *et al.*, 2023b; Nova *et al.*, 2017a; Ren *et al.*, 2021a; Schaack, 1977; Wu *et al.*, 2023a; Xue *et al.*, 2025a), coupling to magnons (Liu *et al.*, 2021f; Ren *et al.*, 2024), or intrinsic magnetic order (Che *et al.*, 2025b; Yang *et al.*, 2025b; Zhang *et al.*, 2026b). Nonetheless, the fundamental distinctions between phonons and electrons could pave the way for the revelation of novel topological phenomena unique to phonons. Thus, topological phonons are endowed with distinctive attributes that set them apart from their counterparts among other topological quasiparticles.

One difference between fermionic electrons and bosonic phonons lies in the robustness of gapless topological states such as nodal-line or nodal-ring states. In electronic systems, these states are often fragile and can be gaped by spin-orbit coupling (SOC), leading to either trivial or topological insulating states. This is, however, not the case for phonons as phonons are spin-0 quasiparticles with orbital degree of freedom, hence precludes SOC effect (Zhang *et al.*, 2019b). Furthermore, in three dimensions, the threefold degeneracy of acoustic phonons at the Γ point, arising from their nature as Goldstone modes, becomes spin-1 Weyl phonons in chiral crystals, where the degeneracy between the two transverse modes is lifted. Phonons can possess additional quantum degrees of freedom, such as pseudospin and angular momentum, which endow them with novel physical properties associated with their topological states (Zhang *et al.*, 2023b, 2018a).

### B. Classification of Topological Phonons based on Intrinsic Symmetries

The "ten-fold way" refers to a universal classification framework in condensed matter physics and quantum field theory, primarily used to categorize the symmetry classes of Hamiltonians, particularly in studies of topological insulators (Ando, 2013; Bernevig, 2013; Fu *et al.*, 2007; Hasan and Kane, 2010; Hasan and Moore, 2011; Khanikaev *et al.*, 2013; Qi and Zhang, 2011; Shen, 2012; Tokura *et al.*, 2019) and topological superconductors (Ando and Fu, 2015; Leijnse and Flensberg, 2012; Sato and Ando, 2017). The classifica-tion is determined by the presence or absence of three fundamental discrete symmetries: time-reversal symmetry ($\mathcal{T}$), particle-hole symmetry ($\mathcal{C}$), and chiral (sublattice) symmetry ($\mathcal{S}$), as defined in Eqs. (45)–(47). Originating from the work of Altland and Zirnbauer (Altland and Zirnbauer, 1997), this scheme establishes that physical systems can be grouped into ten distinct symmetry classes shown in Tab. II, thereby forming the foundation of the modern classification theory of topological phases (Kitaev, 2009; Ryu *et al.*, 2010), and has been extensively applied to gapped phonon spectra.

Mathematically, the ten-fold classification is formulated within the framework of K-theory, which provides a systematic description of gapped Hamiltonians subject to specified symmetry constraints. In this approach, each symmetry class is associated with a corresponding classifying space, while the underlying Clifford algebra structure gives rise to the characteristic Bott periodicity of the classification scheme, as shown in Tab. III. In particular, the topological classification exhibits an eightfold periodicity for the real symmetry classes and a twofold periodicity for the complex classes. This periodic structure originates from the recursive representation properties of Clifford algebras and leads to recurring patterns of topological invariants, such as $\mathbb{Z}$, $\mathbb{Z}_2$, or trivial phases, as functions of spatial dimension.

Physically, the ten-fold classification establishes a direct correspondence between the symmetry and dimensionality of the system. This framework also uncovers the existence of quantized bulk invariants and symmetry-protected gapless boundary states. Therefore, the ten-fold classification table provides a unified theoretical foundation that connects diverse topological phenomena, ranging from the integer quantum Hall effect to topological superconductivity.

Consider a Hamiltonian, $\mathcal{H}(k)$, whose transformations under the symmetry operations $\mathcal{T}$, $\mathcal{C}$, and $\mathcal{S}$ are given by:

$$\mathcal{T}^{-1}H(k)\mathcal{T} = H(-k), \qquad \mathcal{T} = U_T\mathcal{K}, \quad U_T U_T^* = \pm 1, \tag{45}$$

$$\mathcal{C}^{-1}H(k)\mathcal{C} = -H(-k), \quad \mathcal{C} = U_C\mathcal{K}, \quad U_C U_C^* = \pm 1, \tag{46}$$

$$\mathcal{S}^{-1}H(k)\mathcal{S} = -H(k), \qquad \mathcal{S} = U_S, \qquad U_S^2 = 1. \tag{47}$$

$U_T$ and $U_C$ are unitary matrix, and $\mathcal{K}$ is the complex conjugate operator.

In a system exhibiting $\mathcal{T}$, $\mathcal{T}^2 = -1$ or $+1$ are respectively corresponding to half-integer spin systems (including spinful electronic systems) and integer-spin systems (e.g phonons and photons), respectively. When the time-reversal symmetry is broken, phenomena like the quantum anomalous Hall effect (QAHE) emerge (Chang *et al.*, 2023, 2013; Deng *et al.*, 2020; Haldane, 1988; Liu *et al.*, 2008, 2016), and this case is symbolically expressed as $\mathcal{T}^2 = 0$ (where "0" signifies the absence of time-reversal symmetry). Similarly, particle-hole symmetry ($\mathcal{C}$) is characterized by an anti-unitary operator with $\mathcal{C}^2 = \pm 1$, or $\mathcal{C}^2 = 0$ in the absence of particle-hole symmetry, yielding three possible cases. Chiral symmetry (or sublattice symmetry in certain contexts) can be interpreted as a combination of $\mathcal{T}$ and $\mathcal{C}$ ($\mathcal{S} = \mathcal{TC}$), which also results in three possibilities, with $\mathcal{S}^2 = \pm 1, 0$. Thus, we have identified 9 distinct types of systems by considering the possible values of $\mathcal{T}^2$, $\mathcal{C}^2$, $\mathcal{S}^2$, each of which can take values of $-1$, $1$ or $0$. This naturally raises the question: what constitutes the tenth kind in the "ten-fold way"? The answer lies in systems that lack both $\mathcal{T}$ and $\mathcal{C}$ individually, yet still exhibit symmetry under their combined operation, $\mathcal{S} = \mathcal{TC}$. In such systems, invariance is preserved when particles and holes are interchanged and time is reversed, even though neither symmetry exists alone. This unique scenario completes the ten-fold classification.

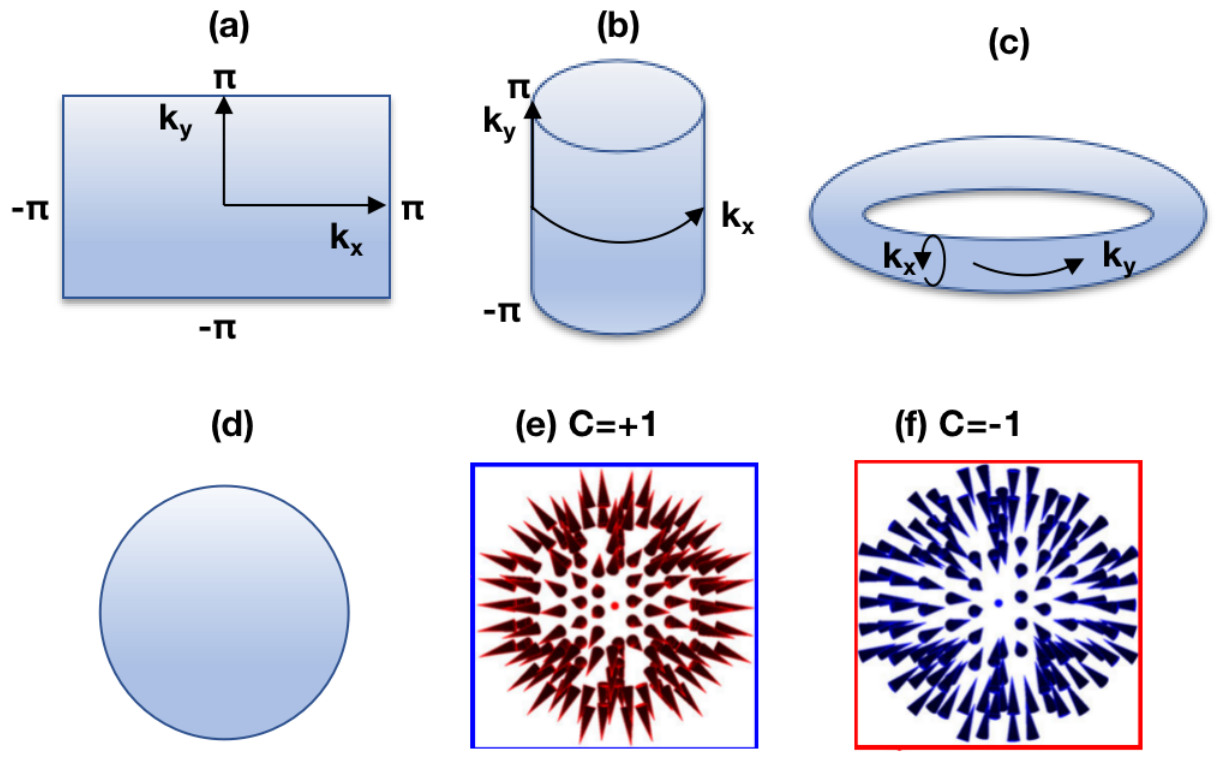

Figure 7 Illustration of the integral curved surface for the Chern number. (a) Depicts a typical Brillouin zone for a two-dimensional system. (b) and (c) Show two configurations of integrated surfaces that are topologically equivalent to the BZ in (a). This equivalence arises due to the periodic boundary conditions of the Brillouin zone, allowing for alternative representations of the integration domain. (d) Illustrates the integrated surface used to calculate the Chern number for Weyl phonons. This surface is typically a sphere enclosing the Weyl phonon, capturing the topological charge associated with it. (e) and (f) display the Berry curvature distributions for Weyl points with Chern numbers $C = +1$ and $-1$, respectively. The integration of the Berry curvature over the enclosing surface yields the Chern number, which characterizes the topological nature of the Weyl phonon. (e) and (f) are adapted from Ref. (Weng *et al.*, 2015)

Up to this point, we have established ten distinct topological classifications based on three symmetries, as summarized in Tab. II. Here, the notation "$\pm 1$" indicates the presence of a symmetry with $\mathcal{T}^2$, $\mathcal{C}^2$, $\mathcal{S}^2 = +1, -1$, while "0" signifies the absence of the corresponding symmetry. The topological classification is not only determined by these symmetries, but also the dimension of the system. Table III shows the "ten-fold way" table, which incorporates the interplay between different symmetries and system dimensions. Within given symmetries and dimensions, topological classes can be distinguished by a certain Abelian group, such as $\mathbb{Z}$, $\mathbb{Z}_2$, or topologically trivial case (labeled by "0"). A closer examination of the table reveals a periodic pattern that repeats every 8 dimensions. Specifically, the behavior of systems in dimension $d = 9$ mirrors that of $d = 1$, and this periodicity continues as the dimensionality increases further.

Table II "Cartan label" for "ten-fold way" classes of single-particle Hamiltonians $H$, based on time-reversal symmetry ($\mathcal{T}$), particle-hole symmetry ($\mathcal{C}$) and chiral symmetry ($\mathcal{S} = \mathcal{T}\mathcal{C}$). Adapted from Ref. (Ryu *et al.*, 2010).

| Cartan label | $\mathcal{T}$ | $\mathcal{C}$ | $\mathcal{S}$ | Hamiltonian | $G/H$ (ferm. NL$\sigma$M) |
|---|---|---|---|---|---|
| A (unitary) | 0 | 0 | 0 | $U(N)$ | $U(2n)/U(n) \times U(n)$ |
| AI (orthogonal) | +1 | 0 | 0 | $U(N)/O(N)$ | $Sp(2n)/Sp(n) \times Sp(n)$ |
| AII (symplectic) | -1 | 0 | 0 | $U(2N)/Sp(2N)$ | $O(2n)/O(n) \times O(n)$ |
| AIII (ch. unit.) | 0 | 0 | 1 | $U(N+M)/U(N) \times U(M)$ | $U(n)$ |
| BDI (ch. orth.) | +1 | +1 | 1 | $O(N+M)/O(N) \times O(M)$ | $U(2n)/Sp(2n)$ |
| CII (ch. sympl.) | -1 | -1 | 1 | $Sp(N+M)/Sp(N) \times Sp(M)$ | $U(2n)/O(2n)$ |
| D (BdG) | 0 | +1 | 0 | $SO(2N)$ | $O(2n)/U(n)$ |
| C (BdG) | 0 | -1 | 0 | $Sp(2N)$ | $Sp(2n)/U(n)$ |
| DIII (BdG) | -1 | +1 | 1 | $SO(2N)/U(N)$ | $O(2n)$ |
| CI (BdG) | +1 | -1 | 1 | $Sp(2N)/U(N)$ | $Sp(2n)$ |

Table III "Ten-fold way" table for topological insulators and superconductors as a function of spatial dimension $d$ and three discrete symmetries, such that the symmetry classes are arranged to reveal a periodic pattern in spatial dimensionality. The "Cartan label" (first column) is shown in Fig. II. These ten classes are subsequently categorized into two groups: the complex and real ones, mainly contingent upon the complexity of the Hamiltonian. $\mathbb{Z}$ and $\mathbb{Z}_2$ represent the distinct phases within a specific symmetry class of topological insulators/superconductors, distinguished by an integer invariant. Within a specific $\mathbb{Z}$ group, the topological invariant is isomorphic to the $\mathbb{Z}$-type Abelian group and can assume any integer value. For the $\mathbb{Z}_2$ class, the topological invariant is isomorphic to the $\mathbb{Z}_2$-type Abelian group, such that the topological invariant can take two classes of values, the odd ones and the even ones. "0" means that no topological insulator/superconductor is present, indicating that all quantum ground states are topologically trivial. Adapted from Ref. (Ryu *et al.*, 2010)

| | $d$ | | | | | | | | | | | | |
|---|---|---|---|---|---|---|---|---|---|---|---|---|---|
| Cartan | 0 | 1 | 2 | 3 | 4 | 5 | 6 | 7 | 8 | 9 | 10 | 11 | … |
| *Complex case:* | | | | | | | | | | | | | |
| A | $\mathbb{Z}$ | 0 | $\mathbb{Z}$ | 0 | $\mathbb{Z}$ | 0 | $\mathbb{Z}$ | 0 | $\mathbb{Z}$ | 0 | $\mathbb{Z}$ | 0 | … |
| AIII | 0 | $\mathbb{Z}$ | 0 | $\mathbb{Z}$ | 0 | $\mathbb{Z}$ | 0 | $\mathbb{Z}$ | 0 | $\mathbb{Z}$ | 0 | $\mathbb{Z}$ | … |
| *Real case:* | | | | | | | | | | | | | |
| AI | $\mathbb{Z}$ | 0 | 0 | 0 | $2\mathbb{Z}$ | 0 | $\mathbb{Z}_2$ | $\mathbb{Z}_2$ | $\mathbb{Z}$ | 0 | 0 | 0 | … |
| BDI | $\mathbb{Z}_2$ | $\mathbb{Z}$ | 0 | 0 | 0 | 0 | $2\mathbb{Z}$ | 0 | $\mathbb{Z}_2$ | $\mathbb{Z}_2$ | $\mathbb{Z}$ | 0 | 0 |
| D | $\mathbb{Z}_2$ | $\mathbb{Z}_2$ | $\mathbb{Z}$ | 0 | 0 | 0 | $2\mathbb{Z}$ | 0 | $\mathbb{Z}_2$ | $\mathbb{Z}_2$ | $\mathbb{Z}$ | 0 | … |
| DIII | 0 | $\mathbb{Z}_2$ | $\mathbb{Z}_2$ | $\mathbb{Z}$ | 0 | 0 | 0 | $2\mathbb{Z}$ | 0 | $\mathbb{Z}_2$ | $\mathbb{Z}_2$ | $\mathbb{Z}$ | … |
| AII | $2\mathbb{Z}$ | 0 | $\mathbb{Z}_2$ | $\mathbb{Z}_2$ | $\mathbb{Z}$ | 0 | 0 | 0 | $2\mathbb{Z}$ | 0 | $\mathbb{Z}_2$ | $\mathbb{Z}_2$ | … |
| CII | 0 | $2\mathbb{Z}$ | 0 | $\mathbb{Z}_2$ | $\mathbb{Z}_2$ | $\mathbb{Z}$ | 0 | 0 | 0 | $2\mathbb{Z}$ | 0 | $\mathbb{Z}_2$ | … |
| C | 0 | 0 | $2\mathbb{Z}$ | 0 | $\mathbb{Z}_2$ | $\mathbb{Z}_2$ | $\mathbb{Z}$ | 0 | 0 | 0 | $2\mathbb{Z}$ | 0 | … |
| CI | 0 | 0 | 0 | $2\mathbb{Z}$ | 0 | $\mathbb{Z}_2$ | $\mathbb{Z}_2$ | $\mathbb{Z}$ | 0 | 0 | 0 | $2\mathbb{Z}$ | … |

To gain a deeper understanding of this table, let us consider the topological class for A class with $d = 2$ as an example. This classification falls under the $\mathbb{Z}$ type, which corresponds to systems such as the Chern insulator case (or the QAHE). The topological invariant in this case is defined as the Chern number (Avron *et al.*, 1983; Hatsugai, 1993; Thouless *et al.*, 1982b):

$$C = \frac{1}{2\pi} \int_{\mathrm{BZ}} \Omega(\boldsymbol{k})\, \mathrm{d}^2 k, \tag{48}$$

where $C$ is the Chern number, $\Omega(\boldsymbol{k})$ is the Berry curvature, and the integral is taken over the whole Brillouin zone (BZ) in 2D systems, as illustrated in Fig. 7 (a). The Chern number can take arbitrary integer values, reflecting the fact that the topological phase may vary across different systems. This variability underscores the richness of topological phases in gapped systems. The classification scheme based on such topological invariants is crucial for understanding the diverse phases of topological matter, enabling the prediction and characterization of novel quantum states.

As discussed earlier in this section, the "ten-fold way" classification can also be extended to gapped phonon spectra in solids. Since phonons are bosons, $\mathcal{T}^2$ can only take values of 0 and 1. As a result, topological classifications for gapped phonon systems are restricted to the classes of A, AI, AIII, BDI, D, C, and CI. In condensed matter physics, the $\mathcal{C}$ and $\mathcal{S}$ are often not well-defined. Therefore, $\mathcal{T}$ plays the central role in classifying topological gapped states for phonons, particularly in the standard Altland-Zirnbauer classes A and AI. On the other hand, focusing only on $\mathcal{T}$ alone can overlook potential topological phases. For instance, if we study a specific phonon branch, one can construct an effective low-energy model where $\mathcal{C}$ or $\mathcal{S}$ symmetry is meaningfully defined and preserved. These symmetries, in turn, enable the emergence of novel topological phases not captured by the $\mathcal{T}$-only classification. Furthermore, when crystalline symmetries (e.g., rotational, mirror, or inversion) are considered, the topological classification of phonons becomes significantly richer (Fu, 2011; Kruthoff *et al.*, 2017; Po *et al.*, 2017; Shiozaki and Sato, 2014; Song *et al.*, 2018). These additional symmetries not only expand the range of possible topological phases but also provide more concrete signatures for their identification and diagnosis in real materials.

To realize topological phases in gapped phonon spectra, specific symmetry conditions must be met. For instance, achieving a Class A topological phase (e.g., characterized by the Chern number, a $\mathbb{Z}$ invariant in two dimensions) requires breaking $\mathcal{T}$ in the phonon system. This can be accomplished through various mechanisms, such as by external magnetic fields (Baydin *et al.*, 2022a; Chaudhary *et al.*, 2024a; Cheng *et al.*, 2020b; Juraschek *et al.*, 2022a; Lujan *et al.*, 2024a; Luo *et al.*, 2023b; Nova *et al.*, 2017a; Ren *et al.*, 2021a; Schaack, 1977; Wu *et al.*, 2023a; Xue *et al.*, 2025a), coupling to magnons (Liu *et al.*, 2021f; Ren *et al.*, 2024), or intrinsic magnetic order (Che *et al.*, 2025b; Yang *et al.*, 2025b; Zhang *et al.*, 2026b). Experimental verification of such phases in phononic systems remains an active and promising frontier of research.

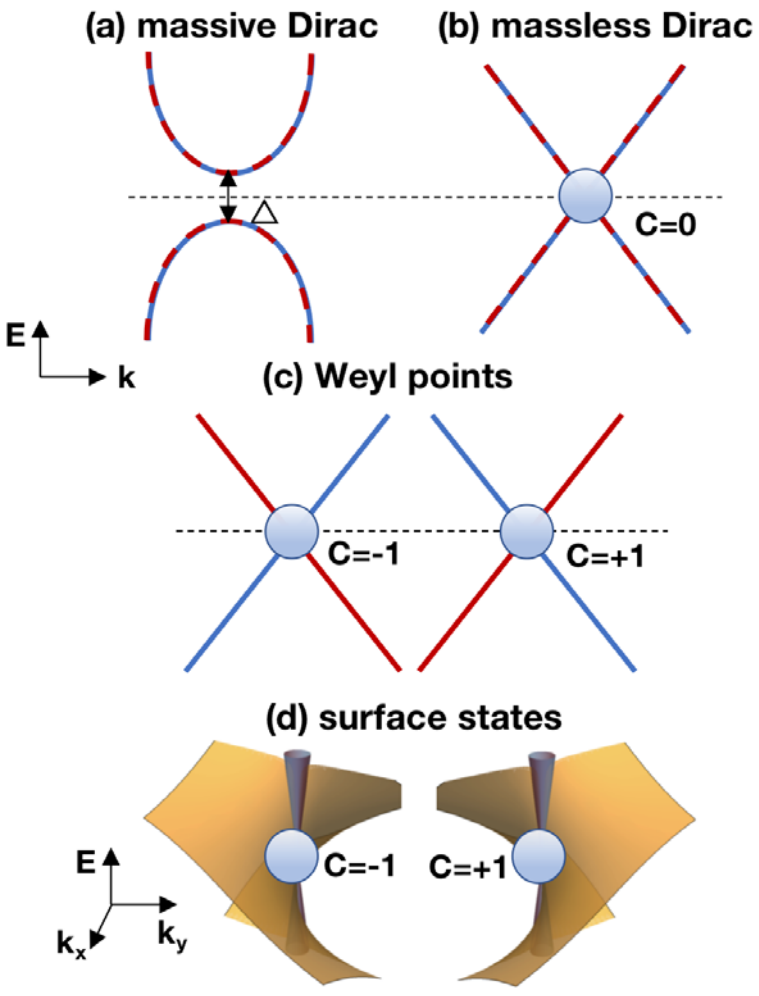

Figure 8 (a) and (b) depict the massive and massless Dirac phonon spectra, respectively. In both cases, the Chern number is zero, indicating the absence of nontrivial topological charge. (c) Weyl phonons with opposite Chern number and chirality. The massless Dirac pononon is composed of two Weyl phonons with $C = \pm 1$. (d) Topological surface states of Weyl phonons are in helicoid shapes. These surface states share the same chirality as the Chern numbers of the corresponding Weyl phonons, highlighting the direct connection between bulk topology and surface topology.

### C. Generalized to Crystalline Symmetries

In the preceding subsections, we reviewed the classification of topological phonons, together with their associated topological invariants and topological surface states. Mathematically, the framework of topological band theory is rooted in the topology of the ground-state wave functions, which is characterized by quantized topological numbers known as topological invariants. These invariants provide a fundamental means of identifying and distinguishing topological phases, thereby serving as essential diagnostic tools for their classification. Since symmetries play a central role in determining topological classifications, the emergence of additional symmetries can, in turn, give rise to new classes of topological states. Consequently, symmetry considerations are not only fundamental to the classification of topological phases but also provide powerful simplifications for the diagnosis and evaluation of topological invariants.

In 2007, (Fu *et al.*, 2007) proposed that the calculation for topological invariant of $\mathbb{Z}_2$ topological insulators can be simplified in systems with both inversion ($\mathcal{P}$) and time-reversal ($\mathcal{T}$) symmetries. This method involves examining the parity of the electron occupation at time-reversal invariant momenta (TRIM). A significant contribution of this work was the illustration of how the utilization of symmetries can simplify the computation of topological invariant, thereby enhancing the understanding and analysis within the field. In conjunction with the "ten-fold way" theory discussed in Sec. II.B, it becomes evident that symmetries play a pivotal role in simplifying the classification and diagnosis of topological states. This insight laid the groundwork for subsequent developments in symmetry-based indicators (Po *et al.*, 2017) and the independent advancement of topological quantum chemistry (Bradlyn *et al.*, 2017).

The symmetry-based indicator theory constitutes a conceptual framework that utilizes symmetries to simplify the calculation and classification of topological states across various symmetry classes, particularly within the context of the 230 space groups, and accounting for the presence or absence of time-reversal ($\mathcal{T}$). The key idea of this theory is the construction of a gapped band structure space, denoted as $\{\mathrm{BS}\}$, for a given specified band system within the realm of 230 space groups. This space $\{\mathrm{BS}\}$ is constructed from the symmetry attributes of the occupied bands at each high-symmetry momenta in the Brillouin zone:

$$\mathrm{BS} = \sum_{i=1}^{d_{\mathrm{BS}}} m_i \boldsymbol{b}_i. \tag{49}$$

Here, $d_{\mathrm{BS}}$ represents a positive integer, $\boldsymbol{b}_i$ serves as the generator of the band insulators, $m_i$ is an integer and is uniquely determined once the basis is chosen. At the same time, an atomic insulator space, denoted as $\{\mathrm{AI}\}$, is also constructed under the 230 space groups. These $\{\mathrm{AI}\}$ are exemplars of band insulators that arise from an initial symmetrical configuration of atomic sites in real space, with the subsequent full occupation of a defined set of orbitals at each atomic position. Consequently, the aggregation of arbitrary sets of atomic insulators will invariably yield another atomic insulator space, as expressed by:

$$\{\mathrm{AI}\} \simeq \mathbb{Z}^{d_{\mathrm{AI}}} = \left\{ \sum_{i=1}^{d_{\mathrm{AI}}} m_i \boldsymbol{a}_i : m_i \in \mathbb{Z} \right\}, \tag{50}$$

Table IV The nontrivial symmetry-based indicator groups, denoted as $X_{BS}$, apply to phonon systems that preserve time-reversal symmetry. $X_{BS}$ represents the quotient group formed by the set of "gapped" band structures and the set of "atomic insulators," a concept originally derived from electronic systems. This notion can be extended to encompass all physical systems described by band theory. Adapted from Ref. (Po *et al.*, 2017).

| $X_{BS}$ | Space groups |
|---|---|
| $\mathbb{Z}_2$ | 3, 11, 14, 27, 37, 48, 49, 50, 52, 53, 54, 56, 58, 60, 66, 68, 70, 75, 77, 82, 85, 86, 88, 103, 124, 128, 130, 162, 163, 164, 165, 166, 167, 168, 171, 172, 176, 184, 192, 201, 203 |
| $\mathbb{Z}_2 \times \mathbb{Z}_2$ | 12, 13, 15, 81, 84, 87 |
| $\mathbb{Z}_2 \times \mathbb{Z}_4$ | 147, 148 |
| $\mathbb{Z}_2 \times \mathbb{Z}_2 \times \mathbb{Z}_2$ | 10, 83, 175 |
| $\mathbb{Z}_2 \times \mathbb{Z}_2 \times \mathbb{Z}_2 \times \mathbb{Z}_4$ | 2 |

where $\boldsymbol{a}_i$ is the generator of the atomic insulators. Hereafter, the quotient group for $\{\text{BS}\}$ and $\{\text{AI}\}$ can be straightforward to evaluate:

$$X_{\text{BS}} = \frac{\{\text{BS}\}}{\{\text{AI}\}}. \tag{51}$$

An element within $X_{\text{BS}}$ denotes a comprehensive collection of band structures that are differentiated solely by the addition of an atomic insulator in each instance. By harnessing the various symmetries of a system, this theoretical framework provides a methodical strategy for identifying and categorizing topological states.

Table IV displays the nontrivial symmetry-based indicator groups $X_{\text{BS}}$ for the 230 space groups in the context of phonon spectra. The notation $\mathbb{Z}_n \times \mathbb{Z}_m$ signifies Abelian groups that are isomorphic to the respective quotient groups (Po *et al.*, 2017). (Song *et al.*, 2018; Zhang *et al.*, 2020a; Zhang and Murakami, 2021) linked the symmetry-based indicator groups to the specific topological invariants based on rigorous derivations, providing both the general proof and the analytical expression to quantify these indicators. Specifically, a nontrivial symmetry-based indicator group, as illustrated in Tab. IV, may correspond to multiple distinct topological invariants. For example, a $\mathbb{Z}_2$ indicator can be associated with either a Berry phase or a $\mathbb{Z}_2$ monopole charge $Q$, reflecting the intrinsically one-to-many correspondence characteristic of symmetry-indicator theories. In the subsequent sections, we will use space group #2, #3 and #82 as case studies to illustrate the implications of the symmetry-based indicators in diagnosing topological phonon.

### 1. Symmetry-based Indicators for Nodal-line Phonons

The symmetry-based indicator group for phonon systems with space group #2 is $\mathbb{Z}_2\times\mathbb{Z}_2\times\mathbb{Z}_2\times\mathbb{Z}_4$. This group indicates a topological semimetal that exhibits nodal-line or nodal-ring band crossings in the Brillouin zone when not all of the four $\mathbb{Z}_n$ components are trivial (to be zero). The explicit formula for the symmetry-based indicator group of space group #2 is provided below:

$$z_{2,1} \equiv \sum_{\boldsymbol{K}\in\text{TRIM at }\{k_1=\pi\}} \frac{N_-(\boldsymbol{K}) - N_+(\boldsymbol{K})}{2} \mod 2, \tag{52}$$

$$z_{2,2} \equiv \sum_{\boldsymbol{K}\in\text{TRIM at }\{k_2=\pi\}} \frac{N_-(\boldsymbol{K}) - N_+(\boldsymbol{K})}{2} \mod 2, \tag{53}$$

$$z_{2,3} \equiv \sum_{\boldsymbol{K}\in\text{TRIM at }\{k_3=\pi\}} \frac{N_-(\boldsymbol{K}) - N_+(\boldsymbol{K})}{2} \mod 2, \tag{54}$$

$$z_4 \equiv \sum_{\boldsymbol{K}\in\text{TRIM}} \frac{N_-(\boldsymbol{K}) - N_+(\boldsymbol{K})}{2} \mod 4, \tag{55}$$

where $N_{\pm}(\boldsymbol{K})$ represents the count of bands that exhibit even/odd parity under inversion symmetry at the four time-reversal-invariant momenta (TRIM) located on the $k_{i\in\{1,2,3\}} = \pi$ plane. The quantities $z_{2,1}z_{2,2}z_{2,3}z_4$ serve as the indicators for the symmetry-based indicator group $\mathbb{Z}_2 \times \mathbb{Z}_2 \times \mathbb{Z}_2 \times \mathbb{Z}_4$. The first three $\mathbb{Z}_2$ indicate the orientation and the location of the node-line/ring band crossings. Specifically, nodal lines are oriented along the reciprocal lattice direction $k_i$ when $z_{2,i} = 1$, or they are situated around $k_i = \pi$. The final $z_4$ index indicates the count of nodal line/ring band crossings. An even number of nodal lines or rings within the Brillouin zone is signified by $z_4 = 0, 2$, while an odd number is indicated by $z_4 = 1, 3$. Combined the information from these four symmetry-based indicators, one can ascertain the entire configuration of nodal lines/rings throughout the BZ (Zhang *et al.*, 2020a; Zhang and Murakami, 2021; Zhang *et al.*, 2022c, 2019b,b).

For instance, when the values for the symmetry-based indicator group $\mathbb{Z}_2 \times \mathbb{Z}_2 \times \mathbb{Z}_2 \times \mathbb{Z}_4$ are $z_{2,1}z_{2,2}z_{2,3}z_4 = (0001)$, this indicates that an odd number of nodal lines/rings will be present around the $\Gamma$ point, corresponding to the most basic configuration featuring a single nodal ring encircling $\Gamma$, as depicted in Fig. 9. For the symmetry-based indicator values of $\mathbb{Z}_2 \times \mathbb{Z}_2 \times \mathbb{Z}_2 \times \mathbb{Z}_4$ given by $z_{2,1}z_{2,2}z_{2,3}z_4 = (1002)$, an even number of nodal lines/rings will be aligned along the reciprocal lattice vector $k_1$, which is represented by the two nodal lines. For the symmetry-based indicator values of $\mathbb{Z}_2 \times \mathbb{Z}_2 \times \mathbb{Z}_2 \times \mathbb{Z}_4$ specified as $z_{2,1}z_{2,2}z_{2,3}z_4 = (0002)$, an even number of nodal lines/rings will be found in the vicinity of the $\Gamma$ point, exemplified by the two nodal rings. It is important to note that nodal lines/rings characterized by $z_{2,1}z_{2,2}z_{2,3}z_4 = (0002)$ possess a nonzero $\mathbb{Z}_2$ monopole charge, and such features can only emerge or disappear in conjunction, that is, in pairs.

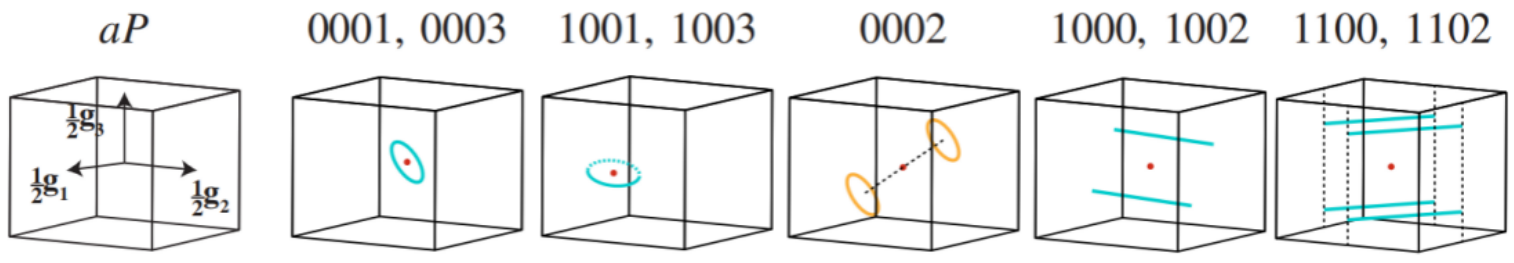

Figure 9 In space group #2, which only preserves the inversion symmetry, the configuration of topological phonon bands is associated with the symmetry-based indicators. Yellow nodal rings with $z_{2,1}z_{2,2}z_{2,3}z_4 = (0002)$ carry a nonzero $\mathbb{Z}_2$ monopole charge. Adapted from Ref. (Song *et al.*, 2018).

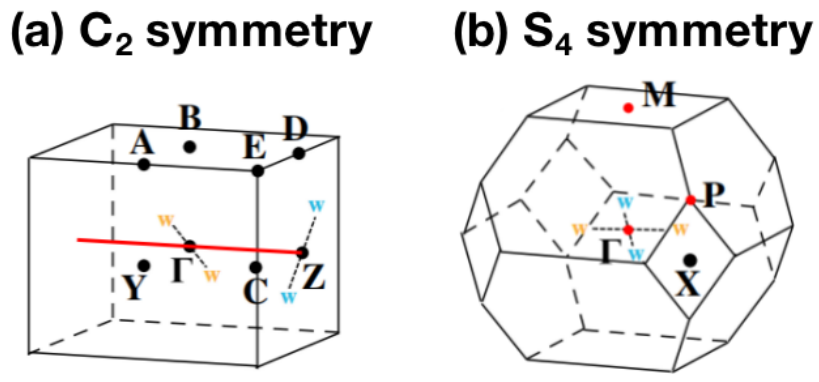

Figure 10 The symmetry-based indicator theory is employed to identify Weyl phonons in space groups #3 with $C_2$ symmetry and #82 with $S_4$ symmetry. The blue and yellow markers denote the positions of Weyl phonons with distinct chirality. This theoretical framework enables the diagnosis of Weyl phonons at the $k = 0$ and $k = \pi$ planes in $C_2$-symmetric systems, and at the $k = 0$ plane in $S_4$-symmetric systems. Adapted from Ref. (Song *et al.*, 2018).

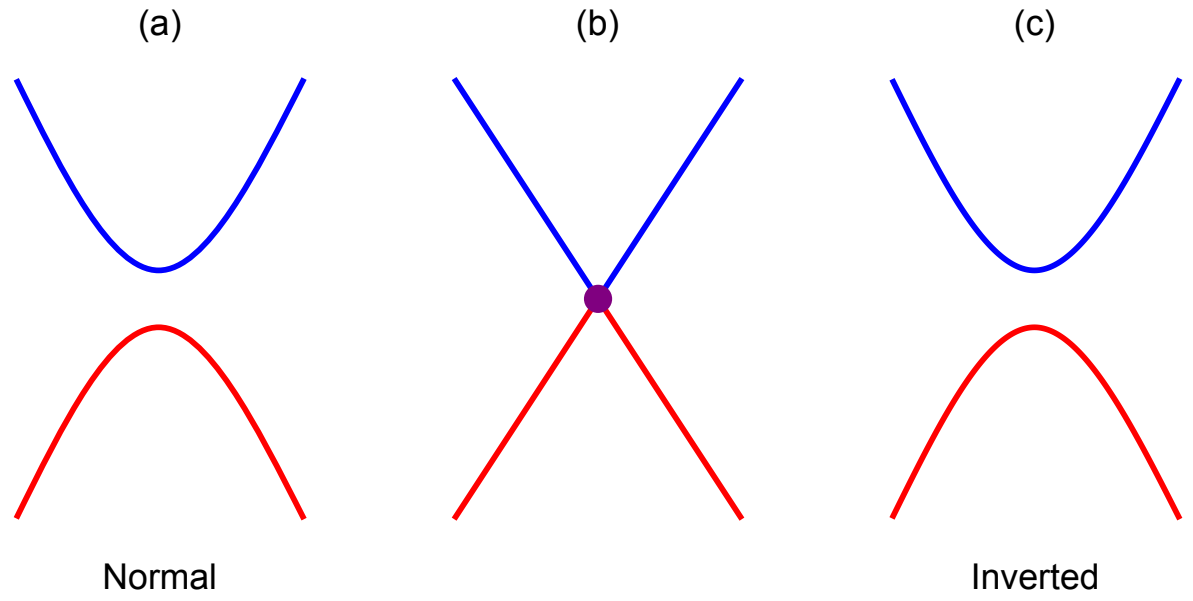

Figure 11 Illustration of the topological band structure featuring different band order, with panels (a) through (c) depicting the conventional, semimetallic, and inverted band configurations, respectively. The inverted band order signifies a nontrivial topological index for the bulk Bloch wave functions.

### 2. Symmetry-based Indicators for Weyl Phonons

In non-centrosymmetric systems, the emergence of Weyl phonons also can be diagnosed by symmetry-based indicators. For example, space group #3 only includes a $C_2$ symmetry. The indicator $\mathbb{Z}_2$ for this space group is linked to the product of the $C_2$ eigenvalues at all TRIMs, which serves to diagnose the occurrence of Weyl phonons within a specific plane. $\mathbb{Z}_2$ is described by the topological invariant $\alpha_2$, as presented below:

$$\alpha_2 \equiv \sum_{K_i=0,\boldsymbol{K}\in\mathrm{TRIM}} N_{\zeta=-1}(\boldsymbol{K}) \mod 2, \tag{56}$$

where $i$ denotes the rotational axis for the $C_2$ operation, and $\zeta$ represents the eigenvalue corresponding to the $C_2$ operator. If $\alpha_2 = 1$, the Berry phase accumulated along a loop that encircles half of the Brillouin zone on the $k_i = 0$ plane is $\pi$. This condition gives rise to 1 modulo 2 Weyl phonons on the half of the $k_i = 0$ plane and 2 modulo 4 Weyl phonons across the entire $k_i = 0$ plane, as depicted in Fig. 10 (a). Because two Weyl phonons on the $k_i = 0$ plane share the same chirality, the conservation of monopole charge necessitates the existence of two additional Weyl phonons with opposite chirality on the $k_i = \pi$ plane.

For space group #82, which exhibits $S_4$ symmetry, symmetry-based indicator $\mathbb{Z}_2$ can be established as follows:

$$\omega_2^0 = N_{\xi=-1}(\Gamma)+N_{\xi=-1}(M)+N_{\zeta=-1}(X) \mod 2, \tag{57}$$

$$\omega_2^\pi = N_{\xi=-1}(Z)+N_{\xi=-1}(A)+N_{\zeta=-1}(R) \mod 2. \tag{58}$$

$\xi$ is the eigenvalue of the $C_4$ operator, the quantities $\omega_2^0$ and $\omega_2^\pi$ represent the counts of Weyl phonons located at the $k_i = 0$ and $k_i = \pi$ planes, respectively. Since space group #82 constitutes a body-centered lattice, $\omega_2^\pi$ is null. Consequently, if $\omega_2^0$ equals 1, it signifies the presence of 4 modulo 8 Weyl phonons on the $k_i = 0$ plane. This is illustrated in Fig. 10 (b), where the blue and yellow letters "w" denote the locations of Weyl phonons with distinct chirality (Song *et al.*, 2018; Zhang *et al.*, 2020a; Zhang and Murakami, 2021).

## III. TOPOLOGICAL PHONONIC MODELS

The band topology is closely related to the concept of geometric phase, first introduced by M. Berry in the context of quantum adiabatic evolution within arbitrary parameter space (Berry, 1984a). In solid-state systems, the Bloch wave vector $\boldsymbol{k}$ offers a natural quantum adiabatic parameter space, and the geometric phase based on $\boldsymbol{k}$ was first introduced by J. Zak (Zak, 1989). Two different kinds of geometric phases can be defined according to different physical dimensions: (i) One is the line integral of Berry connection $\theta_C = \oint_C \mathrm{d}\boldsymbol{k} \cdot \mathbf{A}(\boldsymbol{k})$, for any closed 1D loop $C$ within the Brillouin zone; (ii) The other is the surface integral of the Berry curvature $\theta_S = \iint_S \mathrm{d}\boldsymbol{S} \cdot \boldsymbol{\Omega}(\boldsymbol{k})$ for any 2D surface $S$. $\mathbf{A}(\boldsymbol{k}) = i\langle u_{\boldsymbol{k}} | \boldsymbol{\nabla}_{\boldsymbol{k}} u_{\boldsymbol{k}} \rangle$ and $\boldsymbol{\Omega}(\boldsymbol{k}) = \boldsymbol{\nabla}_{\boldsymbol{k}} \times \mathbf{A}$ are the Berry connection and curvature, respectively (with $|u_{\boldsymbol{k}}\rangle$ being the cell-periodic part of Bloch wave function). The former can be defined in 1D-3D but the latter exist only in 2D and 3D because the cross derivative is ill-defined in 1D.

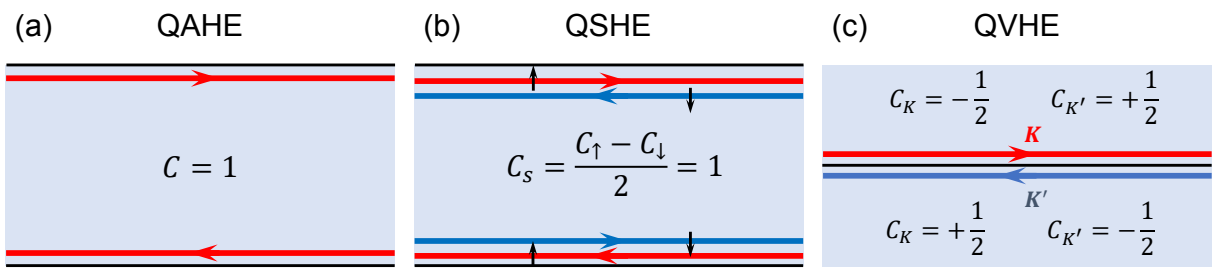

Figure 12 Schematics of various topological states of phonon in 2D lattices. (a)-(c) depict the phononic quantum anomalous Hall effect (QAHE), quantum spin Hall effect (QSHE) and quantum valley Hall effect (QVHE) characterized by nonzero Chern number, spin Chern number, and valley Chern number, respectively.

In 1D systems, the integration of the Berry connection over the entire Brillouin zone yields a topological invariant known as the Zak phase (Xiao *et al.*, 2010; Zak, 1989):

$$\theta_Z = i \int_{-\pi/a}^{\pi/a} \mathrm{d}k \left\langle u_k \middle| \frac{\partial u_k}{\partial k} \right\rangle, \tag{59}$$

where $a$ is the lattice constant. The Zak phase, denoted as $\theta_Z$, is intricately linked to the electronic polarization, forming the cornerstone of the modern theory of electric charge polarization (King-Smith and Vanderbilt, 1993). Due to its oddness under inversion symmetry $\mathcal{P}$, the Zak phase assumes only two possible values, 0 or $\pi$, within systems possessing $\mathcal{P}$ symmetry. A $\theta_Z$ value of $\pi$ (or $\theta_Z = 0$) signifies an inverted (or normal) band order, respectively (confer Fig. 11).

In 2D systems, the integration of Berry curvature over the 2D BZ defines the topological Chern number (Hasan and Kane, 2010; Qi and Zhang, 2011; Thouless *et al.*, 1982a)

$$C = \frac{1}{2\pi} \iint \mathrm{d}^2\boldsymbol{k}\, \Omega_z(\boldsymbol{k}) \tag{60}$$

which describes the quantum Hall (QH) family of topological phases of matter as shown in Fig. 12. The phononic quantum (anomalous/valley/spin) Hall like states are representative gapped states of topological phonons in 2D lattices as shown in Fig. 12. The quantum (anomalous) Hall effect (QAHE) support one-way chiral edge states, distinguished by a nonzero phonon Chern number $C$ within the bulk 2D phonon band structure. The chiral edge states are immune to back scattering thus providing robust phonon transport channels. The quantum spin Hall effect (QSHE) can be viewed as a pair of quantum (anomalous) Hall states with helical (pseudo-)spin-polarized edge states. As the phonon is a spin-zero quasi-particle, a pseudospin degree of freedoms refers to any kind of double band degeneracy which can be effectively characterized by $2 \times 2$ Pauli matrices. By integrating the Berry curvature over a subset of the BZ, valley Chern numbers can be defined, which may take on non-integer values. Consequently, the valley Chern number alone does not ensure the presence of topological edge states at bare open edges. However, for an interface across which the change of the valley Chern number is a nonzero integer, the valley-polarized interface states will appear, which is called quantum valley Hall effect (QVHE).

In this section, we will introduce specific models that are used to realize topological phononic states in 1D and 2D systems. Possible materials realizations are discussed in the end. The theoretical 1D and 2D models described in this section equally apply to crystalline solids and some metamaterials which have been theoretically proposed (Jin *et al.*, 2018b; Kane and Lubensky, 2014; Li *et al.*, 2020a) and experimentally confirmed (Li *et al.*, 2023a; Lu *et al.*, 2018; Xiao *et al.*, 2015).

### A. Model in 1D: Su-Schrieffer-Heeger Model

The phononic Su-Schrieffer-Heeger (SSH) model is a prototypical example of a topological band structure that can be realized in a diatomic chain with alternating nearest-neighbor force constants, as depicted in Fig. 13. Assuming each atomic site has only one vibrational degree of freedom, the dynamical matrix for this system is given by:

$$D(k) = \frac{1}{m} \begin{pmatrix} K_1 + K_2 & -K_1 - K_2 e^{ika} \\ -K_1 - K_2 e^{-ika} & K_1 + K_2 \end{pmatrix}, \tag{61}$$

with $K_1$ and $K_2$ being the intra-cell and inter-cell force constants, respectively, and $m$ is the atomic mass. $k$ and $a$ refer to the Bloch wave vector and the lattice constant, respectively. The model exhibits two distinct phonon branches: an acoustic branch with a lower frequency and an optical branch with a higher frequency. At the $\Gamma$ point, where $k = 0$, the acoustic mode (A) with zero frequency is represented by the state $|u_1(k = 0)\rangle = \frac{1}{\sqrt{2}}(1, 1)^T$, where the superscript $T$ indicates the transpose of the vector. This mode is $\mathcal{P}$-odd, indicating it changes sign under spatial inversion. In contrast, the optical mode (O) at $k = 0$ is given by $|u_2(k = 0)\rangle = \frac{1}{\sqrt{2}}(1, -1)^T$, which is $\mathcal{P}$-even, meaning it remains invariant under spatial inversion. This distinction in parity defines the conventional band ordering in the SSH model. The A mode resembles electronic bonding state which usually has lower energy compared to the O mode resembling electronic antibonding state. However, at $X$ point, *i.e.* $k = \pi/a$, the band order of A and O modes depends on the model parameters $K_1$ and $K_2$. When $K_1 > K_2$ the band order at $X$ is the same as $\Gamma$ which represents a normal order [Fig. 13(a)]; when $K_1 = K_2$ the two modes become degenerate at $X$ [Fig. 13(b)]; when $K_1 < K_2$

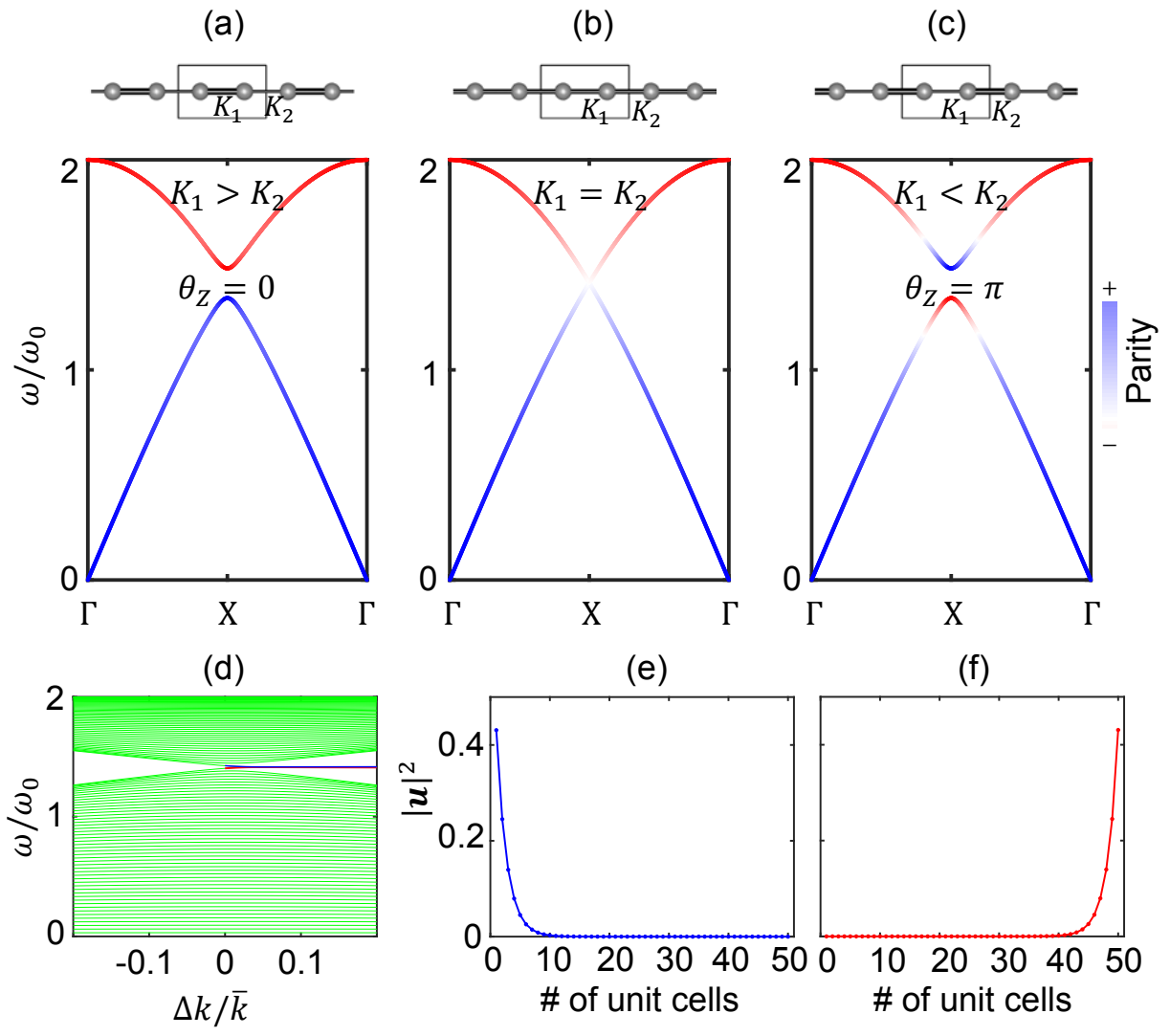

Figure 13 In a diatomic chain with one vibrational degree of freedom per atom, the phononic Su-Schrieffer-Heeger (SSH) model exhibits distinct topological characteristics: (a) For a topologically trivial phonon band structure, the intra-cell force constant $K_1$ exceeds the inter-cell force constant $K_2$. This configuration yields a conventional band ordering where the out-of-phase mode (with parity $+$") is higher in energy than the in-phase mode (with parity $-$"). (b) At the $X$ point, a degeneracy occurs between the in-phase and out-of-phase modes when $K_1$ is equal to $K_2$. (c) A topologically nontrivial phonon band emerges when $K_1$ is less than $K_2$, resulting in an inverted band order between the $\Gamma$ and $X$ points. (d) The eigenmode spectrum $\omega$ is plotted for a finite chain comprising 50 unit cells, varying as a function of $\Delta K/\bar{K}$ ($\Delta K = K_2 - K_1$, $\bar{K} = (K_2 + K_1)/2$) under fixed boundary conditions. The reference frequency is given by $\omega_0 = \sqrt{\frac{\bar{K}}{m}}$. The red and blue curves denote the topological end (edge) states, with their respective wave displacement fields illustrated in figures (e) and (f). Adapted from Ref. (Liu *et al.*, 2020b).

the band order at $X$ are inverted [Fig. 13(c)] indicating a nontrivial bulk topology.

According to Eq. (59), the Zak phase can be calculated as

$$\theta_Z = \int_{-\frac{\pi}{a}}^{\frac{\pi}{a}} \mathrm{d}k \, i\langle u_1(k)|\partial_k u_1(k)\rangle = \begin{cases} \pi & (K_2 > K_1) \\ 0 & (K_2 < K_1) \end{cases} \tag{62}$$

The non-zero topological invariant $\theta_Z$ signifies the presence of edge states within the bulk band gap, as depicted in Fig. 13(d). The red and blue curves represent the in-gap edge states, which are localized at the left and right boundaries of the chain, respectively, as shown in Figs. 13(e)-(f). Notably, although such in-gap edge states often appear in various variants of the SSH model, they are not topological in general and their presence/absence depends on boundary conditions. Namely, this quantized Zak phase does have "bulk-boundary correspondence". Only when the system has chiral symmetry, the bulk-boundary correspondence holds for the Zak phase, which is related to the winding number. The original SSH model for electrons has the chiral symmetry, but the phononic counterpart does not. However, in practice, in-gap edge states often emerge, in the cases of the nontrivial Zak phse, even when the chiral symmetry is absent. The phononic SSH model has been theoretically introduced in isostatic lattices (Kane and Lubensky, 2014) and has been experimentally observed in acoustic systems (Xiao *et al.*, 2015).

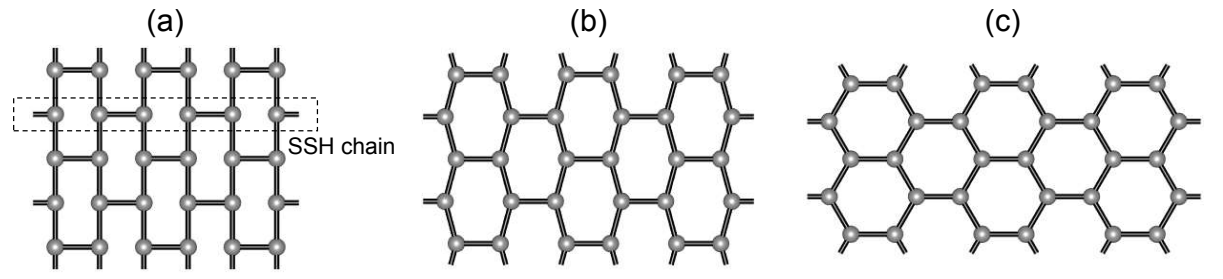

Figure 14 2D lattices that support a Dirac phonon phase include: (a) the brick lattice, which is constructed by stacking SSH chains and is capable of supporting a 2D Dirac phonon phase; (b) the stretched brick lattice, which is a variant of the brick lattice that similarly preserves a Dirac phonon phase; and (c) the honeycomb lattice, another well-known 2D lattice structure that maintains a Dirac phonon phase.

## B. Model in 2D

In 2D systems, twofold Dirac phonons (different from the fourfold Dirac phonons in 3D systems) exemplify gapless topological states of phonons, characterized by a cone-shaped dispersion with a nonzero Berry phase of $\pi$ for a closed trajectory encircling the Dirac phonon. These phononic states are typically protected by $\mathcal{PT}$ symmetry or mirror symmetry. In contrast to the twofold Weyl phonons in 3D systems, the topological nature of Dirac phonons relies on symmetry protection; the breaking of such symmetries can lead to the opening of a gap at the Dirac points. A codimension analysis reveals a closer analogy between 2D Dirac phonons and 3D nodal-line phonons. In this section, we will briefly introduce the main results about Dirac phonons in 2D.

### 1. 2D Dirac Phonons

In his 1928 seminal work, P. Dirac introduced the relativistic equation for electrons, capable of describing both massive and massless particles. The equation is $i\partial_t|\psi(t)\rangle = H|\psi(t)\rangle$ with the Hamiltonian $H = m\beta + \sum_i^d p_i\alpha_i$ ($\hbar = c = 1$) (Dirac, 1928). $m$ and $p_i$ are the mass and momentum, respectively, and $d$ refers to the spatial dimension. In order that the Dirac equation can reproduce the Klein-Gordon equation $(\boldsymbol{p}^2 + m^2 + \partial_t^2)|\psi\rangle = 0$, the introduced $\beta$ and $\alpha_i$ must satisfy $\beta^2 = 1$, $\{\alpha_i, \alpha_j\} = 2\delta_{ij}$, and $\{\beta, \alpha_i\} = 0$, which is the Clifford algebra. The anticommutation relations of $\alpha_i$ and $\beta$ indicate that $\alpha_i$ and $\beta$ must have matrix forms of which the smallest dimension is 2. In 2D case, *i.e.* $d = 2$, the Dirac matrices $\alpha_i$ and $\beta$ have a simple representation: $\alpha_x = \sigma_x$, $\alpha_y = \sigma_y$, and $\beta = \sigma_z$ where $\sigma_{x,y,z}$ are Pauli matrices. Therefore, the 2D massive Hamiltonian is given as $H = k_x\sigma_x + k_y\sigma_y + m\sigma_z$.

The case of $m = 0$ describes the massless Dirac fermions in graphene (Castro Neto *et al.*, 2009) at the Brillouin corners $K$ and $K'$. The cone-shaped band dispersion, stemming from the honeycomb lattice geometry, is not confined to spinless electronic states, prompting extensive exploration of topological physical phenomena in photonic crystals (Lu *et al.*, 2014; Wu and Hu, 2015) and magnonic systems (Zhang *et al.*, 2013), and phonon systems (Liu *et al.*, 2020b).

The topological invariant of 2D Dirac phonons is characterized by the quantized Berry phase accumulated along the loop encircling the Dirac point, *i.e.* $\theta_{\mathrm{B}} = \oint_C \mathrm{d}\boldsymbol{k} \cdot \boldsymbol{\mathcal{A}} = \pi$. The 2D Dirac phonon is protected by the combined symmetry of inversion and time-reversal, denoted $\mathcal{PT}$. Upon symmetry breaking, 2D Dirac phonons can transition into a variety of topologically gapped phases, such as Quantum Anomalous Hall (QAH), Quantum Spin Hall (QSH), and Quantum Valley Hall (QVH), among others.

In electronic systems, the 2D Dirac semimetal with gapless Dirac cones is the parent phase for the QH family (Liu *et al.*, 2020b, 2018) which typically exist in honeycomb lattices. This principle can be extended to phononic systems as well. The 2D Dirac states can be constructed from stacked 1D SSH models, *i.e.* the brick lattice model, as shown in Fig. 14. The honeycomb lattice can be obtained by gradually stretching the brick lattice without closing or opening band gaps.

Figure 15(a) shows a phononic model on a deformed honeycomb lattice with lattice constants $\boldsymbol{a}_1 = (a_x, a_y) = a(1 + \cos\phi, \sin\phi)$ ($a$ is the nearest-neighbor atomic distance) and $\boldsymbol{a}_2 = (a_x, -a_y)$. The conventional honeycomb lattice corresponds to the special case of $\phi = \pi/3$. The dynamical matrix of this lattice model is given by

$$D(\boldsymbol{k}) = \frac{1}{m}\begin{pmatrix} 3K & -K - \tilde{K}e^{-ik_x a_x} \\ -K - \tilde{K}e^{ik_x a_x} & 3K \end{pmatrix}, \qquad \tilde{K} = 2K\cos(k_y a_y). \tag{63}$$

Building upon the SSH model discussed earlier, the system at a constant $k_y$ exhibits a topologically nontrivial (trivial) band structure, characterized by an inverted (normal) band order along each $k_x$ line when $\tilde{K} > K$ ($\tilde{K} < K$), as shown in Fig. 13 (b). The case where $\tilde{K} = K$ represents a critical condition, resulting in a 1D band structure along $k_x$ with a vanishing gap at $k_x = \pi/a_x$, thereby forming a pair of Dirac cones at $(\pi/a_x, \pm\pi/3a_y)$, as shown in Fig. 13(c). These Dirac points are protected by $\mathcal{PT}$, ensuring that the Berry phase accumulated around any loop encircling a Dirac point is $\pi$, as evidenced by the abrupt discontinuity in the Zak phase $\theta_Z^x(k_y) = \int_0^{2b_x} \mathrm{d}k_x\, i\langle u_1(k_x, k_y)|\partial_{k_x} u_1(k_x, k_y)\rangle$ near $k_y = \pm\pi/3a_y$ [Fig. 13(d)]. The nonzero Zak phase can support localized edge modes, as depicted in the edge state spectrum of a 1D periodic ribbon of finite width along the $x$-direction in Fig. 13(e).

Graphene serves as a typical example of a 2D topological Dirac phonon, distinguished by a nonzero Berry phase of $\pi$. The topological nature of the Dirac phonon in 2D differs from that of the Weyl phonon in 3D; whereas the former requires symmetry protection, such as $\mathcal{PT}$ symmetry or mirror symmetry, the latter maintains topological stability irrespective of symmetry, being characterized by a nonzero Chern number. In this context, the 2D Dirac phonons can be conceptualized as an extension of the 1D SSH model, as both are classified by the same topological invariant.

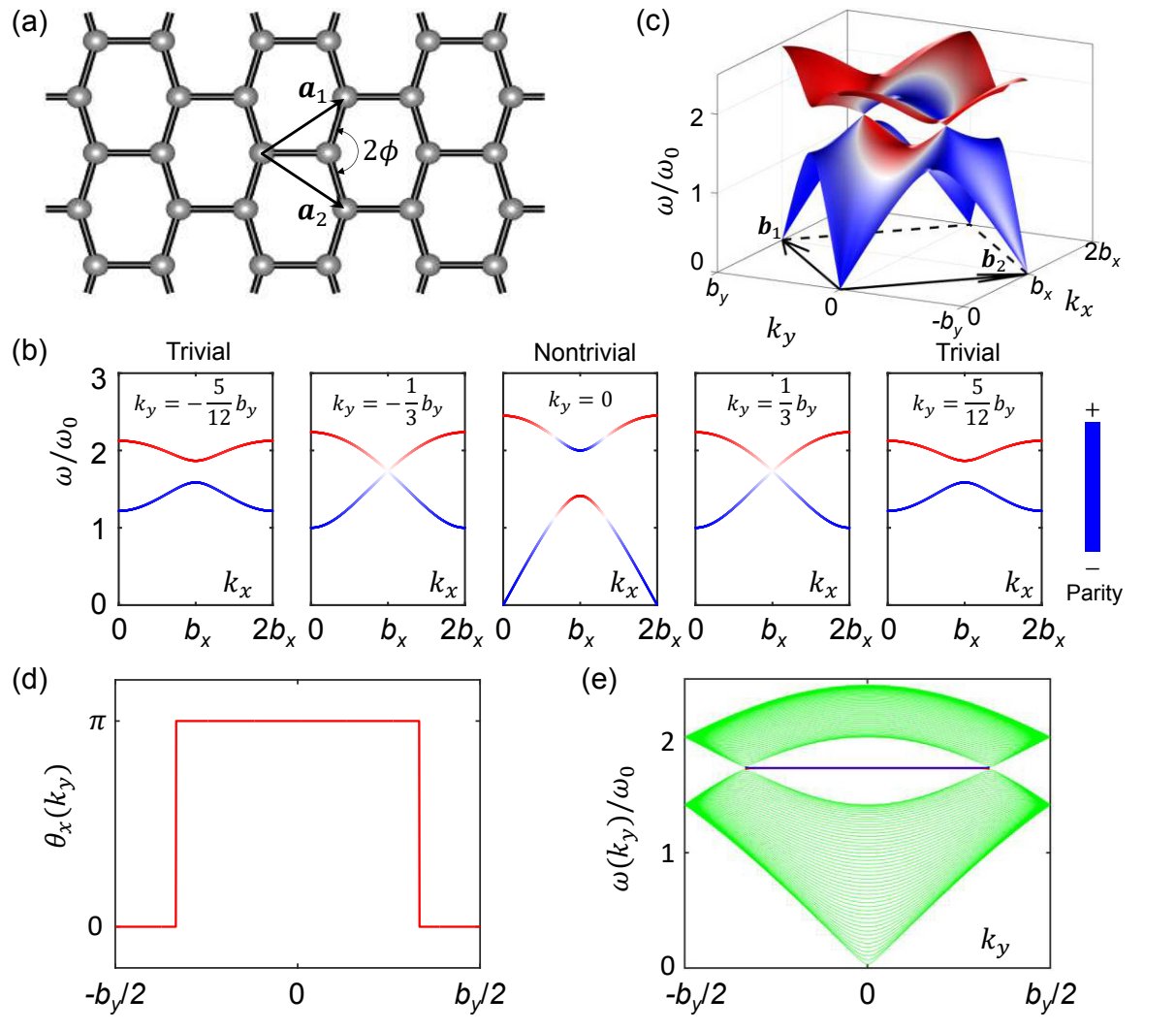

Figure 15 (a) Phononic model on a stretched brick lattice with lattice constants $\boldsymbol{a}_1 = (a_x, a_y) = a(1 + \cos\phi, \sin\phi)$, and $\boldsymbol{a}_2 = (a_x, -a_y)$. Its reciprocal lattice vectors are $\boldsymbol{b}_1 = (b_x, b_y) = (\pi/a_x, \pi/a_y)$ and $\boldsymbol{b}_2 = (b_x, -b_y)$. The honeycomb lattice corresponds to the case of $\phi = \pi/3$. (b) 1D phononic band structure as function of $k_x$ for different $k_y$. The 1D phonon band is topologically nontrivial within the range $-b_y/3 < k_y < b_y/3$, and it reverts to a trivial state outside this interval. (c) 2D phononic band showing a pair of Dirac cones at $(b_x, \pm b_y/3)$. (d) Topological Zak phase $\theta_Z^x(k_y)$ calculated as function of $k_y$ for each Wilson loop parallel to $k_x$ axis. (e) Phononic band structure of a system with finite width along the $y$-axis and periodicity along the $x$-axis. The red and blue curves denote the topological edge states.

## 2. Quantum Hall Family of Phonons

The quantum Hall family of phonons, encompassing QAH-, QSH-, and QVH- like states, can be realized by breaking specific symmetries in a honeycomb lattice (Liu *et al.*, 2020b), as illustrated in Fig. 16. The honeycomb lattice inherently supports a pair of Dirac cones at the Brillouin zone corners $K$ and $K'$, characteristic of a 2D Dirac phonon phase, with each Dirac point associated with a quantized Berry phase of $\pi$ for a loop enclosing it. Through mechanisms such as time-reversal symmetry breaking, inversion symmetry breaking, or Kekulé distortion, the Dirac cones can be gapped, leading to the emergence of topologically distinct phases, including the QAHE, QVHE, and QSHE.

The quantum anomalous Hall (QAH) states in phonon systems, also known as phonon Chern insulators, represent two-dimensional gapped phases distinguished by nonzero phonon Chern numbers. These topological invariants arise from the integration of the Berry curvature, $\Omega_z(\mathbf{k})$, over the Brillouin zone. Crucially, in the presence of time-reversal symmetry $\mathcal{T}$, the Berry curvature must be an odd function of momentum $[\Omega_z(-\mathbf{k}) = -\Omega_z(\mathbf{k})]$. This yields a zero Chern number. Consequently, the realization of a two-dimensional phonon Chern insulator inherently requires breaking $\mathcal{T}$ symmetry in the phonon system. Within the ten-fold way classification, the phonon QAH effect belongs to the "A" class, as illustrated in Figs. II and III.

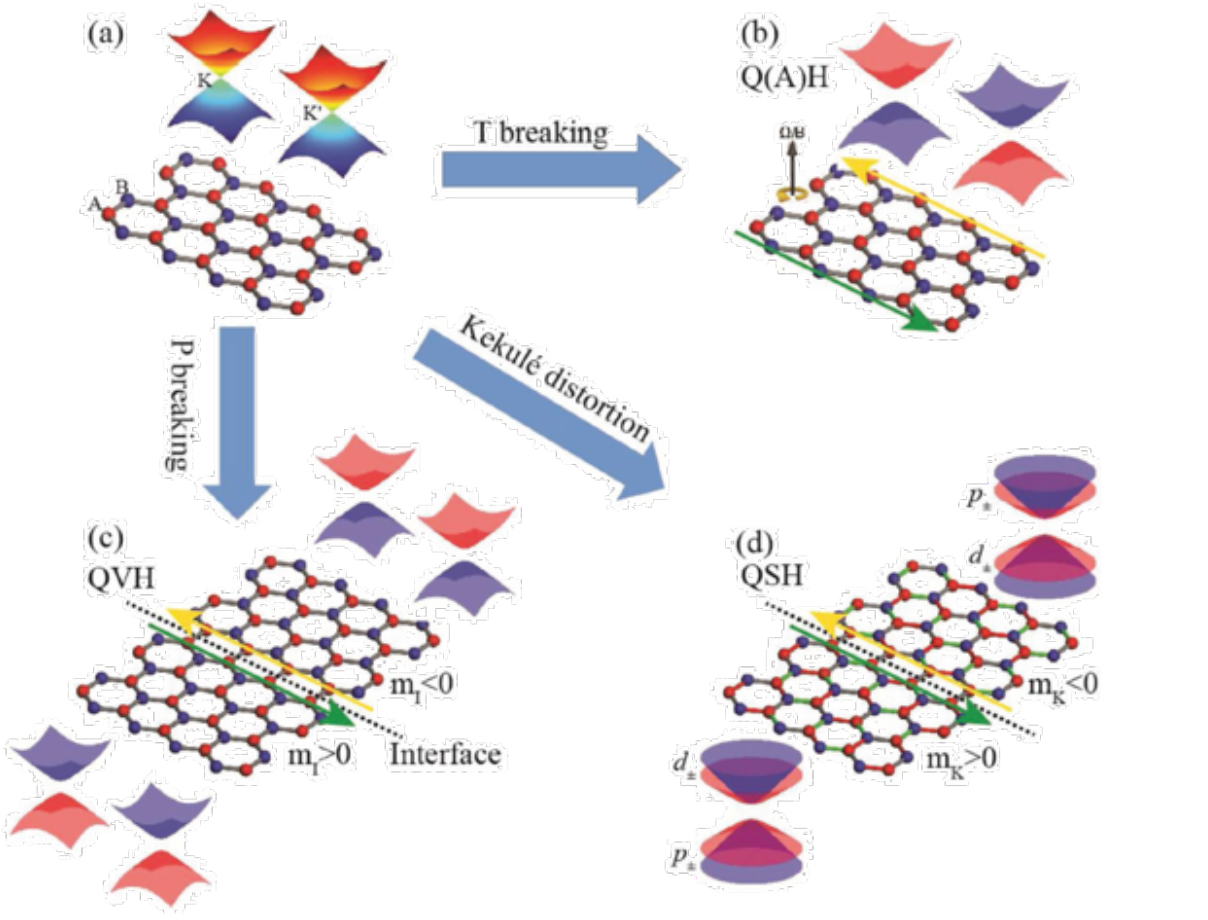

Figure 16 Various topological phonon states in 2D honeycomb lattices. (a) A honeycomb lattice hosts a 2D topological Dirac phonon phase, featuring a pair of Dirac cones at the Brillouin zone corners $K$ and $K'$. (b)-(d) By breaking different symmetries, the Dirac cones can be gapped, leading to the emergence of topologically distinct phases, including (b) quantum (anomalous) Hall [Q(A)H], (c) quantum valley Hall (QVH), and (d) quantum (pseudo-)spin Hall (QSH) states. These phases arise from specific symmetry-breaking mechanisms, illustrating the rich topological landscape of phonons in 2D systems. Adapted from Ref. (Liu *et al.*, 2020b).

The ordinary lattice Hamiltonian comprises a kinetic energy term $T = \sum_i \frac{1}{2}\dot{u}_i^2$ (where $u_i$ is the mass-weighted displacement of the $i$-th vibrational mode) and a harmonic potential energy term $V = \sum_{i,j} \frac{1}{2} K_{ij} u_i u_j$, both explicitly preserving time-reversal symmetry $\mathcal{T}$. To break $\mathcal{T}$ in phonon systems, a Raman-type spin-lattice interaction was found to introduce the velocity-displacement coupling (Sheng *et al.*, 2006) which can lead to nonzero phonon thermal Hall conductivity (Strohm *et al.*, 2005) and topological phonon Hall effect (Zhang *et al.*, 2010). A similar interaction term was studied in microtubles which can also lead to phonon Chern insulator phase with $C = \pm 1$ which support one-way chiral edge states of phonons (Prodan and Prodan, 2009b).

In crystalline solids, such $\mathcal{T}$-breaking effect of phonon can be realized in magnetic materials with Raman-type spin-lattice coupling (Sheng *et al.*, 2006), Coriolis force (Wang *et al.*, 2015a,b), and molecular Berry curvature due to electron-phonon couplings (Qin *et al.*, 2011; Saito *et al.*, 2019; Saparov *et al.*, 2022) which will give rise to phonon Hall effect (Strohm *et al.*, 2005; Zhang *et al.*, 2010). Recently, Refs. (Wang *et al.*, 2025b; Zhang *et al.*, 2026b) has successfully developed an *ab initio* algorithm to calculate the phonon Hall effect in a series of magnetic materials. Moreover, in the presence of such displacement-velocity coupling terms, the lattice dynamics sorely based on ordinary dynamical matrix fails to describe the phonon eigenvector and of course band topology based on it because it only characterize the properties of displacements but not velocities. This is closely related to the fact that the ordinary Newtonian or Lagrangian dynamical equation is of second order in time derivative and both initial displacements and velocities are required to determine the further state of motion. To solve this issue, (Huber, 2016b; Liu *et al.*, 2018, 2017b; Süsstrunk and Huber, 2016; Zhang *et al.*, 2010) introduced the Hamiltonian dynamics which is of first order in time derivative.

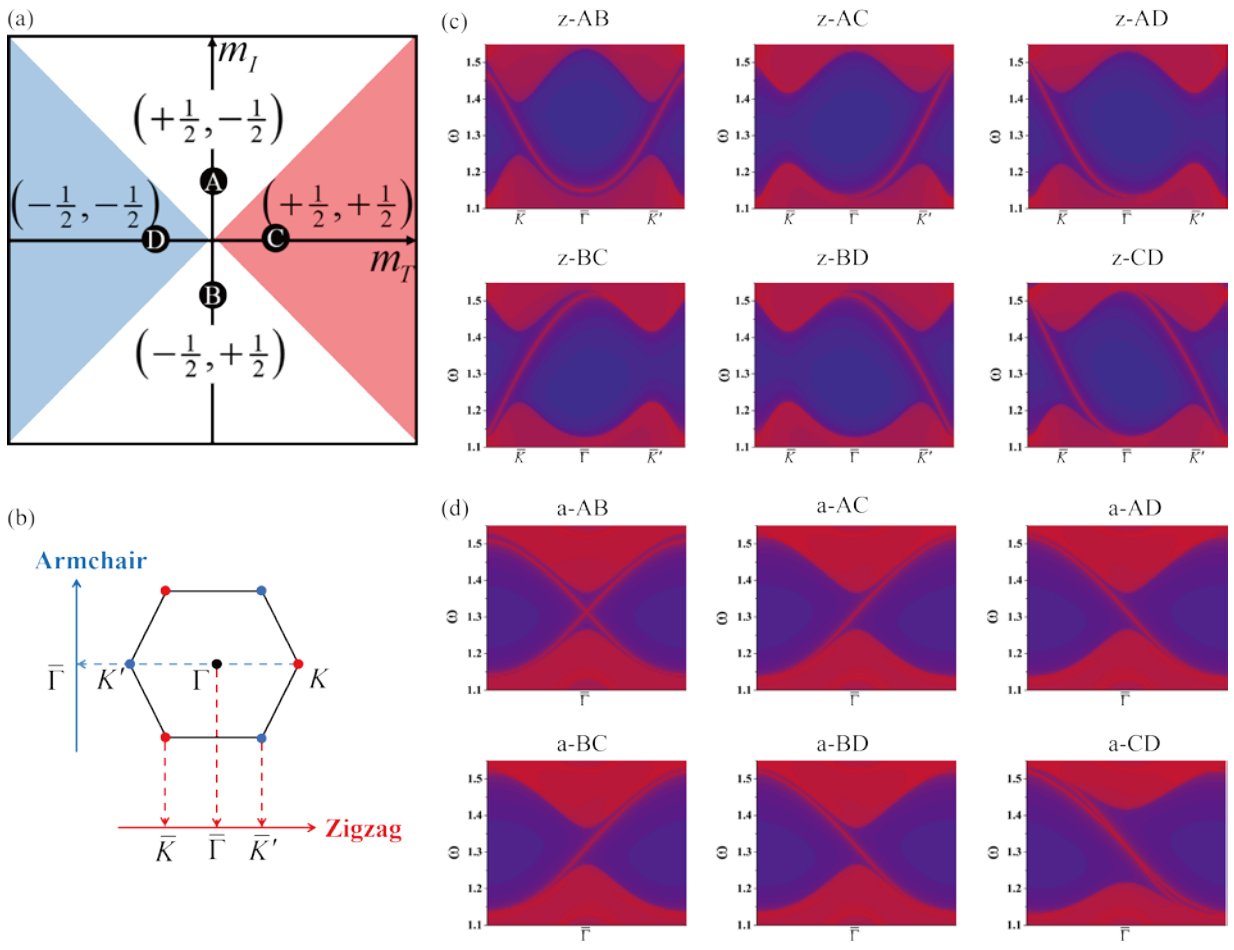

Figure 17 The honeycomb lattice model governed by $H_0+H_I+H_T$ (Eqs. (64), (66), (65)) hosts valley-polarized topological boundary states due to the interplay of symmetry-breaking terms $H_I$ (e.g., inversion-symmetry-breaking mass) and $H_T$ (e.g., time-reversal-symmetry-breaking flux). (a) Topological phase diagram illustrating the valley Chern number pair $(c_K, c_{K'})$ as function of parameters $m_T$ and $m_I$. TThe diagram is partitioned into four distinct regions by the critical lines $|m_I| = |m_T|$. Representative points A, B, C, and D in each region exhibit nearly quantized valley Chern numbers of $\pm\frac{1}{2}$. (b) Schematic of the 2D Brillouin zone and its projection onto the 1D boundaries. For zigzag boundary, the valleys $K$ and $K'$ project to distinct edge $k$ points $\bar{K}$ and $\bar{K}'$, respectively, while for armchair boundary, both valleys coalesce at the high-symmetry point $\bar{\Gamma}$. (c)-(d) Calculated spectral function of states for (c) zigzag and (d) armchair boundaries, respectively.

The honeycomb lattice serves as a prototypical topological system, exhibiting rich tunable topological phenomena. Its in-plane vibrational modes host a pair of Dirac cones at the $K$ and $K'$ points of the Brillouin zone. The effective $k \cdot p$ Hamiltonian near these Dirac points mirrors graphene's low-energy model and is expressed as:

$$H_0 = v_D(k_y \tau_z \sigma_x - k_x \sigma_y), \tag{64}$$

where $v_D$ refers to the group velocity at the phononic Dirac points; $\sigma$ and $\tau$ are Pauli matrices acting on the sublattice and valley subspaces, respectively. The existence of these 2D Dirac points is protected by the combination of inversion and time-reversal symmetries. The topology of each Dirac point is characterized by the quantized Berry phase $\theta_B = \oint_C \mathrm{d}\boldsymbol{k} \cdot \boldsymbol{A}(\boldsymbol{k}) = \pi$ for a closed loop enclosing the Dirac point. Symmetry-breaking mass terms can be introduced into Eq. (64), opening band gaps and driving the system into distinct topological phases.

The quantum anomalous Hall like states of phonons is described by $H_0 + H_T$ with

$$H_T = m_T \sigma_z \tau_z \tag{65}$$

being the Haldane-type mass term (Liu *et al.*, 2018, 2017b).

The quantum anomalous Hall (QAH)-like states of phonon, characterized by the nonzero Chern number $C$, require the breaking of time-reversal symmetry $\mathcal{T}$ in the phonon system. In systems where $\mathcal{T}$ is preserved, the total Chern number must vanish. However, integration of Berry curvature around the Dirac points $K$ and $K'$ can be nonzero when a finite band gap is opened by breaking the spatial inversion symmetry $\mathcal{P}$.

The effective model of QVH of phonons is given by $H_0 + H_I$ with

$$H_I = m_I \sigma_z \tag{66}$$

being the Semenoff-type mass term. For a small mass $m_I$, the Berry curvature distribution becomes predominantly concentrated around the two valleys $K$ and $K'$, which leads to nearly quantized valley Chern numbers $c_K = c_{K'} = \pm\frac{1}{2}$. The contrasting Chern numbers between the valleys facilitate the formation of a pair of valley-polarized topological boundary states. These states are characterized by a sign change in the mass $m_I$ across them, resulting in a shift in the valley Chern numbers $c_K$ and $c_{K'}$ by $\pm 1$, as illustrated in Fig. 12(c). Such valley-polarized topological boundary states have been the subject of extensive research in various systems, including acoustic systems (Xue *et al.*, 2022; Zhang *et al.*, 2018d; Zheng *et al.*, 2020), mechanical lattices (Lu *et al.*, 2018; Zhang *et al.*, 2018b), phononic crystals (Liu *et al.*, 2020b; Xia *et al.*, 2022; Zhang *et al.*, 2022a), and so on (Liu *et al.*, 2021b; Wu *et al.*, 2017).

When both $\mathcal{P}$ and $\mathcal{T}$ symmetries are simultaneously broken, that is, when both $m_I$ and $m_T$ are nonzero, the variation in $m_I$ and $m_T$ results in the valley Chern numbers $c_K$ and $c_{K'}$ being determined as follows:

$$\begin{aligned} c_K &= \frac{1}{2}\mathrm{sgn}(m_T + m_I), \\ c_{K'} &= \frac{1}{2}\mathrm{sgn}(m_T - m_I), \end{aligned} \tag{67}$$

which gives rise to the topological phase diagram shown in Fig. 17.

The quantum anomalous Hall (QAH)-like states of phonons or phonon Chern insulators requires the breaking of $\mathcal{T}$ symmetry for phonons. However, achieving $\mathcal{T}$-symmetry breaking in phonons is inherently challenging due to their charge-neutral nature and weak coupling to external electric or magnetic fields. In contrast, the quantum spin Hall (QSH) effect represents a prominent class of topological states characterized by spin-momentum-locked helical boundary states (Hasan and Kane, 2010; Kane and Mele, 2005; Qi and Zhang, 2010b, 2011). The QSH insulator, or equivalently $\mathbb{Z}_2$ topological insulator, relies on the existence of double-degenerate states on multiple surface time-reversal invariant momentum (Fu *et al.*, 2007). However, unlike electrons whose intrinsic spin-$\frac{1}{2}$ property ensures Kramers doublets due to $\mathcal{T}^2 = -1$, phonons do not have intrinsic Kramers-like double degenerate degree of freedom generically. This motivates some studies on crystalline-symmetry-protected double degeneracy enabled topological phononic lattices (Liu *et al.*, 2017a, 2019c).

The additional spin degree of freedom requires a four-band model as the minimum requirement to realize QSH. In 2015, (Süsstrunk and Huber, 2015) successfully constructed a mechanical topological insulator which mimics the Kane-Mele model (Kane and Mele, 2005). (Wu and Hu, 2015) designed a four-band "double Dirac cone" model within a photonic honeycomb lattice characterized by a Kekulé pattern. This "double Dirac cone" arises from the folding of single Dirac cones originating from the $K$ and $K'$ points. (Liu *et al.*, 2017a) discovered phononic pseudospins in the Kekulé lattice, exhibiting quantized pseud-angular momenta and pseudospin-polarized Berry curvature, which support pseudospin-polarized helical boundary modes that are immune to disorder-induced scattering.

The 2D Dirac phonons provide a natural way of generalization to the 3D Weyl phonons. For example, by adding a $k_z$-dependent mass term $k_z\sigma_z$ into the effective model of 2D Dirac phonon [Eq. (64)] will lead to a 3D Weyl phonon with Chern number $C = \pm 1$, similar to the case of Eq. (74) with $N = \pm 1$. Such kind of 3D Weyl phonons with $N = \pm 1$ and $N = \pm 2$ can be realized in the chiral materials with three-fold skew rotation symmetry like Te (Zhang *et al.*, 2023b) and $\alpha$-quartz (Lange *et al.*, 2024; Wang *et al.*, 2020).

## 3. Transport properties for the phonon quantum Hall states

The QAHE-like states of phonons support gapped bulk states and gapless one-way edge states simultaneously. The bulk states characterized by nonzero phonon Chern number $C$ support phonon Hall effect (Zhang *et al.*, 2010). However, despite the quantization of phonon Chern number, the phonon Hall conductivity $\kappa_{xy}$ is not quantized because the phonon occupation number (Bose-Einstein statistics) is not quantized. For QSHE/QVHE like states, the gapped bulk states exhibit zero phonon Hall conductivity but nonzero phonon spin/valley Hall transport. The edge states of all the quantum Hall family all exhibit superior disorder-scattering-immune property.

Historically, the study of phonon Hall effect is initiated by experimental discovery of a magnetotransverse thermal con-ductivity in $Tb_3Ga_5O_{12}$ (Strohm *et al.*, 2005). Such a Hall effect is not expected to exist for phonons which do not have charge thus Lorentz force is not directly applicable. Generally, a nonzero Hall effect requires the breaking of time-reversal symmetry $\mathcal{T}$. It was soon proposed that the Raman-type spin-lattice interaction (Sheng *et al.*, 2006) which can break $\mathcal{T}$ symmetry of phonons. Later, based on the same interaction, a topological nature of phonon Hall was revealed (Qin *et al.*, 2011, 2012; Zhang *et al.*, 2010) with nonzero phonon Chern number.

### C. Higher-order Topological Phonons

The phononic quantum Hall (QH) family originates from the Berry curvature of phonon band structures. In 2D materials, the combined $\mathcal{PT}$ or $C_2\mathcal{T}$ symmetry enforces a vanishing phononic Berry curvature across the entire Brillouin zone. Nevertheless, a topological invariant—the Stiefel-Whitney (SW) number $w_2$ can still be defined using parity eigenvalues at time-reversal invariant momenta (TRIM) below a phononic band gap. In 3D materials, the second SW number characterize the interlocking topology of $\mathbb{Z}_2$ nodal lines which must exist in pairs (Ahn *et al.*, 2018; Fang *et al.*, 2015). In contrast, the topological nodal lines with Berry flux $\pi$, *i.e.* a nonzero first SW number $w_1 = 1$, can exist alone (Ahn *et al.*, 2019; Bouhon *et al.*, 2019).

Under $\mathcal{PT}$ or $C_2\mathcal{T}$ symmetry, all phononic eigenvectors can be chosen to be real. The topology of such real vectors in Hilbert space is characterized by a $\mathbb{Z}_2$ index, *e.g.*, the second SW number $w_2$, which is determined by

$$(-1)^{w_2} = \prod_{i=1}^{4}(-1)^{\lfloor N^-(\Lambda_i)/2 \rfloor}, \tag{68}$$

where $\Lambda_i$ ($i = 1, 2, 3, 4$) denotes four TRIM in 2D Brillouin zone, and $N^-(\Lambda_i)$ counts the number of negative-parity eigenmodes at $\Lambda_i$ below a given band gap. $\lfloor \cdot \rfloor$ refers to the floor function. The 2D gapped phononic band structures with nonzero $w_2$ is called phononic SW insulator or phononic real Chern insulator (Ding *et al.*, 2024). The value of $w_2$ can alternatively be determined from the spectrum of the Wilson loop operator:

$$\hat{W}(k_\perp) = \exp\left[-i\oint_C \mathrm{d}k_\parallel \, \hat{\mathcal{A}}_\parallel(k_\parallel, k_\perp)\right] \tag{69}$$

where $\hat{\mathcal{A}}$ refers to the nonabelian Berry connection whose matrix elements is $\boldsymbol{\mathcal{A}}_{nm} = i\langle u_n|\boldsymbol{\nabla}_{\boldsymbol{k}} u_m\rangle$ with $n$ and $m$ being the band indexes below the gap. $C$ is the Wilson loop parallel to $k_\parallel$ at fixed $k_\perp$. Since the Wilson operator is unitary, its eigenvalues take the form $\exp[-i\theta(k_\perp)]$. In electronic systems, $\theta(k_\perp)$ corresponds to the electric charge polarization, which is proportional to the Wannier centers of charge. The phase $\theta(k_\perp)$ is only defined modulo $2\pi$, with its principal value within $[0, 2\pi]$. Let the number of times $\Theta = \theta(k_\perp)$ crosses with $\Theta = \pi$ be $N$. The parity of $N$ (even for $w_2 = 0$ and odd for $w_2 = 1$) constitutes a $\mathcal{PT}$ or $C_{2z}\mathcal{T}$-protected topological invariant, yielding the second Stiefel-Whitney number:

$$w_2 = N \bmod 2. \tag{70}$$

The atomic insulator exhibits a trivial band gap with $w_2 = 0$. In contrast, a nontrivial SW insulator can be realized through a double band inversion at certain TRIM as indicated by Eq. (68). Regarding "bulk-boundary correspondence", the SW insulator is a second-order topological insulator rather than first-order. This is because the protecting symmetry, either $\mathcal{P}$ or $C_{2z}$, is typically broken at the 1D open boundary, resulting in gapped edge states. Consequently, a 2D SW insulator with $w_2 = 1$ supports topological corner modes, characteristic of its second-order nature. As an example, (Pan and Huang, 2022) demonstrated the phononic SW phase in two-dimensional Xenes and their ligand-functionalized derivatives. Beyond these examples, $\mathcal{PT}$-symmetric topological phononic modes have also been explored in Kekulé lattices (Liu *et al.*, 2017a), including graphdiyne (Zhang *et al.*, 2023c), hydrogen-substituted graphdiyne (Mu *et al.*, 2022), and $C_3N$ compound (Huang *et al.*, 2023).

The concept of Stiefel-Whitney (SW) insulator in 2D materials can be generalized to 3D by (Wang *et al.*, 2024c). For a generic 3D material, the band structure on two planes with $k_z = 0$ and $k_z = \pi$ which preserve $\mathcal{PT}$ symmetry can be effectively described by second SW number $w_2(k_z = 0) = \nu_0$ and $w_2(k_z = \pi) = \nu_\pi$. The value of $\nu_0$ and the relative sign between $(-1)^{\nu_0}$ and $(-1)^{\nu_\pi}$ determine the topological phases of 3D materials having "real" wavefunctions. The 3D phononic real Chern insulator (PRCI) is characterized by $(\nu_0, \nu_\pi) = (1, 1)$, whereas $(\nu_0, \nu_\pi) = (1, 0)$ or $(0, 1)$ will indicate a phononic real semimetal (PRSM) phase. The PRSM can be characterized by a $\mathbb{Z}_2$ index $\nu_{\text{PRSM}}$ as

$$(-1)^{\nu_{\text{PRSM}}} = (-1)^{\nu_0}(-1)^{\nu_\pi}. \tag{71}$$

A nonzero $\nu_{\text{PRSM}}$ indicate a topological distinction between $k_z = 0$ and $k_z = \pi$ planes so that there must exist gap closing points between $k_z = 0$ and $k_z = \pi$. Such gap closing points can be either four-fold degenerate (Dirac type), triply degenerate, or two-fold degenerate nodal lines [1]. In the PRSM phase, the degenerate $\boldsymbol{k}$-nodes (nodal points or lines) must exist in pairs.

### D. Models in 3D

In contrast to the topological states in gapped phonon spectra discussed in the previous subsection, topological phonons in gapless systems have attracted considerably greater attention. The classification of gapless topological phonons is primarily determined by two key characteristics of the band crossings: their degeneracy and codimension (e.g., point or line nodes). The interplay between these two aspects gives rise to a rich variety of topological phonon excitations (Jin *et al.*, 2022; Li *et al.*, 2021a; Liu *et al.*, 2020b; Miao *et al.*, 2018; Qin *et al.*, 2024; Wang *et al.*, 2022; Yang *et al.*, 2024b; Zhang *et al.*, 2022b, 2023b, 2020a; Zhang and Murakami, 2021, 2023; Zhang *et al.*, 2018a, 2020b, 2019b), which are typically stabilized by symmetries, as illustrated in Figs. 8 and 22. In this subsection, we review various types of topological phonons in gapless spectra, together with their associated topological invariants and standard diagnostic approaches.

[1] The two-fold degenerate points cannot be Weyl points because they are prohibited by $\mathcal{PT}$ symmetry.

### 1. Dirac and Weyl Phonons

In particle physics, elementary particles serve as the fundamental building blocks of matter and the universe. The standard model categorizes these particles into three primary categories: quarks, leptons, and gauge bosons (Burgess and Moore, 2007). Dirac fermions play a central role in the standard model of particle physics, encompassing all the known quarks and charged leptons. They are crucial for describing the properties and interactions of these fundamental particles. Beyond particle physics, Dirac fermions also emerge in condensed matter physics, where they refer to quasiparticles that are governed by the Dirac equation. This equation unites quantum mechanics and special relativity, providing a framework for describing spin-$\frac{1}{2}$ particles.

In the context of condensed matter physics, Dirac phonons are analogous to spinless Dirac fermions in solids. They are described within the framework of band theory, with their universal effective Hamiltonian expressed as:

$$H_{\mathrm{D}}(\boldsymbol{k}) = \begin{pmatrix} v_0\boldsymbol{\sigma}\cdot\boldsymbol{k} & m(\boldsymbol{k}) \\ m^*(\boldsymbol{k}) & -v_0\boldsymbol{\sigma}\cdot\boldsymbol{k} \end{pmatrix}, \tag{72}$$

where $\boldsymbol{\sigma} = (\sigma_x,\ \sigma_y,\ \sigma_z)$ represent the Pauli matrix, $\boldsymbol{k} = (k_x,\ k_y,\ k_z)$ is the crystal momentum vector, $v_0$ is the group velocity and $m(\boldsymbol{k})$ is the mass term of the Dirac Hamiltonian. Eigenvalues for Eq. (72) are $E_\pm(\boldsymbol{k}) = \pm\sqrt{m(\boldsymbol{k})^2 + v_0^2\boldsymbol{k}^2}$, with each branch of $E_\pm(\boldsymbol{k})$ doubly degenerated in inversion-symmetric systems, as shown by the phonon spectra in Figs. 8 (a) and (b). When $m(\boldsymbol{k}) = 0$, the Dirac point will be fourfold degenerate and can be decomposed into two Weyl phonons with opposite chirality. The chiral Weyl phonon can be described by:

$$H^{\pm}_{\mathrm{Weyl}}(\boldsymbol{k}) = v_0\cdot\begin{pmatrix} k_z & k_x \mp ik_y \\ k_x \pm ik_y & -k_z \end{pmatrix}, \tag{73}$$

where the sign $\pm$ indicates the chirality. For chirality "$\pm$", the Chern number of a closed surface enclosing the Weyl point is $C = \pm 1$. Thus, the nonzero Chern number indicates chirality, or more specifically, the topological chirality. We note that a massless Dirac fermion has a vanishing total Chern number ($C$=0) as a result of the cancellation between two Weyl points carrying opposite Chern numbers. While phonons are analogous to spinless fermions in solids, under certain symmetry the Dirac and Weyl phonons emerge, similar to spin-$\frac{1}{2}$ fermions, as shown in Fig. 8.

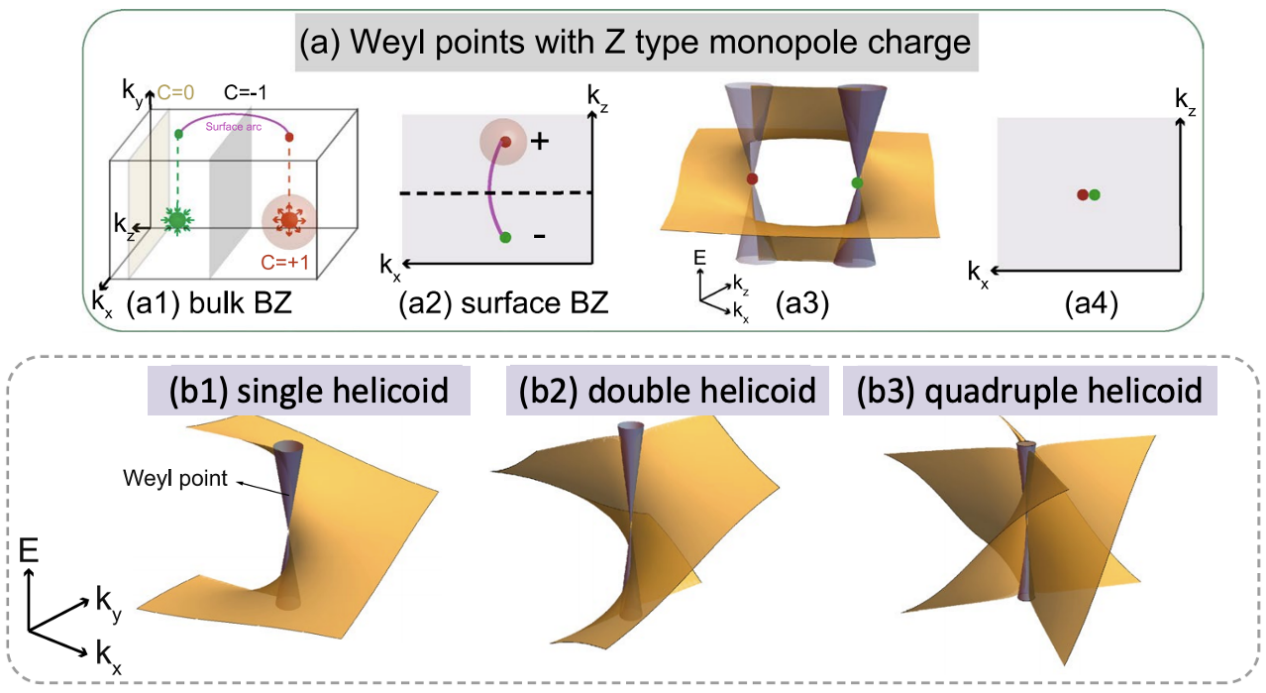

Figure 18 (a) Surface states and surface arcs for Weyl phonons. The Chern number of Weyl phonons can be calculated either on a sphere enclosing the Weyl phonons or the 2D colored plane shown in (a1). In (a2), when two Weyl phonons with opposite chirality project onto different momenta in the surface Brillouin zone (BZ), a surface arc connecting the two Weyl phonons emerges. (a3) depicts the topological surface states linking two Weyl phonons, which exhibit different chiralities near each Weyl phonon. In (a4), when two Weyl phonons with opposite chirality project onto the same momentum in the surface BZ, the surface arc disappears. (b1-b3) Topological surface states for Weyl phonons with Chern number of $C$ = +1, +2 and +4, respectively, all of them are in helicoid/spiral shapes. (a) is adapted from Ref. (Zhang *et al.*, 2022b), (b1)-(b3) are adapted from Ref. (Zhang and Murakami, 2023).

The topological invariant for Weyl phonons is defined in a manner analogous to the Chern number, as expressed in Eq. (48). However, instead of integrating over a two-dimensional plane in the Brillouin zone, the integral is performed over a sphere enclosing the Weyl point in three-dimensional momentum space, as shown in Figs. 8 (c) and 7 (a)-(d). Figures 7 (e)-(f) are the Berry curvature distribution on the integral sphere enclosing the Weyl phonon with $C = +1$ and $-1$, respectively. When $C \neq 0$, the Weyl phonon exhibits a topological nature, which is closely linked to the emergence of topological surface states. These surface states typically exhibit a helicoid/spiral shape, as illustrated in Fig. 8 (d). The chirality of these surface states follows the sign of the Chern number. Consequently, in systems containing a pair of Weyl phonons, the surface states connecting them display opposite chirality near each Weyl phonon, as shown in Fig. 18 (a3).

If an isoenergy surface is considered, analogous to the Fermi surface in fermionic systems, a surface arc will emerge in the surface Brillouin zone when a pair of Weyl phonons project onto distinct momenta. This behavior is illustrated in Figs. 18 (a1) and (a2), where the surface arc connects the projections of the Weyl phonons, reflecting their topological nature. However, if a pair of Weyl phonons with opposite chirality project onto the same momentum in the surface Brillouin zone, the surface arc vanishes, as depicted in Fig. 18 (a4). Consequently, massless Dirac phonons do not exhibit surface arcs, since the two Weyl phonons that constitute a Dirac phonon always map to identical momenta on the surface BZ. This absence of surface arcs is a direct result of the cancellation of chirality and the overlapping projections of the Weyl phonons in the Dirac system.

In Eq. (72), Dirac points are represented using gamma matrices that satisfy the anti-commutation relations of the Clifford algebra. As a result, Dirac points can exhibit two distinct behaviors: they can be massive, associated with a gapped band dispersion, or massless, associated with a gapless band dispersion, depending on the coefficients of the gamma matrices. On the other hand, Eq. (73) involve two bands and are represented using the Pauli matrices. Since the number of the Pauli matrices is three, Weyl points in 3D are topological and robust, as is distinct from the ones in 2D. This feature is common to any quasiparticles, such as electrons and phonons.

In 3D systems, Dirac phonons and Weyl phonons can be distinguished by two key features: (1) their band degeneracy and (2) the topological invariant of the system. Dirac phonons are typically topologically trivial due to the combination of two Weyl points with opposite topological charge, while Weyl phonons are characterized by their chiral nature and non-zero topological charge, as reflected in their distinct topological invariants. However, in a special case, Dirac phonons can carry a nonzero monopole charge, known as charge-2 Dirac phonons (Jin *et al.*, 2022; Miao *et al.*, 2018; Zhang *et al.*, 2018a) or $\mathbb{Z}_2$ Dirac phonons (Fang *et al.*, 2016; Kargarian *et al.*, 2016; Kim *et al.*, 2021; Le *et al.*, 2018; Yukitake *et al.*, 2025a,b; Zhang *et al.*, 2022b; Zhang and Murakami, 2023), which will be reviewd in Sec. III.D.3. This occurs when they are protected by a combination of non-symmorphic symmetries and time-reversal symmetry.

Though conventional Dirac phonons are topologically trivial and do not possess surface states, they play a crucial role as the critical phase that governs topological phase transitions, as emphasized in the electronic systems by Ref. (Murakami, 2007). For example, when $m(\boldsymbol{k}) \neq 0$, the Dirac point transitions into a massive one, acquiring an energy gap, as shown in Fig. 8 (a). The topology of two massive Dirac systems differs based on the sign of $m(\boldsymbol{k})$, a consequence of the evolution process, as depicted in Fig. 19. This highlights the significance of Dirac and Weyl phonons as versatile platforms for modulating and exploring a wide range of topological states.

### 2. Unconventional Weyl Phonons

In contrast to the real universe, the “material universe” refers to the effective theory of low-energy, long-wavelength excitations in solids. In 2007, (Murakami, 2007) proposed the idea of discovering Weyl fermions, a concept originally from high-energy physics, in condensed matter systems. This groundbreaking idea was later realized in 2015 when (Huang *et al.*, 2015; Weng *et al.*, 2015) independently predicted the existence of Weyl fermions in real materials, a prediction that was experimentally confirmed in TaAs (Huang *et al.*, 2015; Lv *et al.*, 2015a). Building on this, in 2016, (Bradlyn *et al.*, 2016) suggested that crystals could host a variety of “unconventional Weyl fermions” with Chern numbers greater than 1, i.e., $|C| > 1$. Since then, theoretical research into these “new quasiparticle excitations” has broadened to include bosonic systems, such as phonon spectra, with a focus on their topological invariants (Chern number, $\mathbb{Z}_2$ monopole charge, etc.), degeneracy and dispersion properties. This expansion has opened new avenues for exploring topological phenomena in bosonic systems within the material universe.

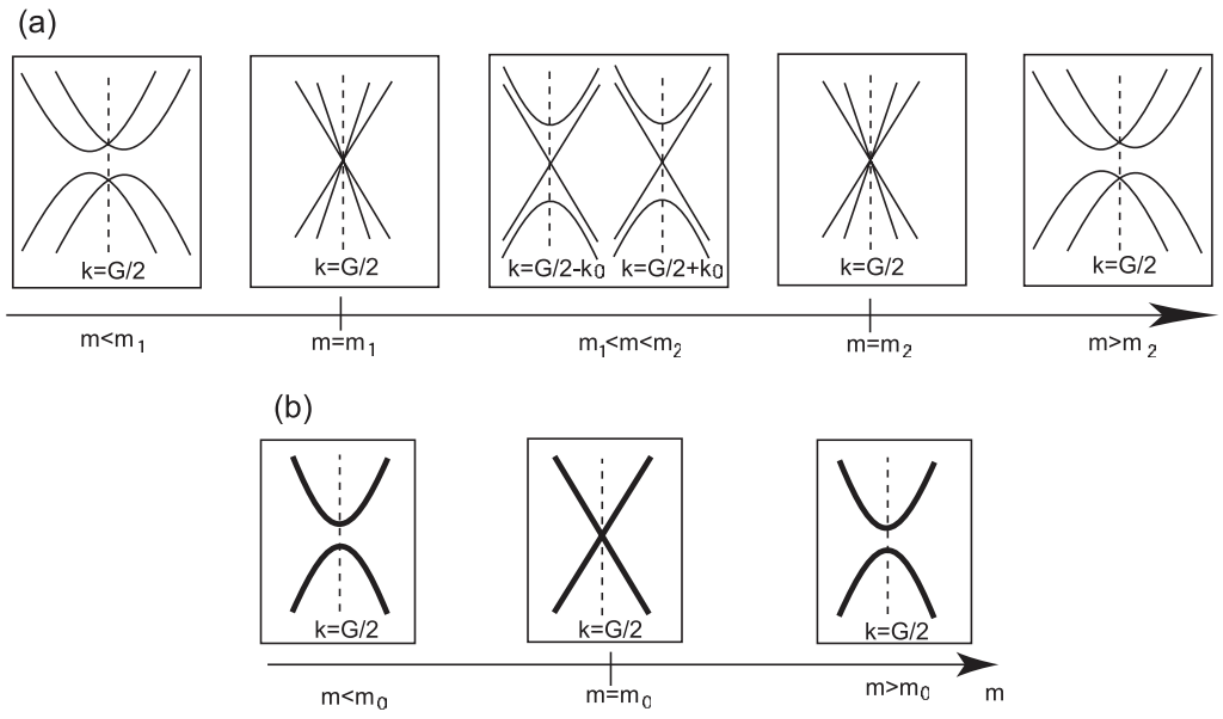

Figure 19 A phase transition can occur between two gapped systems, one topologically nontrivial and the other topologically trivial, by tuning the mass term of the system. This figure shows spin-$\frac{1}{2}$ systems with time-reversal symmetry as an example. The critical point of this transition is marked by a massless Dirac point, which serves as the boundary between the two phases. In Fig. (a), the system lacks inversion symmetry, resulting in non-degenerate bands, while in Fig. (b), the system possesses inversion symmetry, leading to doubly degenerate bands. These scenarios illustrate how symmetry and the mass term influence the topological properties and band structure of the system during the phase transition. The figure is adapted from Ref. (Murakami, 2007).

For example, the low-energy effective $k \cdot p$ model for a twofold Weyl phonon with Chern number of $C$=$N$ ($> 0$) can be written as:

$$H_N(\boldsymbol{k}) = \begin{pmatrix} Ak_z & B(k_x - ik_y)^N \\ B(k_x + ik_y)^N & -Ak_z \end{pmatrix}, \tag{74}$$

where $A$ and $B$ are real constants. The Chern number $N$ is determined by the rotational symmetry of the system. For example, unconventional Weyl phonons with Chern numbers $C$=$\pm 2$ can be obtained in systems with $C_{3,4,6}$ rotational symmetries (Fang *et al.*, 2012; Tsirkin *et al.*, 2017), associated with helicoid/spiral surface states shown in Fig. 18 (a2).

Since phonons are spin-0 bosons, such property can give rise to a new type of twofold Weyl phonon with the protection of both time-reversal symmetry and chiral cubic symmetry. The equivalence of $k_x/k_y/k_z$ under these symmetries leads to a low-energy effective $k \cdot p$ Hamiltonian :

| Space group | Two vertical glide mirrors | Location | Momenta |
|---|---|---|---|
| #73 | $\{M_x\|\frac{1}{2},\frac{1}{2},0\}$; $\{M_y\|\frac{1}{2},0,0\}$ | W | Γ, T |
| #110 | $\{M_x\|\frac{1}{2},\frac{1}{2},0\}$; $\{M_y\|\frac{1}{2},\frac{1}{2},0\}$ | P | Γ, X |
| #142 | $\{M_x\|\frac{1}{2},\frac{1}{2},0\}$; $\{M_y\|\frac{1}{2},0,0\}$ | P | Γ, X |
| #206 | $\{M_x\|\frac{1}{2},\frac{1}{2},0\}$; $\{M_y\|\frac{1}{2},0,0\}$ | P | Γ, N |
| #228 | $\{M_x\|\frac{1}{2},\frac{3}{4},0\}$; $\{M_y\|\frac{3}{4},\frac{1}{2},0\}$ | W | Γ, X |
| #230 | $\{M_x\|\frac{1}{2},\frac{1}{2},0\}$; $\{M_y\|\frac{1}{2},0,0\}$ | P | Γ, N |

Table V  Phonon systems (or any spinless systems) where $\mathbb{Z}_2$ Dirac points associated with quad-helicoid surface states (QHSSs) can be obtained. Each column represents for the space group number, two vertical glide mirrors which protect the $\mathbb{Z}_2$ Dirac points together with $\mathcal{T}$, the momentum where $\mathbb{Z}_2$ Dirac points are located. Notably, all the listed space groups exhibit a wallpaper group of $p2gg$ on the (001) surface with two vertical glide mirror symmetries. Conversely, surfaces with a wall group of $p4gg$ is forbidden to have QHSSs.

$$H_{\mathrm{TQW}}(\boldsymbol{k}) = -\begin{pmatrix} Ak_xk_yk_z & B(k_x^2+\omega k_y^2+\omega^2 k_z^2) \\ B(k_x^2+\omega^2 k_y^2+\omega k_z^2) & -Ak_xk_yk_z \end{pmatrix}, \tag{75}$$

where $\omega{=}e^{-\frac{2\pi i}{3}}$. In this case, an unconventional Weyl phonon with each band carrying the highest Chern number of $C=\pm4$ can emerge in solids. Since the Weyl phonon described in Eq. (75) possesses a monopole charge of 4, it is referred to as a "twofold quadruple Weyl phonon (TQW)" in Ref. (Li *et al.*, 2021a; Zhang *et al.*, 2020b). The existence of such twofold quadruple Weyl points was overlooked by topological band theory prior to 2020, as earlier studies did not account for the Bravis lattice degrees of freedom. The twofold quadruple Weyl phonon requires the protection of chiral cubic symmetry and $\mathcal{T}$. As a result, it can only occur in specific scenarios, such as being located at the $\Gamma$ for all chiral cubic space groups or at the $(\frac{\pi}{a},\frac{\pi}{a},\frac{\pi}{a})$ point for certain chiral cubic space groups, as detailed in Tab. V. Here, $a$ represents the lattice constant of the cubic lattice.

Figures 18 (a1)-(a3) show three distinct twofold Weyl phonons with the Chern numbers of $C$ = +1, +2, and +4 respectively. According to the "bulk-surface correspondence" (Hasan and Kane, 2010; Qi and Zhang, 2011), the number and the chirality of the topological surface states associated with Weyl phonons must match the value and sign of the Chern number. Additionally, threefold and fourfold unconventional Weyl phonons can also exist in solids, stabilized by different crystalline symmetries. These cases will be explored in detail in Sec. IV.A.

### 3. Unconventional Dirac Phonons

As discussed in the previous subsection, surface arcs arise when a pair of Weyl phonons with opposite Chern numbers project to distinct momenta in the surface Brillouin zone, as illustrated in Figs. 18 (a1)-(a3). Conversely, if two Weyl points with opposite chirality project to the same momentum, the surface arcs vanish, as shown in Fig. 18 (a4). When such a pair of Weyl phonons carrying opposite Chern numbers approaches each other under $\mathcal{PT}$ symmetry, they coalesce into a Dirac node accompanied by two counter-propagating helicoidal surface states. Crucially, this $\mathcal{PT}$-protected Dirac phonon carries a net monopole charge of zero, rendering it topologically trivial and devoid of protected surface states in Dirac semimetals (Kargarian *et al.*, 2016; Le *et al.*, 2018). Introducing a symmetry-preserving perturbation hybridizes the anti-parallel surface states, lifting their degeneracy and generating gapped, topologically trivial surface states (Kargarian *et al.*, 2016; Le *et al.*, 2018). Despite this straightforward theoretical picture, two critical challenges remain unresolved before the year of 2018:

(p1). In electronic systems, double Fermi arcs have predominantly been proposed in Dirac semimetals like $Na_3Bi$ (Liu *et al.*, 2014b; Wang *et al.*, 2012; Xiong *et al.*, 2015) and $Cd_3As_2$, (Borisenko *et al.*, 2014; He *et al.*, 2014; Li *et al.*, 2016; Liu *et al.*, 2014a; Neupane *et al.*, 2014; Wang *et al.*, 2013; Yi *et al.*, 2014), they serve as critical experimental signatures of their topological nature. These arcs arise from the hybridization of two anti-parallel surface states, which remains incomplete due to their weak hybridization. A key unresolved question is identifying the conditions that lead to fully gapped surface states (and the absence of Fermi arcs). In other words, what is the microscopic mechanism for obtaining a strong hybridization of two anti-parallel surface states in Dirac phonon systems?

(p2). While the Dirac phonon is not topological in general and associated with gapped surface states, is it possible to obtain topologically protected Dirac phonons associated with topological surface states?

To address (p1), (Le *et al.*, 2018) pointed out that the deformation of the anti-parallel surface states (termed double Fermi arcs in their study) arises due to a significant $k^3$ term that respects all crystal symmetries. They illustrated this with first-principle calculations on $\beta$-CuI, for which the effective Hamiltonian at the Dirac point can be formulated as $H_{\text{eff}}(\boldsymbol{k})$:

$$\begin{aligned} H_{\text{eff}}(\boldsymbol{k}) &= H_0+H_1+H_2, \\ H_0 &= \epsilon(\boldsymbol{k})+M(\boldsymbol{k})\sigma_0\tau_3-A(\boldsymbol{k}_{\parallel})\left(k_x\sigma_3\tau_2+k_y\sigma_0\tau_1\right), \\ H_1 &= \left(D_2+D_3k_z^2\right)\left(-k_x\sigma_1\tau_2+k_y\sigma_2\tau_2\right), \\ H_2 &= -D_1k_z\left[\left(k_x^2-k_y^2\right)\sigma_1\tau_2+2k_xk_y\sigma_2\tau_2\right]. \end{aligned} \tag{76}$$

Hence, the parameter $D_1$ plays a crucial role in the formation of double Fermi arcs in Dirac semimetals, and it corresponds to a three-step hybridization process. First, two I atoms exhibit strong coupling due to substantial atomic spin-orbit coupling (SOC), denoted by $\lambda_1$. Subsequently, a robustly hybridized $\sigma$ bond between I and Cu atoms, denoted by $t_1$. Lastly, the proximity of two Cu atoms contributes to a significant coupling, $t_2$. Usually, the cubic term is negligible in Dirac semimetals; however, in $\beta$-CuI, its relevance stems from the unique crystal structure, where all three parameters are substantial. Consequently, $\beta$-CuI exhibits a large $D_1$ parameter, leading to considerable hybridization of the antiparallel surface states.

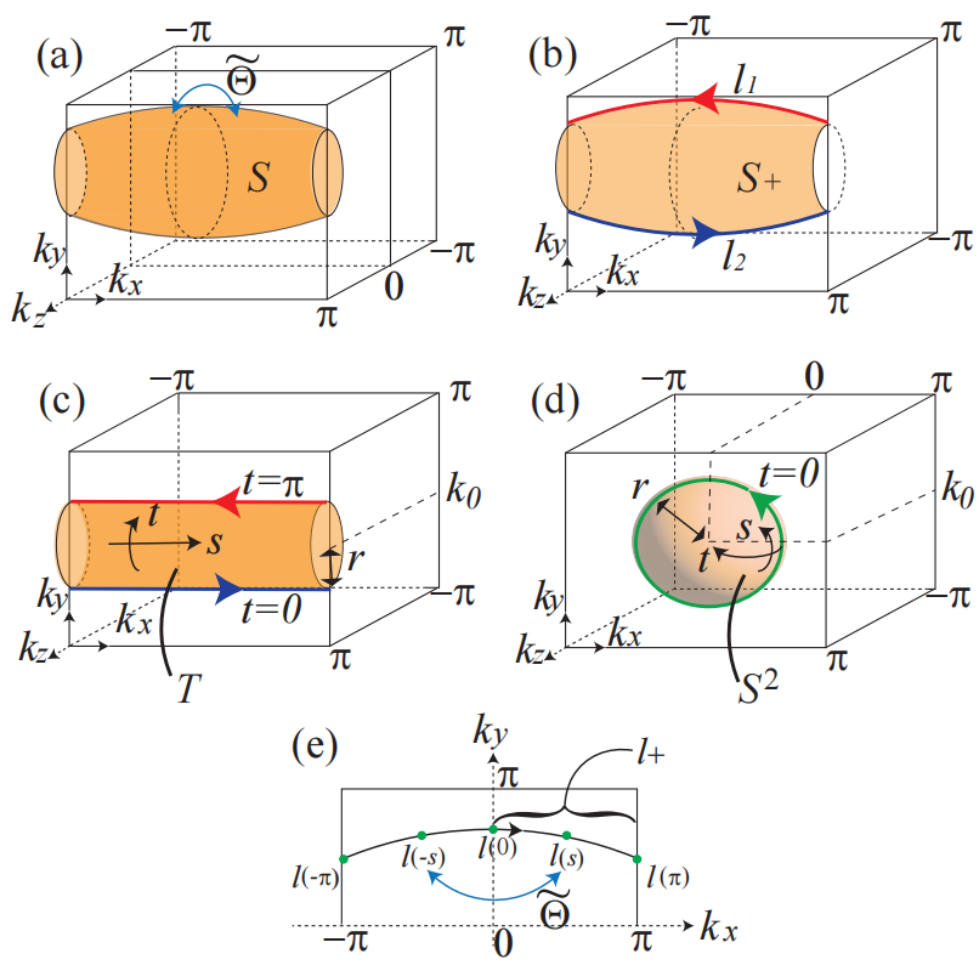

Figure 20 The $\mathbb{Z}_2$ monopole charge $Q$ is defined using specific surfaces and lines. In (a), $Q$ is defined on a surface $S$ that does not intersect with the plane $k_z = 0$. This surface $S$ is closed under $\hat{\Theta}$, with the half part $S_+$ ($k_z \geq \pi$) shown in (b). The boundary of $S_+$, denoted as $\partial S_+ = l_1 \cup l_2$, consists of two lines. Examples of such surfaces $S$ are provided in (c) and (d), where $S$ is defined by a torus $T$ and a sphere $S^2$, respectively. (e) illustrates a closed line $l$ used to define the quantity $P[l]$. Adapted from Ref. (Yukitake *et al.*, 2025b).

Despite the SOC parameter $\lambda_1$ being even more pronounced in Dirac semimetals such as $Na_3Bi$ and $Cd_3As_2$ compared to $\beta$-CuI, the second and third parameters, $t_1$ and $t_2$, are significantly smaller because of the weak bonding between cations and anions. This suggests a correspondingly small value for the parameter $D_1$ in these materials.

In Dirac phonon systems, the SOC parameter $\lambda_1$ is zero due to the spin-zero nature of phonons. However, a large parameter $D_1$ can be achived in systems that exhibit strong chemical bond hybridization or proximity effects.

To address (p2), a new type of topological invariant termed $\mathbb{Z}_2$-type monopole charge $Q$ must be introduced for topological Dirac phonons, since the $\mathbb{Z}$-type monopole charge $C$ is zero. (Fang *et al.*, 2016; Yukitake *et al.*, 2025b; Zhang *et al.*, 2022b; Zhang and Murakami, 2023) have highlighted that in systems exhibiting $\mathcal{TG}$ symmetry, where $\mathcal{G}$ signifies the glide mirror symmetry, this $\mathbb{Z}_2$ monopole charge $Q$ can be defined (Cai *et al.*, 2020; Cheng *et al.*, 2020d; Fang *et al.*, 2016; Kim *et al.*, 2021; Lu *et al.*, 2016a; Qian *et al.*, 2023; Su *et al.*, 2022; Yukitake *et al.*, 2025a,b; Zhang *et al.*, 2022b; Zhang and Murakami, 2023). When $Q \neq 0$, double helicoid surface states emerge, as shown in Fig. 21 (a3). Conversely, if $Q = 0$, the Dirac phonon is trivial with gapped surface states. Details are as follows.

(Yukitake *et al.*, 2025b) considered a system that possess a single glide symmetry operation, denoted as $\mathcal{G}_y = \{M_y|00\frac{1}{2}\}$, as shown in Fig. 21 (a). When the time-reversal symmetry, $\mathcal{T}$, is a symmetry of the system, the operation $\tilde{\Theta}_y = \mathcal{T}\mathcal{G}_y$ is also conserved. It is found that $\tilde{\Theta}_y^2 = e^{-ik_z}$, indicating that the $k_z = \pi$ ($-\pi$) plane is distinctive, as it satisfies the condition $\tilde{\Theta}_y^2 = -1$. In this context, (Yukitake *et al.*, 2025b) examined the $\mathbb{Z}_2$-type Dirac points on the $k_z = \pi$ plane in the 3D BZ. These two Dirac points are connected by $\mathcal{T}$ symmetry, precluding their occurrence at time-reversal-invariant momenta (TRIM). Consequently, $\mathbb{Z}_2$ Dirac points typically emerge along high-symmetry lines in systems possessing a single glide mirror symmetry, and at non-TRIM high-symmetry points in systems with two vertical glide mirror symmetries (Fang *et al.*, 2016; Kim *et al.*, 2021; Yukitake *et al.*, 2025a,b; Zhang *et al.*, 2022b; Zhang and Murakami, 2023). Subsequently, we will review the convention-independent $\mathbb{Z}_2$ monopole charge $Q[S]$ associated with an oriented closed surface $S$, which adheres to the following criteria as established by (Yukitake *et al.*, 2025b): (i) The surface $S$ is closed under the operation $\hat{\Theta} : (k_x, k_y, k_z) \mapsto (-k_x, k_y, -k_z)$, (ii) The surface $S$ does not intersect with the plane defined by $k_z = 0$ (iii) The system exhibits a bulk energy gap over the entire surface $S$.

Analogous to $\mathbb{Z}$-type monopole charge $C$, the $\mathbb{Z}_2$-type monopole charge $Q$ is defined in terms of wavefunctions over a sphere $S$ that encloses the Dirac point, as described in (Fang *et al.*, 2016). Thus, $Q$ can be expressed as:

$$Q = \frac{1}{2\pi}\int_{S_+} d\boldsymbol{S}\cdot\nabla_{\boldsymbol{k}}\times\boldsymbol{A}(\boldsymbol{k}) - 2\sum_{l\in\partial S_+} P[l] \pmod 2, \quad (77)$$

where $S_+$ is the divided part with $k_z \geq \pi$ shown in Fig. 20 (b) and $\partial S_+$ is the boundary of $S_+$ on the plane $k_z = \pi$. $P[l]$ is defined as $P[l] = \frac{1}{2\pi}\gamma^\alpha[l] \pmod 1$, where $\gamma^\alpha[l]$ represents the Berry phase when the occupied bands are divided into two groups $\alpha$ and $\beta$, such that $\tilde{\Theta}$ transforms the two groups mutually. Since the Berry curvature is gauge-independent and $P[l]$ is gauge-independent up to an integer, the definition of $Q$ in Eq. (77) maintains gauge invariance. Moreover, when a smooth gauge choice is adopted over the surface $S$, it can be shown that $Q$ assumes integer values, as expressed by:

$$(-1)^Q = \prod_{l\in\partial S_n} \frac{\mathrm{Pf}\,\omega[l(0)]}{\sqrt{\det\omega[l(0)]}}\frac{\mathrm{Pf}\,\omega[l(\pi)]}{\sqrt{\det\omega[l(\pi)]}}. \quad (78)$$

When the integration surface is chosen to be a torus $T$, as depicted in Fig. 20 (c), the $\mathbb{Z}_2$ monopole charge $Q[T]$ can be expressed as follows:

$$Q[T] = P_{\tilde{\Theta}}(t = \pi) - P_{\tilde{\Theta}}(t = 0) \pmod 2. \quad (79)$$

At $t = \pi$ and $t = 0$, we encounter two distinct lines on the $k_z = \pi$ plane, which are positioned on either side of the Dirac point. The operator $P_{\tilde{\Theta}}$ is defined as:

$$P_{\tilde{\Theta}}(l_i) = \frac{1}{2\pi}(\gamma^{\alpha}[l_i] - \gamma^{\beta}[l_i]) \tag{80}$$

Thus, $Q[T]$ is the difference of $\tilde{\Theta}$-polarization, and it serves to establish the bulk-surface correspondence of $\mathbb{Z}_2$ Dirac points by considering the surface $T$ as a 2D BZ $[-\pi, \pi] \times [-\pi, \pi]$ in the $(s, t)$ parameter space, as shown in Fig. 21 (a1). Given that $\tilde{\Theta}^2 = -1$ along the lines where $t = 0$ and $t = \pi$, the anti-unitary operator $\tilde{\Theta}$ behaves analogously to time-reversal symmetry in $\mathcal{T}$-symmetric $\mathbb{Z}_2$ topological insulators (Fu *et al.*, 2007). This implies that the energy bands, both from the bulk and the surface, will exhibit Kramers's degeneracy at the points at $t = 0$ and $t = \pi$. Upon projection onto the (100) surface BZ, the torus $T$ is projected onto a circle path $c$, as marked by the orange circle in Fig. 21 (a1). If the $\mathbb{Z}_2$ charge $Q[T]$ is nontrivial, an odd number of $\mathbb{Z}_2$ Dirac points are enclosed within the torus. Consequently, a pair of helical surface states will emerge on the circle $C$, as shown by the red and blue bands in Fig. 21 (a2). These two surface states are degenerate at $t = 0$, resulting in a degenerate line along the boundary of the surface BZ. Adjusting the radius of circle $C$ allows for the emergence of two helical surface states, as shonw in Fig. 21 (a3). The intersection of these two surface states is protected by the symmetry $\mathcal{GT}$, leading to the formation of double helical (anti-parallel) surface states. Figure 21 (a4) illustrates the Fermi surface of a system hosting two $\mathbb{Z}_2$ Dirac points, featuring two Fermi arcs that link the projection of the Dirac points.

$\mathbb{Z}_2$-type Dirac points may reside at various momenta, subject only to the constraint of satisfying the $\tilde{\Theta}_y$ symmetry, which is associated with a high-symmetry line in the BZ. When two Dirac points are projected onto distinct momenta within the surface BZ, a pair of anti-parallel helical surface states emerges. These states are topologically protected by the charge $Q$, as shown in Fig. 21 (b2). When two $\tilde{\Theta}_{i,j}$ symmetries are present, with $i, j \in \{x, y, z\}$, Dirac points are confined to the high-symmetry points within the BZ, as listed in Tab. 21. In this case, if two Dirac points are projected onto the identical momentum within the surface BZ, the quadruple surface states appear, as shown in Fig. 21 (b3).

In certain instances, a $\mathbb{Z}_2$ Dirac point may split into a pair of Weyl dipoles or transform into a $\mathbb{Z}_2$ nodal ring, a transformation that depends on the crystalline symmetries inherent to the system. All of the topological band crossings mentioned here can be cataloged by the $\mathbb{Z}_2$-type monopole charge $Q$, associated with topologically protected surface states (Cai *et al.*, 2020; Cheng *et al.*, 2020d; Fang *et al.*, 2016; Kim *et al.*, 2021; Lu *et al.*, 2016a; Qian *et al.*, 2023; Su *et al.*, 2022; Yukitake *et al.*, 2025a,b; Zhang *et al.*, 2022b; Zhang and Murakami, 2023).

### 4. Nodal line Phonons

The gapless topological phonon states discussed above are characterized by band crossings occurring at discrete zero-dimensional points in momentum space. In certain cases, however, these band crossings can extend into one-dimensional manifolds in the three-dimensional Brillouin zone, forming so-called nodal lines or nodal rings (Burkov *et al.*, 2011; Fang *et al.*, 2016; Li *et al.*, 2025; Zhang *et al.*, 2019b). Nodal lines or rings typically require the protection of additional symmetries, such as mirror symmetry, product of spatial inversion and time-reversal symmetry ($\mathcal{PT}$), etc. Perturbations that conserve these symmetries cannot eliminate the crossing line and thereby cannot open a complete energy gap. Consider a two-band phonon system that preserves mirror symmetry $M_z$; the low-energy effective Hamiltonian for this system can be expressed as:

$$H_{\mathrm{NL}}(\boldsymbol{k}) = (m - k_x^2 - k_y^2)\sigma_z + k_z\sigma_x, \tag{81}$$

with the eigenvalues of $E(\boldsymbol{k}) = \pm\sqrt{(m - k_x^2 - k_y^2)^2 + k_z^2}$. Hence, Eq. (81) captures the characteristics of nodal line/ring phonon constrained on the $k_z = 0$ plane (a mirror-invariant plane), and the configuration of the nodal line/ring being influenced by the system's parameters, as shown by the nodal ring in Fig. 22. As $m$ approaches zero, the nodal ring will shrink to a gapless point, evolving into a quadratic Weyl phonon.

The topological invariant characterizing node-line/ring phonons is the Berry phase, denoted as $\gamma$, which is defined by an integral on a closed loop as shown in Fig. 22 (b).

$$\gamma = i\oint_C \langle u(\boldsymbol{k})|\nabla_{\boldsymbol{k}}|u(\boldsymbol{k})\rangle \cdot \mathrm{d}\boldsymbol{k} \mod 2\pi, \tag{82}$$

wherein the phonon bands are entirely separated by a gap along the $k$-loop. If the loop encircles the nodal line/ring, then $\gamma = \pi$; conversely, if the loop does not encircle the nodal line/ring, $\gamma = 0$.

In addition to the mirror-protected node-line/ring phonons, which are fixed to the mirror-invariant plane, the combined spatial inversion and time-reversal symmetry ($\mathcal{PT}$) can also safeguard nodal lines/rings. These nodal lines/rings may exist in various configurations and at any momentum within the Brillouin zone (Zhang *et al.*, 2019b). Independent of the specific symmetries of the system, node-line/ring phonons with $\gamma = \pi$ host topological surface states. These correspond to the drumhead surface states that link the nodal line/ring band crossings on the surface BZ, as depicted by the blue curve in Fig. 22 (c).

## IV. TOPOLOGICAL PHONONIC MATERIALS

The application of first-principles calculations in condensed matter physics and materials science has expanded significantly with the routine implementation of phonon calculations over the past decade. Density functional theory (DFT) for solids provides second-order force constants, enabling the determination of phonon spectra and topological properties.

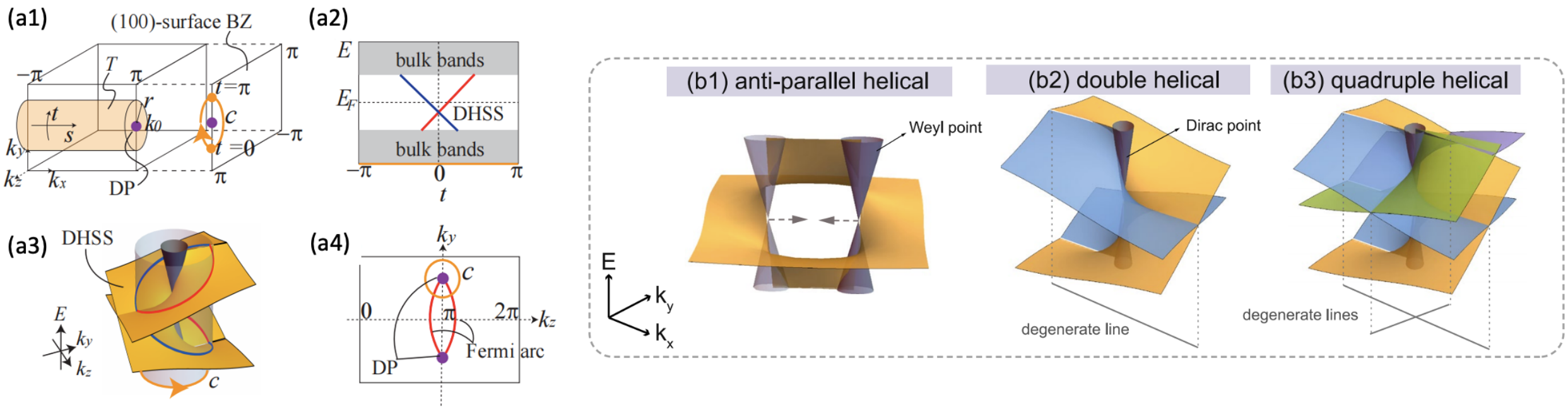

Figure 21 "Bulk-surface correspondence" for the $\mathbb{Z}_2$ monopole charge $Q[T]$, defined on the torus $T$. (a1) Bulk and (100) surface Brillouin Zones. A $\mathbb{Z}_2$ Dirac point is situated within the $k_z = \pi$ plane of the torus $T$ and is projected onto the circle $\mathcal{C}$. The $k_z = \pi$ plane is characterized by $\mathcal{TG}^2 = -1$. The torus may be conceptualized as a 2D rillouin Zone in the parameter space of $(s, t)$, where Krammers' degeneracy appears at $t = \pi, 0$. (a2) When a $\mathbb{Z}_2$ Dirac point is enclosed within $T$, the surface states along the loop $\mathcal{C}$ exhibit helical characteristics and are degenerate at $t = 0$ ($t = \pi$). (a3) By altering the radius of the loop $\mathcal{C}$, double-helicolid surface states is obtained, associated with two Fermi arcs shown in (a4). The charge $Q[T]$ is applicable to systems with $\mathbb{Z}_2$-charged nodal ring or a Weyl dipole, which can interconvert with Dirac points. (b1) Represents the topological surface state connecting a pair of Weyl phonons with opposite chirality. (b2) and (b3) are the double and quadruple helical surface states, respectively, which are attributed to the $\mathbb{Z}_2$ Dirac phonons. (a) is adapted from Ref. (Yukitake *et al.*, 2025b) and (b) is adapted from Ref. (Zhang and Murakami, 2023).

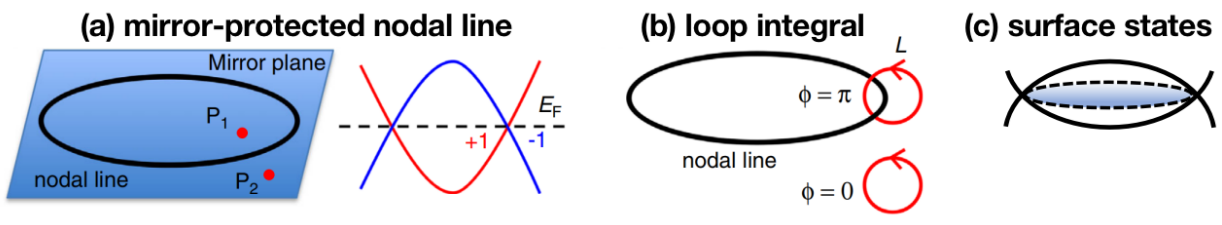

Figure 22 Illustration of node-line phonons. (a) The mirror-protected nodal ring phonon is constrained to the mirror plane and consists of two bands with distinct mirror eigenvalues. (b) The Berry phase is $\pi$ when the integration loop encircles the nodal ring, and it is zero when the integration loop is positioned away from the nodal ring. (c) The drumhead surface state associated with the nodal ring.

However, prior to 2018, research on topological phonons remained largely theoretical, with limited exploration of real materials in solid-state phonon systems. This gap arose because the calculation of phonon spectra requires solving real-space force constants (typically truncated at second order), a far more computationally demanding task than electronic structure calculations.

In electronic systems, topological surface states are commonly computed using tight-binding models constructed from localized Wannier functions, where the surface Green's function of a semi-infinite system is evaluated iteratively. Noting the formal analogy between real-space force-constant matrices and $p$-orbital hopping parameters, Ref. (Zhang *et al.*, 2018a) extended the first-principles tight-binding framework to phonon spectra. This approach enables efficient calculations of topological phonon surface states and facilitates the prediction of topological phonon materials with substantially reduced computational cost.

In realistic solids, additional care is often required when diagnosing the topological indices of phonon band structures, as the conventional force-constant approach may become insufficient. A notable example is the longitudinal-optical–transverse-optical (LO–TO) splitting effect in polar semiconductors. In polar crystals, the long-range macroscopic electric field induced by the LO mode can induce a nonlocal term in the dynamical matrix as $\mathsf{D}(\boldsymbol{q}) = \mathsf{D}^{\text{local}}(\boldsymbol{q}) + \mathsf{D}^{\text{nonlocal}}(\boldsymbol{q})$ where the local term $\mathsf{D}^{\text{local}}(\boldsymbol{q})$ is ordinary dynamical matrix based on the second-order force constants given by Eq. (4) and the nonlocal term $\mathsf{D}^{\text{nonlocal}}(\boldsymbol{q})$ is given by (Gonze and Lee, 1997; Wang *et al.*, 2010)

$$\mathsf{D}^{\text{nonlocal}}_{i\alpha,j\beta}(\boldsymbol{q}) = \begin{cases} \dfrac{4\pi}{V_{\text{uc}}\sqrt{M_i M_j}} \dfrac{(\mathsf{Z}_i \cdot \boldsymbol{q})_\alpha (\mathsf{Z}_j \cdot \boldsymbol{q})_\beta}{\boldsymbol{q} \cdot \epsilon_\infty \cdot \boldsymbol{q}} & (\boldsymbol{q} \to 0) \\ 0 & (\boldsymbol{q} = 0) \end{cases} . \tag{83}$$

$V_{\text{uc}}$ is the volume of unit cell, and $M_{i,j}$ and $\mathsf{Z}_{i,j}$ refer to the atomic mass and Born effective charge matrix of $i, j$-th atom, respectively. $\epsilon_\infty$ refers to the dielectric matrix in the high-frequency limit. The nonlocal term split the degeneracy of LO and TO modes at $\boldsymbol{q} = 0$ and is non-analytical. (Xu *et al.*, 2024c) show that this non-analytical correction (NAC) can induce additional band inversion which may change the topological index of phonon spectrum. $Dy_2O_3$ is such an example. Without NAC, the lowest 11 phonon bands of $Dy_2O_3$ form a topologically nontrivial subset which is totally gapped from the other bands. However, if NAC is taken into account the band order at Γ is further switched which makes the subset topologically trivial.

Another interesting aspect of polaritonic effect is about the topological phonon polaritons (Guddala *et al.*, 2021; Hu *et al.*, 2020; Karzig *et al.*, 2015; Wang and Zhao, 2023). When coupling a transverse optical (TO) phonon branch with a photon mode, the "flat" TO mode can hybridize with the photon branch to open a band gap as shown in Fig. 23. In this hybrid phonon-photon system, the "phonon" and "photon" sectors can be viewed as two different orbitals of the effective Hamiltonian and the band inversion is clearly seen in Fig. 23(a)

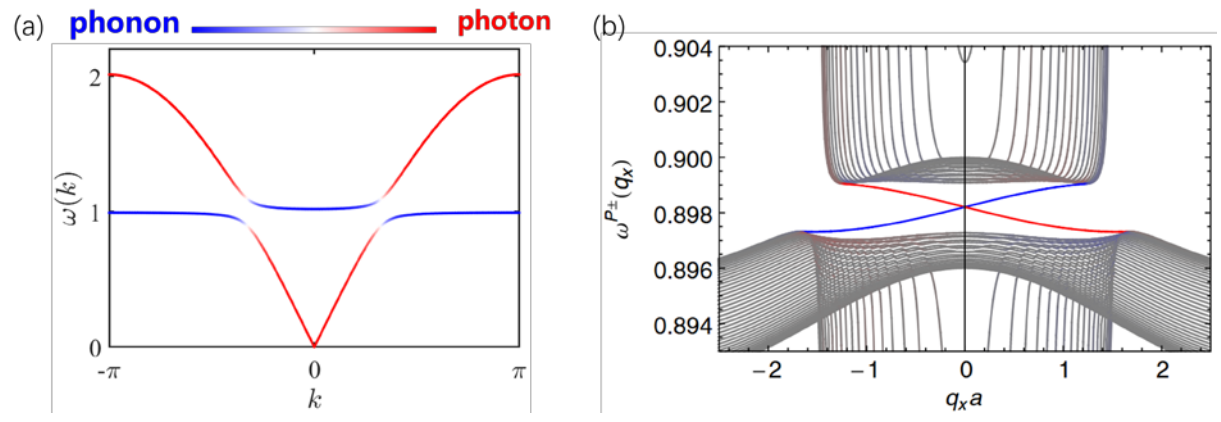

Figure 23 Schematic band structure of phonon polariton. (a) If we treat the "phonon" part and "photon" part as two different orbitals of the hybrid Hamiltonian, the band structure shows a clear band inversion between $k = 0$ and $k \to \infty$. (b) Topological polaritonic edge states, adapted from (Karzig *et al.*, 2015).

between $k = 0$ and $k \to \infty$. The effective Hamiltonain can be expressed as

$$H = \begin{pmatrix} \omega_{\text{phonon}}(\boldsymbol{k}) & \Delta \\ \Delta^{\dagger} & \omega_{\text{photon}}(\boldsymbol{k}) \end{pmatrix}, \tag{84}$$

where $\omega_{\text{phonon}}$ and $\omega_{\text{photon}}$ represent the dispersion of bare phonon and photon, respectively; $\Delta$ represent the phonon-photon coupling. If the coupling parameter $\Delta$ is carefully designed with a nonzero winding number i.e. $\Delta = |\Delta| \exp(\pm i\theta_{\boldsymbol{k}})$ ($\theta_{\boldsymbol{k}}$ refers to the polar angle of $\boldsymbol{k}$ in a 2D system), the effective Hamiltonian can be directly mapped to the topological Qi-Wu-Zhang model with Chern number $C = \pm 1$ (Qi *et al.*, 2006). Figure 23(b) shows the edge states of topological polaritonic system with chiral edge states. Although historically it is first proposed in the exciton polariton system, in principle, it can be generalized to phonon polaritons.

This section reviews representative materials exhibiting distinct topological phases in their phonon spectra. All considered systems preserve time-reversal symmetry $\mathcal{T}$, displaying gapless states with varying band degeneracies, typically protected by crystalline symmetries, as well as gapped topological phases analogous to "obstructed atomic insulators". We systematically classify these topological phonons in Secs. IV.A and IV.B according to the codimension, dispersion and configuration of the band degeneracy. For each category, we discuss corresponding topological invariants, characteristic surface states and material realizations from first-principles calculations. The subsequent sections will present topological phonons with "obstructed atomic insulators" anologous states in Sec. IV.C and high-throughput computational results and the topological phonon materials database in Sec. IV.D.

## A. Topological Nodal Point Phonons

The simplest topological phonon with a point degeneracy is the Weyl phonon, conventionally characterized by a Chern number of $C = \pm 1$, as shown in Fig. 24 (a). Such a Weyl phonon can be described by the two-band Hamiltonian of $H(\boldsymbol{k}) = \boldsymbol{k} \cdot \boldsymbol{S} = \frac{\hbar}{2} \boldsymbol{k} \cdot \boldsymbol{\sigma}$, where $\sigma_i$ are the Pauli matrices and $\boldsymbol{S}$ is the spin-1/2 rotation generator, hence the term spin-1/2 Weyl phonon. When the leading-order $k$-dependence is

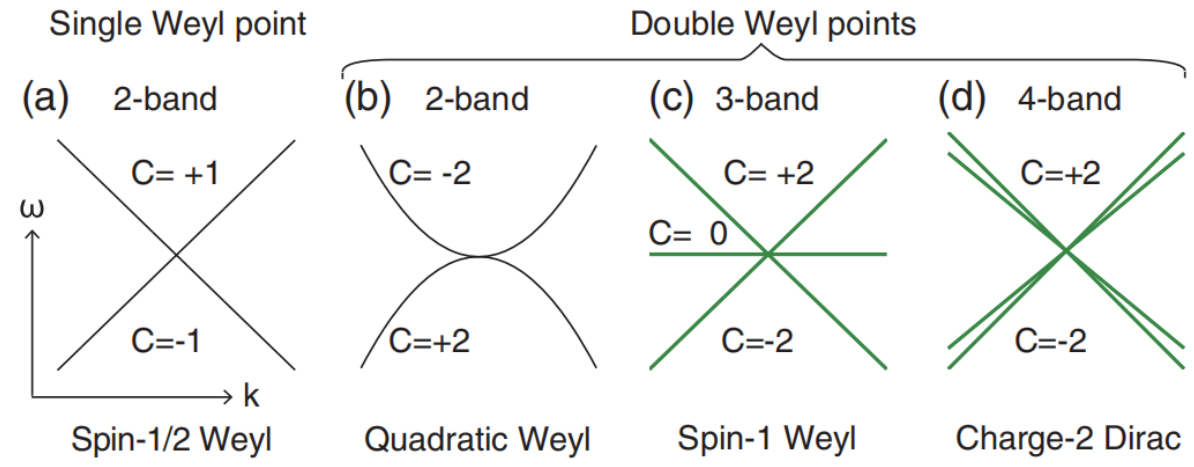

Figure 24 (a) Conventional Weyl phonon. (b-c) Three types of double-Weyl phonons with $|C| = 2$, where (b) is the twofold quadratic Weyl point carrying Chern numbers of $\pm 2$ for each band, (c) is the threefold spin-1 Weyl point with Chern numbers of 0, $\pm 2$ for each band, and (d) is the fourfold charge-2 Dirac point exhibiting Chern numbers of $\pm 2$ for each phonon band pair. (Zhang *et al.*, 2018a) refer to the quadratic Weyl, spin-1 Weyl, and charge-2 Dirac phonons as double-Weyl phonons. Adapted from Ref. (Zhang *et al.*, 2018a).

quadratic, unconventional Weyl phonons with Chern numbers of $C = \pm 2$ emerge, as described by Eq. (74) for $N = 2$. Due to their quadratic dispersion in the $(k_x, k_y)$ plane, these are referred to as quadratic Weyl phonons, illustrated in Fig. 24 (b).

In addition to the case described by Eq. (74) with $N = 2$, double Weyl phonons can emerge in a three-band Hamiltonian $H(\boldsymbol{k}) = \boldsymbol{k} \cdot \boldsymbol{L}$, where $\boldsymbol{L}$ represents the spin-1 rotation generators. The eigenvalues of $L_k = \boldsymbol{L} \cdot \boldsymbol{k} = \hbar, -\hbar, 0$ correspond to Chern numbers $+2, -2, 0$, doubling those of the spin-1/2 Weyl phonon. This motivates the term spin-1 Weyl phonon, as illustrated in Fig. 24 (c).

Another approach to generate double Weyl phonons with is by coupling two identical spin-1/2 Weyl phonons, described by the block-diagonal Hamiltonian of $H(k) \sim \begin{pmatrix} \boldsymbol{k} \cdot \boldsymbol{\sigma} & 0 \\ 0 & \boldsymbol{k} \cdot \boldsymbol{\sigma} \end{pmatrix}$, resulting in a fourfold degenerate Weyl phonon, as shown in Fig. 24 (d). Unlike a conventional 3D Dirac point, composed of two Weyl points with opposite Chern numbers, this Dirac point exhibits a monopole charge of 2, thus it is also referred to "charge-2 Dirac phonon".

### 1. B-20 Type Materials: Double Weyl Phonons

The first realizations of spin-1 Weyl phonons and charge-2 Dirac phonons were theoretically predicted in the phonon spectra of the $M$Si family ($M$ = Fe, Co, Mn, Re, Ru) (Zhang *et al.*, 2018a), as shown in Fig. 25, with experimental confirmation following in the same year (Miao *et al.*, 2018). $M$Si family belongs to the B-20 structure, a simple cubic crystal structure with space group $P2_13$ (#198), as shown in Fig. 25 (a). Figure 25 (b) is the BZ and surface BZ (green plane) along (001) direction. Using FeSi as a representative example, (Zhang *et al.*, 2018a) identified two types of double Weyl phonons in the $M$Si family. The phonon spectrum shown in Fig. 25 (c) reveals that the three acoustic branches form a spin-1 Weyl phonon. The acoustic phonons inherently exhibit threefold degeneracy at the $\Gamma$ point. This degeneracy is lifted in chiral crystals. Thus, in 3D chiral crystals, all acoustic phonons exhibit spin-1 Weyl character, where the longitudinal acoustic (LA) phonon carries a zero Chern number, and the two transverse acoustic (TA) phonons have Chern numbers of $\pm$ 2. However, the topological surface states of the spin-1 Weyl phonon remain unobservable due to the absence of a bulk gap between the LA phonon branches.

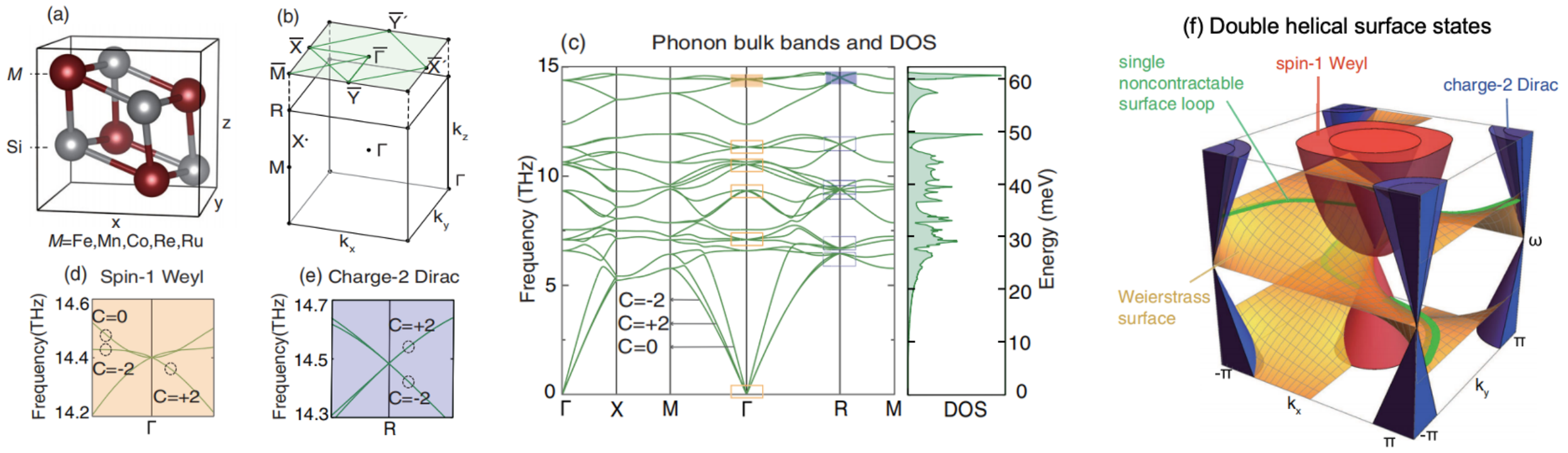

Figure 25 Crystal structure, phonon spectra, and topological phonons in the $M$Si family ($M$ = Fe, Mn, Co, Re, Ru). (a) Crystal structure of the $M$Si family. (b) Bulk and (001) surface Brillouin zones. (c) Phonon dispersion of FeSi along high-symmetry lines in the rillouin zone. The threefold acoustic phonons are spin-1 double Weyl phonons with Chern numbers of 0 and $\pm 2$. Orange and purple boxes highlight the spin-1 Weyl phonon at $\Gamma$ and the charge-2 Dirac phonon at R, respectively. (d) The $\Gamma$ point hosts a spin-1 Weyl phonon. (e) The R point exhibits a charge-2 Dirac phonon. (f) Double-helicoid surface states, described by the Weierstrass elliptic function. The spin-1 Weyl phonon (red cone) corresponds to a double zero point, while the charge-2 Dirac phonon (purple cone) represents a double pole. Two yellow sheets depict topological surface states rotating around these double-Weyl phonons. Adapted from Ref. (Zhang *et al.*, 2018a).

In the optical phonon spectra of FeSi, all threefold-degenerate modes at the $\Gamma$ point are spin-1 Weyl phonons, while all fourfold-degenerate modes at the R point are charge-2 Dirac phonons, as highlighted by orange and purple boxes in Figs. 25 (c)–(e). The spin-1 Weyl phonons are stabilized by chiral cubic symmetry, whereas the charge-2 Dirac phonon is protected by both crystalline symmetries and time-reversal symmetry $\mathcal{T}$. (Zhang *et al.*, 2018a) offered an intuitive perspective to understand the charge-2 Dirac phonon at R. $M$Si family materials possesses twofold screw rotations along each axis, such as $\{C_{2x}|(\frac{1}{2},\frac{1}{2},0)\}$ along $x$ axis, and the other two along $y$ and $z$ axes. These operations anti-commute with each other and satisfy $C_{2x/y/z}^2 = -1$ at R. Since such behavior is analogous to half-integer spin rotation, all irreducible representations at R must have even dimensions. Moreover, R is a time-reversal-invariant momentum and preserves $\mathcal{T}$, the screw rotations must also commute with $\mathcal{T}$, enforcing real matrix representations. The smallest allowed representation under these constraints is 4D, forcing all bands at R to form charge-2 Dirac phonons, a direct consequence of the interplay between crystalline symmetries and $\mathcal{T}$.

According to the "bulk-surface correspondence", two surface states must connect these double Weyl phonons, topologically equivalent to helicoid noncompact Riemann sheets (Fang *et al.*, 2016). This allows the surface state dispersion near the Weyl phonon projections to be mapped to Riemann surfaces of analytic functions, treating the surface momentum as a complex variable. As an illustrative example, consider the highest optical branch in FeSi ($\sim$14.5 THz). Figures 25 (d) and (e) show zoomed-in dispersions of the spin-1 Weyl phonon at $\Gamma$ (monopole charge +2) and the charge-2 Dirac phonon at R (monopole charge $-2$), respectively. In the $M$Si family, Weyl phonons carry Chern numbers of $C = \pm 2$, making their (001) surface state dispersion topologically equivalent to the winding phase of the Weierstrass elliptic function ($\wp$). Specifically, the function has an order-$C$ zero at $\boldsymbol{k}_- = (0,0)$, and order-$C$ pole at $\boldsymbol{k}_+ = (\pi,\pi)$:

$$\omega(k_x,k_y) \sim \wp(z;2\pi,2\pi) = \mathrm{Im}\{\log[\frac{1}{z^2} + \sum_{n,m\neq 0}(\frac{1}{(z+2m\pi+2n\pi i)^2} - \frac{1}{(2m\pi+2n\pi i)^2})]\}, \tag{85}$$

where $z$ represents the planar momentum relative to the Weyl phonon projection, defined as $z \equiv k-(1+i)\pi$. As illustrated in Fig. 25 (f), the spin-1 Weyl phonon (red cone) projects to the center of the (001) surface Brillouin zone, while the charge-2 Dirac phonon (blue cone) projects to the Brillouin zone corner. The double-helicoid surface states of the $M$Si family are depicted as yellow sheets. The isoenergetic contour reveals surface arcs (green curves) with distinct topological features. Unlike conventional Weyl materials where surface arcs are open and terminate at bulk Weyl projections, the $M$Si family exhibits a single non-contractible loop that rotates around both double Weyl phonons as the energy contour varies.

It's worthy to note that all the twofold degeneracies for B-20 type materials at $\Gamma$ are another type of Weyl phonons with $C = \pm 4$, referred to as twofold quadruple Weyl phonon (Li *et al.*, 2021a; Zhang *et al.*, 2020b), as mentioned in Sec. III.D.1. In solids, there are three distinct mechanisms to realize Weyl points with a monopole charge of 4, i.e., double spin-1 Weyl point, spin-3/2 Weyl point and twofold quadruple Weyl point, as shown in Figs. 26 (a), (b) and (c), respectively.

The sixfold double spin-1 Weyl node consists of two identical spin-1 Weyl points, with its monopole charge arising from the two lowest bands, as shown in Fig. 26 (a). The spin-3/2 Weyl point is described by a four-band Hamiltonian of $H(\boldsymbol{k}) = \boldsymbol{k} \cdot \boldsymbol{L}$, with $\boldsymbol{L}$ representing the spin-3/2 rotation generators. This yields Chern numbers of +1, +3, $-1$, $-3$ for each band. Regarding twofold Weyl point, crystals can host three additional types with Chern numbers $\pm1$, $\pm2$ and $\pm3$, respectively. The twofold quadruple Weyl point with Chern numbers of $\pm3$ remained unrealized for an extended period (before 2020) due to the oversight of chiral cubic symmetry in topological band theory analyses. We will discuss the twofold Weyl quadruple phonon in detail in Sec. VI.A.

Xiao *et al.* pointed out that some kinds of nodal surface can carry integer-valued topological charge similar to Weyl monopoles, *i.e.* a topological phase transition can occur when the topological charge of the nodal surface changes by simultaneously emitting or absorbing integer numbers of Weyl points (Xiao and Fan, 2017). In realistic materials, crystals with two-fold screw rotation symmetry $S_{2\alpha}$ ($\alpha = x, y, z$) can host nodal surface phonons at $k_\alpha = \pi$ when combined with time-reversal symmetry (Liu *et al.*, 2021d; Xie *et al.*, 2022, 2021). Consider a lattice invariant under the nonsymmorphic operation $S_{2x} = \{C_{2x}|[\frac{1}{2}00]\}$. Since $(S_{2x})^2 = \{E|[100]\}$ represents a primitive lattice translation along $x$ (assuming lattice constant $a$ = 1), Bloch's theorem gives $(S_{2x})^2 = e^{ik_x}$. At the Brillouin zone boundary $k_x = \pi$, the anti-unitary operator $S_{2x}\mathcal{T}$ satisfies $(S_{2x}\mathcal{T})^2 = (S_{2x})^2 = -1$, enforcing Kramers-like degeneracy across all bands and consequently generating a nodal surface phonon.

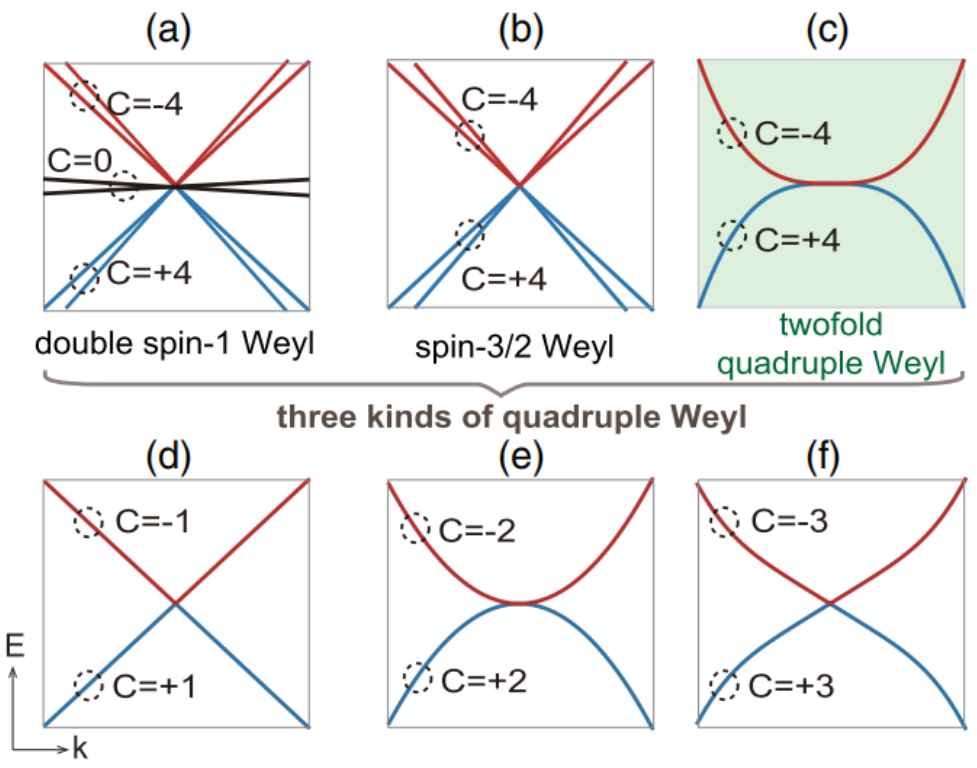

Figure 26 (a–c) Three kinds of Weyl points with the monopole charge of 4. (a) Sixfold double spin-1 Weyl phonon, which is a composite of two identical spin-1 Weyl points. (b) Fourfold spin-3/2 Weyl phonon with Chern numbers of +1, +3, $-1$, and $-3$ for each band. (c) Twofold quadruple Weyl phonon with Chern numbers of +4 and $-4$ for each band. (d–f) Three types of twofold Weyl phonons with distinct Chern numbers that can exist in crystals, with Chern numbers of $\pm1$, $\pm2$ and $\pm3$, respectively. Adapted from Ref. (Zhang *et al.*, 2020b).

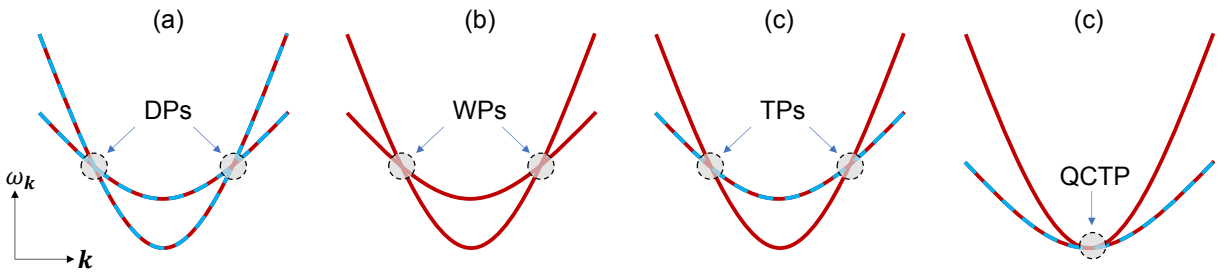

Figure 27 Schematic of phononic topological nodal points: (a) Dirac points (DPs), arising from the band crossings of two sets of doubly degenerate bands, (b) Weyl points (WPs), formed at the band crossing of two non-degenerate bands, (c) Triple points (TPs), formed by the crossing between one non-degenerate band and two doubly degenerate bands, and (d) Quadratic contact triple point (QCTP), characterized by a quadratic dispersion relation in the vicinity of the three-band touching point.

### 2. Face-centered Silicon: Triple Point Phonon

By definition, a topological Dirac point in a 3D system is a fourfold degenerate point formed by the crossing of two pairs of doubly degenerate bands, as illustrated in Fig. 27 (a). In contrast, a topological Weyl phonon (with Chern number of $C = \pm1$) arises from the crossing of two non-degenerate bands, shown in Fig. 27 (b). Between these two cases lies the topological triple point (TP) phonon, which emerges from a three-band crossing involving a doubly degenerate band and a single non-degenerate band [Fig. 27 (c)]. When these three bands exhibit quadratic touching, the degenerate point is termed a quadratic contact triple point (QCTP), as depicted in Fig. 27 (d).

In phonon systems, TPs typically emerge along high-symmetry lines with rotational symmetry $C_n$ ($n \geq 3$). Here, the transverse (T) phonon modes form a doubly degenerate pair (corresponding to a 2D irreducible representation of the point group), while the longitudinal (L) phonon mode remains non-degenerate (a 1D representation). The crossing between the doubly degenerate T bands and the L band results in a symmetry-protected TP, which remains robust against gap opening as long as the rotational symmetry is preserved. In contrast, the QCTP is stabilized by cubic crystalline symmetry at certain high-symmetry points, where the L and T modes collectively form a 3D irreducible representation.

For TPs with the rotation axis aligned along the $k_z$-direction, the L mode can be described by a $|p_z\rangle$-like orbital state, while the two degenerate T modes correspond to the chiral states $|p_x \pm ip_y\rangle$ states. In basis of $\{|p_x + ip_y\rangle, |p_x - ip_y\rangle, |p_z\rangle\}$, the 3-band effective Hamiltonian near a TP can be expressed as:

$$H_{\mathrm{TP}} = \begin{pmatrix} v_T k_z & 0 & ck_+ \\ 0 & v_T k_z & ck_- \\ c^* k_- & c^* k_+ & v_L k_z \end{pmatrix} + O(k^2). \tag{86}$$

$v_T$ and $v_L$ refer to the group velocity of T and L modes, respectively, along $k_z$ axis. $c$ is a complex coupling constant.

We define $k_\pm = k_x \pm ik_y$, and $k_z$ are referenced to the TP. Without loss of generality, we suppose $v_T > v_L$, leading to the formation of a topological nexus (Fig. 28), where a nodal line arises between the lower (upper) two bands along the positive (negative) $k_z$-axis. Unlike conventional closed nodal lines, the nexus features open nodal line segments that terminate at the TPs. Within the degenerate subspace, the two-band effective Hamiltonian can be derived using the Löwdin partitioning method, yielding:

$$H_2 = \begin{pmatrix} 0 & Ak_-^2 \\ Ak_+^2 & 0 \end{pmatrix} = A[(k_x^2 - k_y^2)\sigma_x + 2k_xk_y\sigma_y], \quad (87)$$

up to an irrelevant band bending term and $A = \frac{|c|^2}{(v_T - v_L)k_z}$. This Hamiltonian exhibits a chiral symmetry expressed by the anti-commutation relation $\{\sigma_z, H_2\} = 0$, which enables the definition of the topological winding number:

$$n_W = \oint_C \frac{\mathrm{d}l}{4\pi i} \operatorname{Tr}\left[\sigma_z H_2^{-1} \partial_l H_2\right] = \pm 2. \quad (88)$$

The winding number $n_W$ can be defined for any closed loop $C$ encircling the $k_z$-axis at fixed $k_z$. A nonzero $n_W$ directly reflects the topological protection of the nodal structure and is further linked to the Euler characteristic of the band crossing, as demonstrated in (Park *et al.*, 2021; Peng *et al.*, 2022a,b).

The topology of TPs can also be understood by its topological phase transitions via symmetry breaking terms. When time-reversal symmetry $\mathcal{T}$ is broken while rotational symmetry $C_n$ is preserved, an additional term $H_2' = \delta_T\sigma_z$ modifies the effective Hamiltonian $H_2$. This splits each TP into Weyl point pairs, giving rise to double Fermi-arc-like surface states (Liu *et al.*, 2022). On the other hand, if the rotation symmetry $C_n$ is broken (e.g., by external strain) while $\mathcal{T}$ remains intact, a perturbation $H_2'' = \delta_S\sigma_x$ emerges. Unlike the $\mathcal{T}$-breaking case, $H_2''$ preserves line degeneracy but fragments the nexus into a set of nodal lines, whose positions are determined by

$$\begin{aligned} A(k_x^2 - k_y^2) + \delta_S &= 0, \\ k_xk_y &= 0. \end{aligned} \quad (89)$$

These nodal lines are characterized by a winding number $n_W = \pm 1$ and organize into interlocking Hopf-link chains in momentum space.

In cubic lattices with either $T_d$ or $O_h$ point group symmetry, the L and T modes at Γ point jointly form a three-dimensional irreducible representation. The symmetry-constrained effective Hamiltonian for this system takes the form:

$$H_{\text{QCTP}} = \omega_0 + A\boldsymbol{k}^2 + \begin{pmatrix} Bk_x^2 & Ck_xk_y & Ck_xk_z \\ Ck_yk_x & Bk_y^2 & Ck_yk_z \\ Ck_zk_x & Ck_zk_y & Bk_z^2 \end{pmatrix} \quad (90)$$

with $A$, $B$, and $C$ being $\boldsymbol{k}$-independent constants. Here, $\omega_0$ represents the frequency at $\boldsymbol{k} = 0$, which vanishes for acoustic branches due to the acoustic sum rule, resulting in a linear phonon dispersion near the zone center. This three-band crossing constitutes a topological acoustic triple point (TATP) (Park *et al.*, 2021). For optical branches, $\omega_0$ remains finite, leading to quadratic $\boldsymbol{k} = 0$-dependence of the dispersion and the formation of a quadratic contact triple point (QCTP). Remarkably, the QCTP hosts extended topological surface phonon states that persist across the entire Brillouin zone (Zhong *et al.*, 2021).

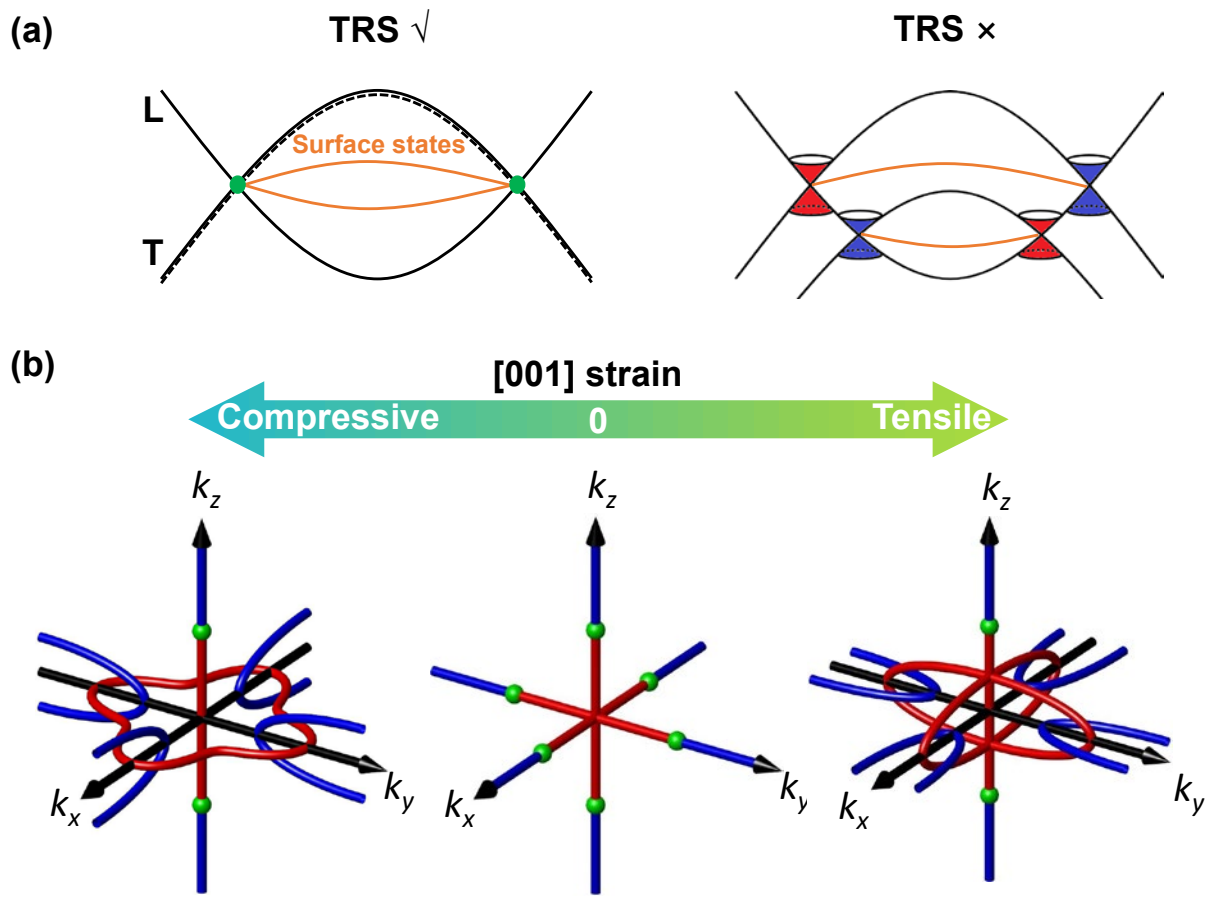

Figure 28 Triple point phonons in diamond-structure silicon. (a) Diamond-structure silicon with space group #227. (b) Phonon band structure of silicon with phononic TPs formed by band crossings between longitudinal optical (LO) and transverse optical (TO) branches along Γ-X. (c) Distribution of topological nodal lines (red solid) and TPs in momentum space. (d) Schematics of TPs and double-fermi-arc-like surface states under time-reversal symmetry (TRS). By breaking the TRS, the TPs are splitted into pairs of WPs which support single-fermi-arc surface states. (e) Under [001] uniaxial strain, TPs transform into interlocking nodal Hopf links. Adapted from Ref. (Liu *et al.*, 2022).

To date, numerous materials hosting phononic TPs and QCTPs have been theoretically predicted through first-principles calculations. (Li *et al.*, 2018) identified three-component Weyl phonons in WC-type compounds (TiS, ZrSe, HfTe) within the THz frequency range. Subsequent work by (Singh *et al.*, 2018) further explored these systems, emphasizing their thermoelectric properties. In CsCl, (Park *et al.*, 2021) demonstrated acoustic TPs with distinct topological invariants: the longitudinal mode carries a Skyrmion number, while the transverse modes are characterized by an Euler number. Phononic QCTPs have been reported in cubic materials such as $Zr_3Ni_3Sb_4$ (Zhong *et al.*, 2021), $Ta_3Sn$ (Yang *et al.*, 2022), etc.

### 3. Body-centered Silicon: $\mathbb{Z}_2$ Dirac Phonon

As discussed in Sec. III.D, conventional Dirac phonons are topologically trivial because they consist of two Weyl phonons with opposite Chern numbers, which do not guarantee gapless helicoid surface states. To provide an intuitive understanding of $\mathbb{Z}_2$ Dirac phonons and their characteristic anti-parallel helical surface states, we will take an experimentally synthesized material candidate as an example in this subsection, showing their topological properties (Zhang and Murakami, 2023).

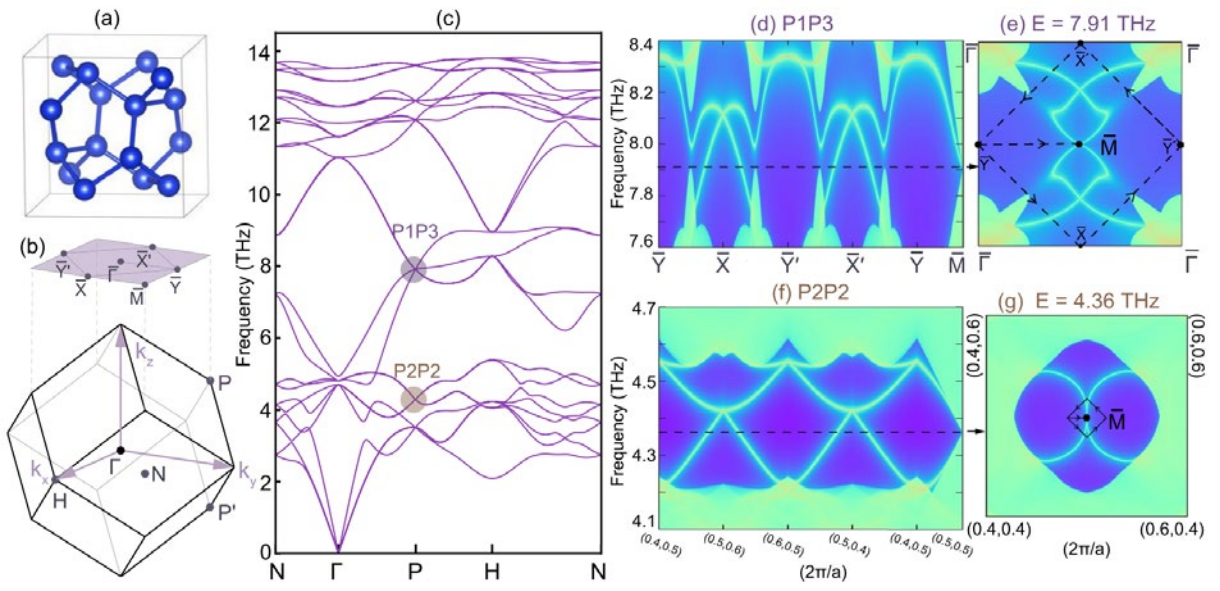

Figure 29 $\mathbb{Z}_2$ Dirac phonon in body-centered silicon with space group $Ia\bar{3}$. (a) Crystal structure of silicon. (b) Bulk Brillouin zone (BZ) and (001) surface BZ. (c) Phonon band structure, where the colored dots denote two distinct $\mathbb{Z}_2$ Dirac phonons, $P_1P_3$ and $P_2P_2$, characterized by different irreducible representations (irreps). (d)-(e) Surface states and Fermi arcs associated with the $\mathbb{Z}_2$ Dirac phonons of irreps $P_1P_3$. (f)-(g) Surface states and Fermi arcs corresponding to the $\mathbb{Z}_2$ Dirac phonons of irreps $P_2P_2$. Adapted from Ref. (Zhang and Murakami, 2023).

Figures 29 (a)–(c) show the crystal structure, Brillouin zone, and phonon spectra of silicon in its experimentally synthesized phase (space group #206). The system hosts two perpendicular glide mirrors, i.e., $G_x = \{M_x|0\frac{1}{2}0\}$ and $G_y = \{M_y|\frac{1}{2}00\}$, which enforce fourfold degenerate bands at the $\mathcal{T}$-related high-symmetry points $P$ and $P'$. All the fourfold degenerate phonons shown in Fig. 29 (c) are $\mathbb{Z}_2$ Dirac phonons with irreducible representations of either $P_2P_2$ or $P_1P_3$, carrying a nonzero monopole charge $Q$.

Figure 29 (d) displays the (001) surface states for the $P_1P_3$ Dirac points identified in Fig. 29 (c), with the corresponding $k$-path shown in the surface arc calculation. The (001) surface Brillouin zone possesses two $\mathcal{T}\mathcal{G}$ symmetries, where $(\mathcal{T}\mathcal{G}_x)^2 = -1$ and $(\mathcal{T}\mathcal{G}_y)^2 = -1$ along the BZ boundaries $\bar{M}$–$\bar{X}$ and $\bar{M}$–$\bar{Y}$. These symmetries enforce Kramers-like degeneracy, resulting in doubly degenerate surface states along these boundaries. The $\mathbb{Z}_2$ Dirac phonons at $P$ and $P'$ project onto the BZ corner ($\bar{M}$), generating two pairs of anti-parallel surface states along the path $\bar{Y}$–$\bar{X}$–$\bar{Y}'$–$\bar{X}'$–$\bar{Y}$ in Fig. 29 (d). Four anti-parallel helical surface states cross along $\bar{X}$–$\bar{M}$–$\bar{X}'$, with their gapless intersection protected by $\mathcal{T}\mathcal{G}_x$. All topological surface states ultimately merge into the $\mathbb{Z}_2$ Dirac cone at $\bar{M}$.

Similarly, Figs. 29 (f) and (g) show the surface states and arcs for the $P_2P_2$ Dirac phonon, which also exhibit two pairs of anti-parallel helical surface states around $\bar{M}$. Thus, both the $P_2P_2$ and $P_1P_3$ Dirac phonons in silicon's phonon spectrum are $\mathbb{Z}_2$ Dirac phonons, each hosting two pairs of anti-parallel helical surface states (termed anti-parallel quad-helical surface states in (Zhang and Murakami, 2023)). The (anti-)parallel quad-helical surface states represent the maximal number of surface states achievable for (Dirac) Weyl phonons in solids, as constrained by onsite and crystalline symmetries.

## B. Topological Nodal Line Phonons

Conceptually, phonons exhibit topological properties analogous to those of spinless electrons. However, in electronic systems, topological degeneracies are typically lifted by finite spin-orbit coupling (SOC) unless protected by additional symmetries. In contrast, phononic systems can host robust topological nodal lines/rings because SOC is inherently absent. In this subsection, we take $MoB_2$ as a prototypical example to illustrate nodal line phonons and their unique features in solids. Other nodal line/ring phonons, as well as their extensions, will be discussed in the next subsection.

$MoB_2$ crystallizes in the centrosymmetric space group $R\bar{3}m$ (# 166), with its crystal structure and primitive cell illustrated in Fig. 30 (a). The system features two distinct boron layers: $B_1$ forms a planar quasi-2D honeycomb lattice, while $B_2$ constitutes a buckled honeycomb network. The phonon spectrum and density of states (DOS), presented in Fig. 30 (b), confirm the dynamical stability of $MoB_2$, as evidenced by the absence of imaginary frequencies. While band crossings appear throughout the spectrum, two particularly interesting features emerge at approximately 7.25 THz and 23 THz (marked by black dashed lines). These crossings correspond to $\mathcal{P}\mathcal{T}$-protected nodal lines carrying a $\mathbb{Z}_2$ monopole charge, forming six helical nodal lines that extend through the Brillouin zone along the $k_z$ direction. Notably, the phonon DOS exhibits local minima at these characteristic frequencies, consistent with the expected nodal dispersion feature.

(Zhang *et al.*, 2019b) focus on the band crossings at 7.25 THz (30 meV) to reveal their topological nature and full configuration. Following the symmetry-based indicator theory and related frameworks (Po *et al.*, 2017; Song *et al.*, 2018; Zhang *et al.*, 2020a) introduced in Sec. II.C, we analyze these features as shown in Fig. 30 (c). For space group No. 166 (R$\bar{3}$m), which possesses a nontrivial symmetry-based indicator group, we first determine the irreducible representations (irreps) of all five phonon bands below 7.25 THz at high-symmetry points $\Gamma$, $T$, $F$, and $L$ through first-principles calculations. However, the compatibility relationships for these irreps in space group # 166 are not satisfied, indicating band inversions between the 5th and 6th phonon modes. This renders the symmetry-based indicator theory inapplicable for the parent space group.

(Zhang *et al.*, 2019b) therefore identify subgroups of # 166 that satisfy two key criteria:

- The irreps must fulfill the compatibility relations within the subgroup
- The subgroup must have a nontrivial symmetry-based indicator group

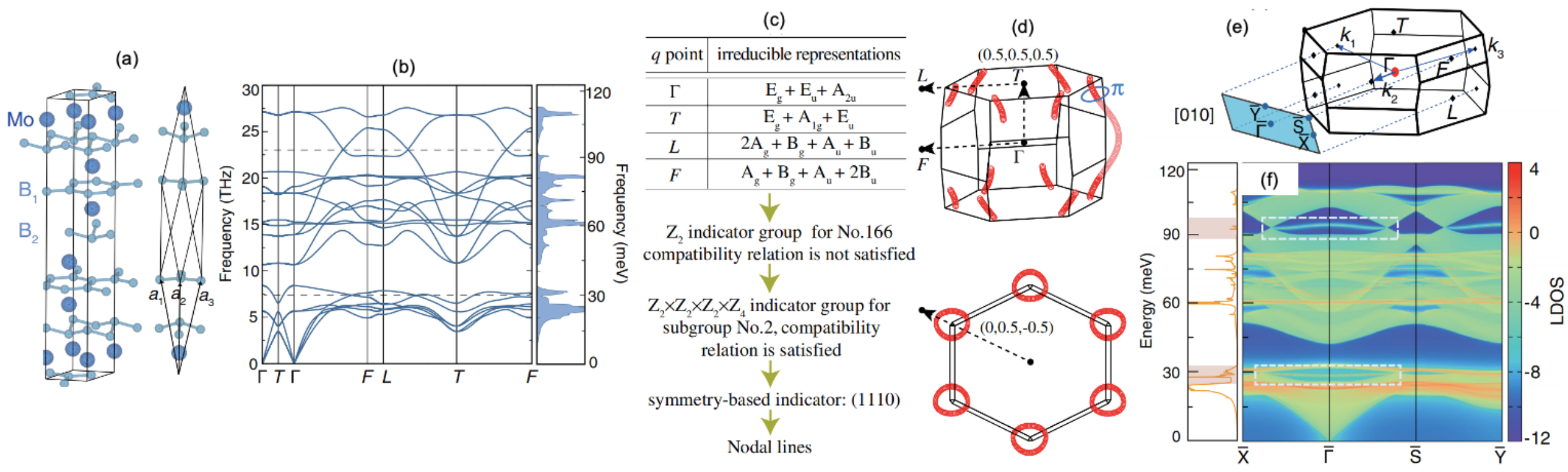

Figure 30 (a) Conventional and primitive unit cells of $MoB_2$. (b) Phonon spectra of $MoB_2$. (c) Topological nodal line identification in $MoB_2$, determined through compatibility relations and symmetry-based indicator analysis. (d)-(e) Side view and [111]-oriented top view of the nodal lines near 30 meV in $MoB_2$, exhibiting a quantized $\pi$ Berry phase. (f) Bulk Brillouin zone and [010]-projected surface BZ of $MoB_2$. (010) surface local density of states (LDOS), showing both bulk-derived modes and surface-unique modes (left panel: integrated intensity). Drumhead surface phonon states (within white dashed rectangles) connecting nodal line crossings display maximal LDOS, as evidenced by sharp intensity peaks at nodal-line energies (shaded regions). Adapted from Ref. (Zhang *et al.*, 2019b).

Space group # 2 ($P\bar{1}$) meets these requirements, allowing us to apply its symmetry-based indicator formulas:

$$z_{2,i} = \sum_{q\in \text{TRIM at } \{q_i=\pi\}} \frac{N_-(q) - N_+(q)}{2} \bmod 2, i = 1,2,3 \quad (91)$$

$$z_4 = \sum_{q\in \text{TRIM}} \frac{N_-(q) - N_+(q)}{2} \bmod 4 \quad (92)$$

For $MoB_2$, symmetry analysis yields the indicator $z_{2,1}z_{2,2}z_{2,3}z_{2,4}z_4 = (1110)$. Following the theory outlined in Sec. II.C (Eqs. 91 and 92) (Song *et al.*, 2018; Zhang *et al.*, 2020a), this indicator predicts:

1. The existence of 2 mod 4 nodal lines along the [111] direction in the Brillouin zone

2. The $\pi$ Berry phase of each nodal line, as shown in Fig. 30 (d)

3. The $\mathbb{Z}_2$ monopole charge of each nodal line

These topological features manifest as nodal lines that extend throughout the Brillouin zone. Crucially, their topological protection ensures robustness against any small perturbations that preserve $\mathcal{PT}$ symmetry.

To demonstrate the topological surface states associated with these $\mathcal{PT}$-protected nodal lines, we present the local density of states (LDOS) along the (010) direction in Fig. 30 (f), corresponding to the surface Brillouin zone shown in Fig. 30 (e). Unlike Weyl points that exhibit helicoid surface modes (Fang *et al.*, 2016; Miao *et al.*, 2018; Zhang *et al.*, 2022b; Zhang and Murakami, 2023; Zhang *et al.*, 2018a, 2020b), the nodal lines generate characteristic drumhead-shaped surface states, as indicated by the white dashed box. Notably, while the bulk phonon density of states shows minima near ∼7.25 THz and ∼23 THz, the surface modes display a pronounced peak in LDOS at these energies (Fig. 30 b). The flat surface states could induce electronic anomalies through electron-phonon coupling, potentially leading to unique surface phenomena.

For any generic two-band system, the low-energy effective Hamiltonian can be expressed in terms of Pauli matrices as:

$$H_{\text{2band}} = \varepsilon(\boldsymbol{k}) + h_1(\boldsymbol{k})\sigma_x + h_2(\boldsymbol{k})\sigma_y + h_3(\boldsymbol{k})\sigma_z, \quad (93)$$

where $h_{1,2,3}$ are arbitrary real functions of $\boldsymbol{k}$. For phononic systems with $\mathcal{PT}$ symmetry, the dynamical matrix and eigenvectors can be made real at any $\boldsymbol{k}$ point, which means $h_2(\boldsymbol{k}) = 0$. The location of nodal line is thus determined by the condition:

$$h_1(\boldsymbol{k}) = h_3(\boldsymbol{k}) = 0 \quad (94)$$

In 2D systems, Eq. (94) typically defines isolated $\boldsymbol{k}$-points, corresponding to 2D Dirac semimetals (Jin *et al.*, 2018b; Li *et al.*, 2020a). In 3D systems, however, the same condition can be satisfied along closed $\boldsymbol{k}$-space contours, giving rise to topological phononic nodal lines. Numerous material candidates hosting such nodal lines have been theoretically predicted and experimentally observed. Based on their geometric configurations, these nodal lines can be categorized into several distinct types: helical nodal lines (Zhang *et al.*, 2019b), straight nodal lines (Li *et al.*, 2020b; Liu *et al.*, 2021a), nodal rings (Jin *et al.*, 2018a), nodal chains (Chen *et al.*, 2021b), and nodal net (Liu *et al.*, 2021c). Phononic topological nodal lines exist ubiquitously in centrosymmetric solids (Liu *et al.*, 2022).

Straight nodal lines along high-symmetry lines (HSLs) with $C_n$ rotational symmetry can be categorized into two fundamental types based on their transverse dispersion: linear and quadratic types (Liu *et al.*, 2021c, 2022; Qin *et al.*, 2024). This classification applies particularly to systems with threefold rotational symmetry ($n = 3$), where the low-energy effective Hamiltonian takes the general form:

$$H = \begin{pmatrix} 0 & Ak_- + Bk_+^2 \\ A^*k_+ + B^*k_-^2 & 0 \end{pmatrix} \tag{95}$$

where $k_\pm = k_1 \pm ik_2$ with $k_1$ and $k_2$ being the wave vector perpendicular to the rotational axis. $A$ and $B$ are complex constants that do not depend on $k_1$ and $k_2$. The location of nodal line is determined by $Ak_- + Bk_+^2 = 0$ which give rise to 4 nodal lines: One located at the rotation axis along $k_3$ with Berry phase $+\pi$, while the other three located at generic positions which are related by $C_3$ rotation and have Berry phase $-\pi$, *i.e.*,

$$\begin{aligned} (0, 0, k_3) &: \text{Berry phase } +\pi \\ \left(k_0 \cos\frac{\theta_0}{3}, k_0 \sin\frac{\theta_0}{3}, k_3\right) &: \text{Berry phase } -\pi \\ \left(k_0 \cos\frac{\theta_0 + 2\pi}{3}, k_0 \sin\frac{\theta_0 + 2\pi}{3}, k_3\right) &: \text{Berry phase } -\pi \\ \left(k_0 \cos\frac{\theta_0 - 2\pi}{3}, k_0 \sin\frac{\theta_0 - 2\pi}{3}, k_3\right) &: \text{Berry phase } -\pi \end{aligned} \tag{96}$$

where $k_0 = |-\frac{A}{B}|$ and $\theta_0 = \arg\left(-\frac{A}{B}\right)$. Nodal lines can be classified into two fundamental types based on their transverse dispersion: (i) linear nodal lines with $\pi$ Berry phase (topologically nontrivial) and (ii) quadratic nodal lines with $2\pi$ Berry phase (Liu *et al.*, 2022; Qin *et al.*, 2024). The topological character is determined by the dominant term in the effective Hamiltonian: when the linear term dominates ($|A/B| \gg 1$), the system exhibits a $\pi$ Berry phase, whereas dominance of the quadratic term ($|A/B| \ll 1$) yields a $-2\pi$ Berry phase. For systems with $C_4$ or $C_6$ symmetry, the linear term in Eq. (95) vanishes, allowing only a single straight nodal line along the rotational axis. In these cases, quadratic nodal lines emerge with Berry phase $\pm 2\pi$, corresponding to a winding number $n_W = \pm 2$ (Liu *et al.*, 2022). This distinct topological signature arises from the doubled winding of the pseudospin texture around the nodal line.

### C. Obstructed Atomic Insulator-like Topological Phonons

Topological phonons can be defined not only in gapless systems but also in gapped systems under specific symmetry conditions, as classified by the "ten-fold way" table in Tab. III and discussed in Sec. III.B.2. Unlike conventional topological gapped states, typically arising from band inversions in phonon spectra, a distinct class termed "obstructed atomic insulator-like phonon modes" exists in gapped systems. These modes exhibit a mismatch between the atomic center and the vibrational center of phonon modes, leading to localized vibrations at boundaries. In this section, we review recent progress on obstructed phonon modes, with emphasis on the prototypical material tellurium (Te), where such modes have been experimentally observed.

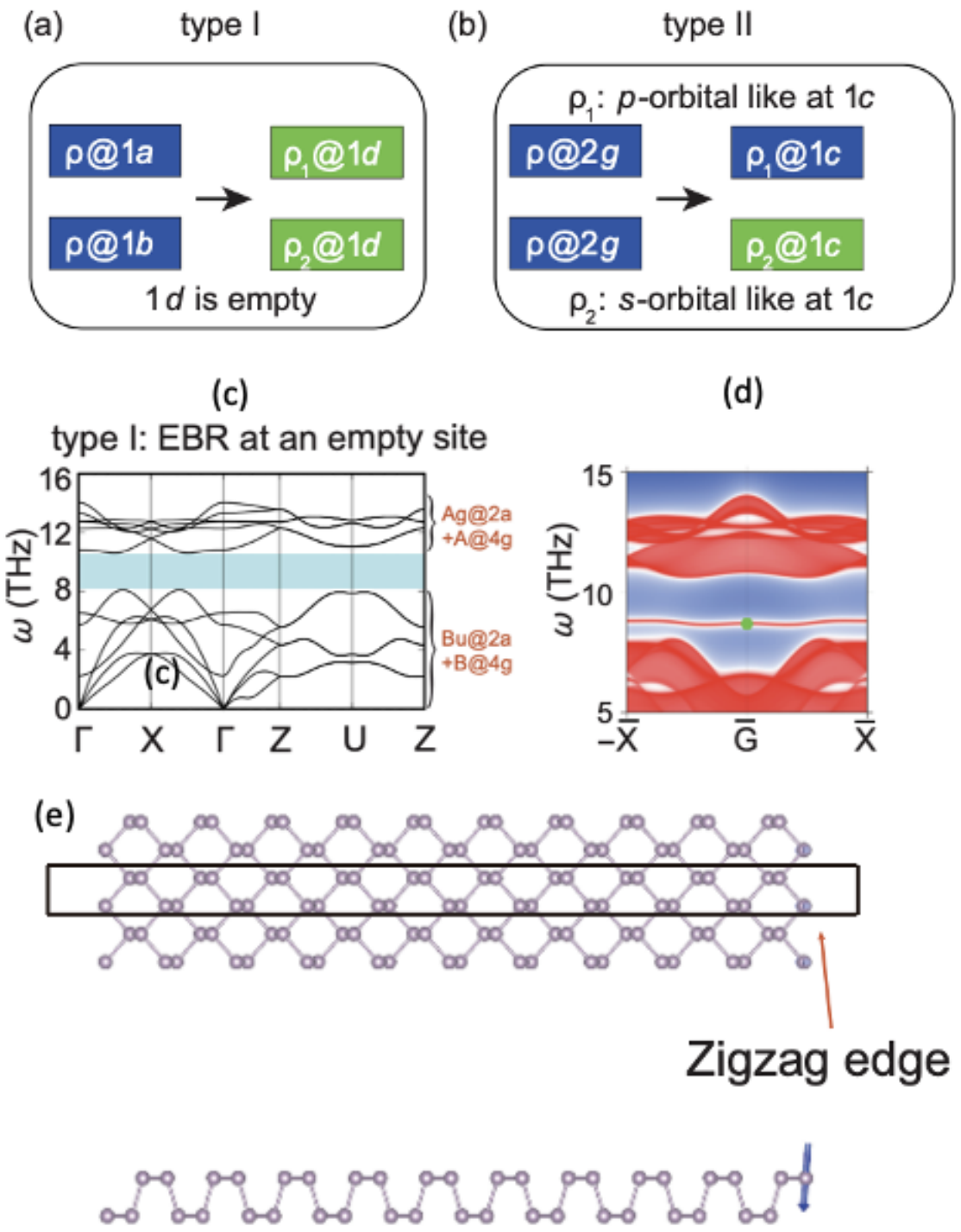

Figure 31 Two types of unconventionality. Atomic vibrations act as $p$ orbitals. (a) Phonon modes contributed by the vibration center at an empty site. (b) Phonon modes contributed by the vibration center at an atom site, yet show a non-atomic vibration orbitals (non-$p$ orbital symmetry). (c) Phonon spectra of the monolayer black phosphorous. (d) Phonon spectra for the monolayer black phosphorous with an open boundary condition, where there are floating obstructed surface phonon modes $\sim 9$ THz. The real space vibration for the surface phonon mode marked by the green hexagon is shown in (e). Adapted from Ref. (Zhang *et al.*, 2023a).

Phonons represent quantized collective vibrations of atomic lattices, where each mode corresponds to a distinct pattern of atomic displacements. However, certain phonon modes exhibit unconventional origins: they may arise from vibrations localized at empty atomic sites or involve non-atomic orbital symmetries (e.g., non-$p$-orbital-like vibrations), as illustrated in Figs. 31 (a)–(b). These anomalous modes, termed obstructed phonon modes (Gao *et al.*, 2022; Peri *et al.*, 2020; Song *et al.*, 2020; Xu *et al.*, 2021, 2024a,c) (Ma *et al.*, 2023; Zhang *et al.*, 2023a,b), defy conventional descriptions of lattice dynamics. The unconventional nature of these modes can be identified through either real-space invariants or open-boundary calculations (Zhang *et al.*, 2023a,b). The former one quantifies the mismatch between atomic positions and vibrational centers, while the latter one reveals localized surface modes ("floating states") detached from bulk phonon spectra.

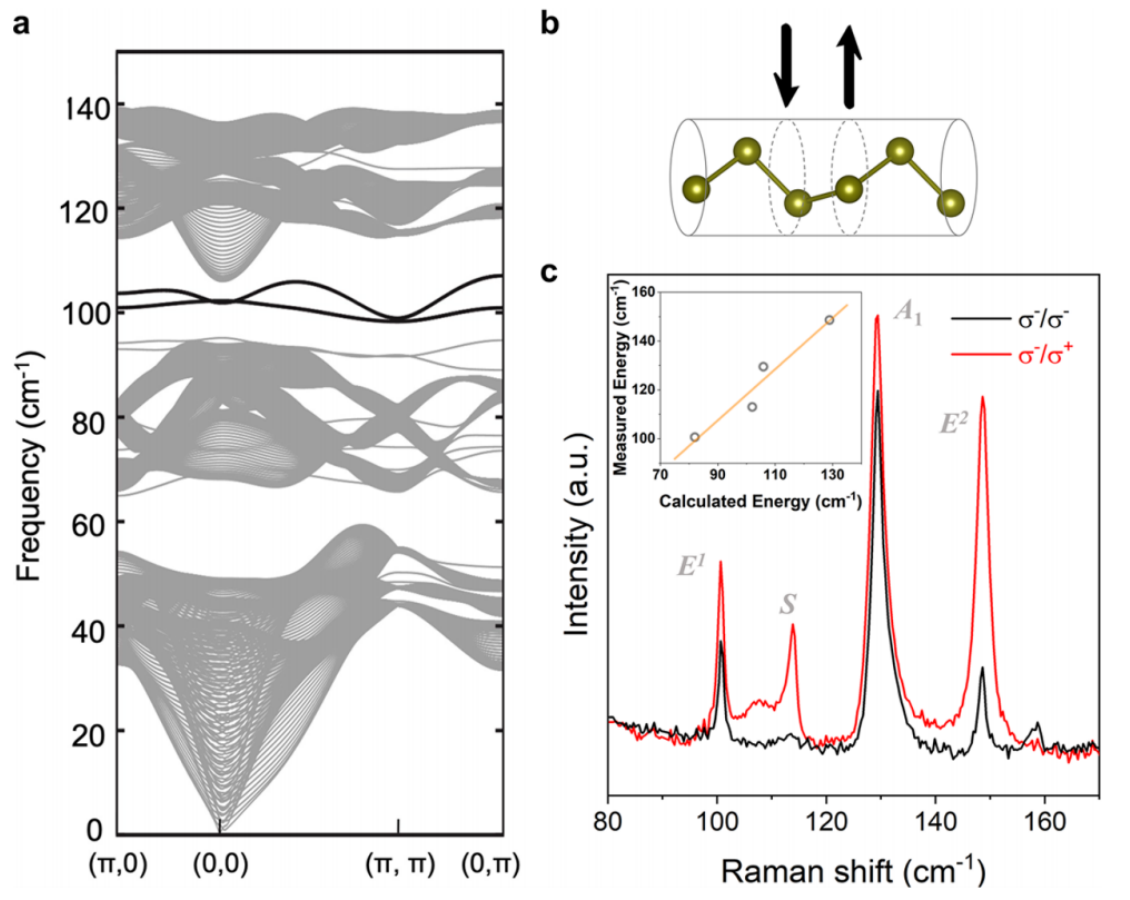

Figure 32 (a) First-principles calculations of right-handed tellurium (Te) with open-boundary conditions along [010] direction. The black solid lines $\sim$100 cm$^{-1}$ correspond to obstructed atomic-insulator-like phonon modes, arising from dangling-bond vibrations at the surface. (b) Experimental configuration for circularly polarized Raman spectroscopy in backscattering geometry. (c) Raman spectrum of Te showing an additional surface (S) mode. The inset displays the fitting results, confirming the S-mode's origin from surface dangling-bond vibrations. Adapted from Ref. (Zhang *et al.*, 2023b).

Such modes typically emerge from dangling-bond vibrations on crystal surfaces and are intimately linked to the material's symmetry and electronic structure. Their spectral isolation and spatial localization make them promising for controlling phonon-mediated phenomena at interfaces or defects.

As a concrete example, Figs. 31 (c) and (d) display the phonon spectra of monolayer black phosphorus under periodic and open boundary conditions, respectively (Zhang *et al.*, 2023a). Notably, the open-boundary spectrum reveals floating obstructed surface phonon modes near 9 THz, which are absent in the periodic calculation. The real-space vibrational pattern of one such surface mode (indicated by the green hexagon in Fig. 31 (d) is shown in Fig. 31 (e), clearly demonstrating its localized character at the edge.

Figure 32 (a) shows the phonon spectra of tellurium under open boundary conditions. The black solid lines near $\sim$100 cm$^{-1}$ correspond to obstructed surface phonons, arising from dangling bonds on the [010] surface. Although the symmetry data (irreducible representations at high-symmetry momenta) of the lowest six phonon bands match those of an "atomic insulator", these modes originate from empty atomic positions, classifying Te as an obstructed phonon material. The obstructed surface phonons persist across the entire surface Brillouin zone and lie within the bulk phonon gap, making them detectable via helicity-resolved Raman scattering.

Figure 32 (c) presents the Raman shifts of Te under circularly polarized incident/scattered light. In addition to the three Raman-active modes ($E^1$, $E^2$, and $A_1$), an extra S mode appears. The frequencies of all four modes align well with first-principles calculations (inset), confirming the S mode as the obstructed surface phonon (Zhang *et al.*, 2023b).

### D. Topological Phonon Materials Database

The method of symmetry-based indicator theory and topological quantum chemistry (TQC) (Bradlyn *et al.*, 2017; Elcoro *et al.*, 2021; Po *et al.*, 2017) provides a powerful framework for identifying topological materials. Its classification hinges on the elementary band representations (EBRs) (Cano and Bradlyn, 2021; Michel and Zak, 2001), which distinguish topologically trivial and nontrivial phases: A band structure adiabatically connected to an "atomic insulator" is trivial, whereas the absence of such a connection typically signals nontrivial topology (Cano and Bradlyn, 2021). This classification is rigorously formalized using symmetry-based indicators (Po *et al.*, 2017; Song *et al.*, 2018; Tang *et al.*, 2019b,c,d; Tang and Wan, 2024; Vergniory *et al.*, 2019a; Zhang *et al.*, 2019a).

To date, extensive high-throughput searches for topological materials, encompassing both electronic and phononic systems, have been conducted using first-principles methods (Chen *et al.*, 2021a; Li *et al.*, 2021c; Tang *et al.*, 2019b; Vergniory *et al.*, 2019a; Xu *et al.*, 2020, 2024b; Zhang *et al.*, 2019a). From the symmetry perspective, phonons behave analogously to spinless electrons, with their three vibrational degrees of freedom resembling $p_{x,y,z}$ orbitals (Liu *et al.*, 2018). Consequently, the complete classification of topological phononic materials can be achieved within the framework of single-valued representations of the 230 space groups. Pioneering this effort, (Li *et al.*, 2021c) performed a high-throughput screening of topological phononic materials, identifying systems hosting phononic nodal lines and Weyl points, as shown in Fig. 33 (a). By computing phononic band structures along high-symmetry paths, they systematically classified materials with these topological features. More recently, (Xu *et al.*, 2024b) expanded this approach by screening phononic obstructed atomic bands using topological quantum chemistry, paving the way for further experimental studies, as shown in Fig. 33 (b).

## V. FROM PHONON TOPOLOGICAL CHIRALITY TO PHONON ROTATIONAL CHIRALITY

As mentioned in Sec. I.B, phonons admit a real-space representation in terms of atomic displacements, where phase-shifted orthogonal components naturally give rise to elliptical or circular motion. Accordingly, vibrational modes can be classified as linearly or circularly polarized (Johnson, 1982; Pine and Dresselhaus, 1969; Rebane, 1983), a distinction that can be quantitatively characterized by the phonon angular momentum operator [AM, $l_{n\boldsymbol{q}}$, see Eq. (13)] (Johnson, 1982; McLellan, 1988; Rebane, 1983). To our knowledge, the term "chiral phonon" was first used by Bermudez and Martin-Delgado 2008; Bermudez *et al.* 2008b to describe phonon modes carrying nonzero angular momentum (AM), namely the right- and left-handed circular motion of a Dirac oscillator, in analogy with circularly polarized light. The concept was subsequently extended to crystalline solids within a band-theoretic framework by Zhang and Niu 2014, 2015, who showed that phonon modes at the $\mathcal{T}$-breaking $K/K'$ valleys of two-dimensional hexagonal lattices carry quantized, nonzero AM, as illustrated in Fig. 34(a). Moreover, they introduced pseudo-angular momentum (PAM, encode the eigenvalues of the rotational operator) as a symmetry-based descriptor of phonon modes, which is useful in analyzing phonon selection rules and in interpreting polarization-resolved spectroscopic measurements (Yao *et al.*, 2008).

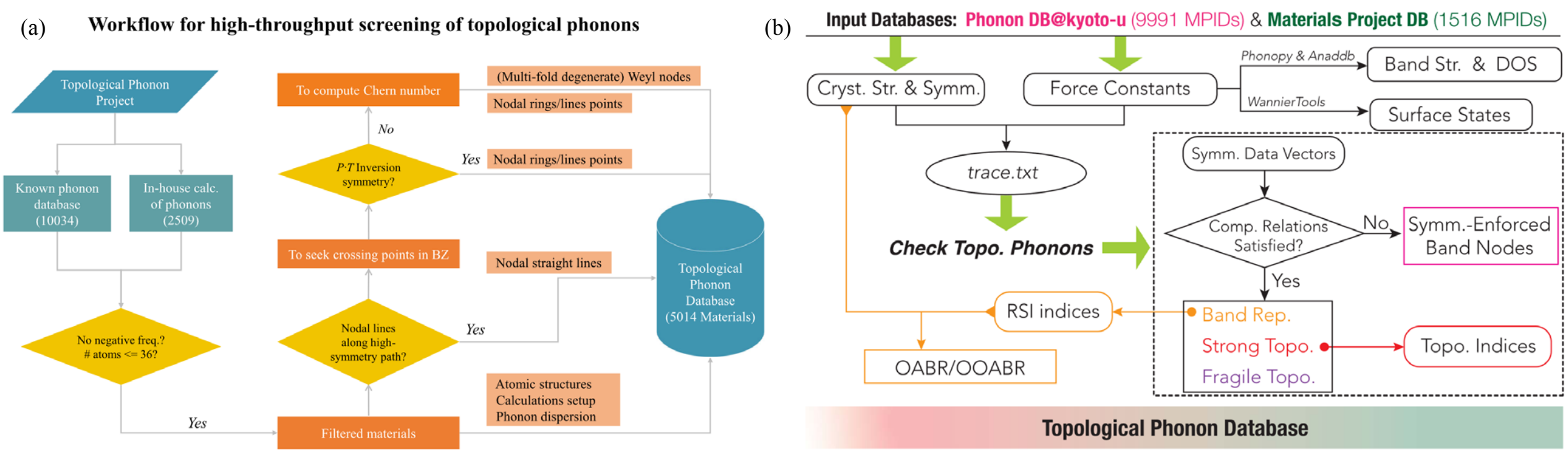

Figure 33 Construction of database for topological phononic materials. (a) High-throughput computational screening of topological phononic nodal lines/rings/points. Adapted from Ref. (Li *et al.*, 2021c). (b) Topological phonon database for obstructed atomic band representations (OABRs) and orbital-selected OABRs (OOABR). Adapted from Ref. (Xu *et al.*, 2024b).

Unlike topological phonons, which are rigorously defined by topological invariants, the notion of "chirality" [2] and "chiral phonons" can vary across different research contexts. For example, in topolgical materials, the chirality of Weyl phonons is defined by the sign of the Chern number, corresponding to topological chirality that has been experimentally confirmed (Jin *et al.*, 2022; Li *et al.*, 2021a; Miao *et al.*, 2018; Zhang *et al.*, 2023b, 2018a, 2020b). Another broadly used quantity is the phonon angular momentum mentioned above. It characterizes the circular motion of atoms in real space and corresponds to the rotational chirality of phonons (Bermudez *et al.*, 2008a; Bermudez and Martin-Delgado, 2008; Bermudez *et al.*, 2008b; Hamada *et al.*, 2018; Hamada and Murakami, 2020b; McLellan, 1988; Ueda *et al.*, 2023; Zhang and Niu, 2014, 2015; Zhang *et al.*, 2024). In the literature, these phonons are also referred to as circularly polarized phonons (Johnson, 1982; Pine and Dresselhaus, 1969; Rebane, 1983) and, more recently, as axial phonons (Chaudhary *et al.*, 2026; Juraschek *et al.*, 2025; Kato *et al.*, 2026), among other terminologies.

In many cases, the circularly polarized phonons are not truly chiral in a strict sense because phonon AM is a axial vector thus is even under spatial inversion $\mathcal{P}$. However, quite a lot of circularly polarized-phonon-related phenomena were observed in materials with chiral structures like, to name a few, by (Funato *et al.*, 2024; Ishito *et al.*, 2023b; Ueda *et al.*, 2023; Zhang *et al.*, 2025b). This is because such phonon is prohibited by the combined $\mathcal{PT}$ symmetry. In nonmagnetic materials with $\mathcal{T}$ symmetry, $\mathcal{P}$ must be broken to induce net nonzero phonon AM. Otherwise, $\mathcal{T}$ must be broken by for example internal magnetism to show signitures of circularly polarized phonons (Che *et al.*, 2024; Yang *et al.*, 2025a).

In this review, we use "circularly polarized phonons" to describe phonon modes that carry nonzero angular momentum. This terminology is also intuitive and emphasizes the circular motion of atoms in real space. In the following, we review several fundamental concepts of circularly polarized phonons and phonon modes exhibiting both topological and rotational chirality, along with their interrelations and representative experimental realizations in specific materials. We further review physical phenomena driven by circularly polarized phonons and outline their potential applications across various contexts.

## A. Fundamentals of Phonon Wavefunction

### 1. Phonon Angular Momentum

Phonon angular momentum ($\boldsymbol{l}_{n\boldsymbol{q}}$) is defined from the collective motion of atoms, arising from phase-shifted (elliptical or circular) vibrational components of the displacement field. One can define phonon AM, i.e., $l^{\alpha}_{n\boldsymbol{q}}$, for a specific phonon mode $|\varepsilon_{n\boldsymbol{q}}\rangle$ at momentum $\boldsymbol{q}$ in the equilibrium state as (Kishine *et al.*, 2020; McLellan, 1988; Zhang and Niu, 2014; Zhang *et al.*, 2026a):

[2] In some references, e.g. (Barron, 2021), the concept of chirality is classified into two catogies: the true chirality and false chirality. The true chirality refers to the existence of two enantiomer states that are interchanged only by spatial inversion operation $\mathcal{P}$ but NOT by time-reversal operation $\mathcal{T}$; however, for the false chirality, the two enantiomer states are interchanged by both $\mathcal{P}$ and $\mathcal{T}$. The distinction between true and false chirality is essential to the problem of homochirality origin of our natural world including DNA. The homochiral phenomenon must have a true chiral origin.

$$l_{n\boldsymbol{q}}^{\alpha} = \hbar\langle\varepsilon_{n\boldsymbol{q}}|M_\alpha|\varepsilon_{n\boldsymbol{q}}\rangle$$
$$= \sum_i^r l_{n\boldsymbol{q}}^{i\alpha} = \sum_i^r \hbar\langle\varepsilon_{n\boldsymbol{q}}^i|\mathfrak{M}_\alpha|\varepsilon_{n\boldsymbol{q}}^i\rangle, \tag{97}$$

where $M_\alpha$ is a $3r \times 3r$ matrix whose matrix form is given by $M_\alpha = \oplus_{i=1}^r \mathfrak{M}_\alpha$. $\mathfrak{M}_\alpha$ ($\alpha \in \{x, y, z\}$) are the generators of Lie algebra of the $O(3)$ group:

$$\mathfrak{M}_x = \begin{pmatrix} 0 & 0 & 0 \\ 0 & 0 & -i \\ 0 & i & 0 \end{pmatrix}, \mathfrak{M}_y = \begin{pmatrix} 0 & 0 & i \\ 0 & 0 & 0 \\ -i & 0 & 0 \end{pmatrix}, \mathfrak{M}_z = \begin{pmatrix} 0 & -i & 0 \\ i & 0 & 0 \\ 0 & 0 & 0 \end{pmatrix}. \tag{98}$$

Consequently, a phonon mode with $l_{n\boldsymbol{q}}^{\alpha} \neq 0$ corresponds to atomic motions where the net circular polarization of all atoms in the unit cell is nonzero. The total AM of the system, denoted as $\boldsymbol{J}_{\text{ph}}$, is defined as the sum of ensemble average of the AM of each phonon mode across the entire Brillouin zone. Mathematically, this can be expressed as: $J_{\text{ph}}^{\alpha} = \sum_{n\boldsymbol{q}}\langle l_{n\boldsymbol{q}}^{\alpha}\rangle = \sum_{n\boldsymbol{q}} l_{n\boldsymbol{q}}^{\alpha} f_{n\boldsymbol{q}}$, where $f_{n\boldsymbol{q}}$ represents the average occupation number of the phonon mode. This framework provides a comprehensive understanding of the phonon Edelstein effect and phonon angular momentum in crystalline materials (Hamada *et al.*, 2018; Hamada and Murakami, 2020b; Xu *et al.*, 2018; Zhang *et al.*, 2025a).

Several different definitions of phonon chirality have been proposed in the literature, including circularly polarized phonons (Johnson, 1982; Pine and Dresselhaus, 1969; Rebane, 1983), phonon helicity (Hu *et al.*, 2021; Liu, 2022; Zhang *et al.*, 2026a), and axial phonons (Juraschek *et al.*, 2025), among others. These formulations are useful in different contexts and emphasize complementary aspects of lattice dynamics rather than representing a single unified or universally accepted definition. For example, phonon helicity, defined as $h_{n\boldsymbol{q}} = \boldsymbol{q} \cdot \boldsymbol{l}_{n\boldsymbol{q}}$, has been used to describe rotational properties of lattice vibrations and has been discussed in connection with chiral charge density waves (Luo and Dai, 2023; Romao and Juraschek, 2024; Yang *et al.*, 2024a; Zhang *et al.*, 2024). However, its interpretation and relation to other definitions remain an active area of discussion, particularly regarding its dependence on symmetry conventions and representation choices. In this review, we adopt the original and most widely used notion of phonon chirality, namely circularly polarized phonons, as introduced in earlier works.

## 2. Phonon Pseudo-angular Momentum

Both "topological chirality" and "rotational chirality" are symmetry-dependent quantities. For instance, the topological invariant, i.e., the phonon Chern number, can be inferred from rotation eigenvalues, thereby encoding the full topological information, including the associated chirality. Since direct observation of phonon AM remains experimentally challenging (Zhang *et al.*, 2025a), the pseudo-angular momentum (PAM), which corresponds to the eigenvalue of the rotational operator, has been introduced as a practical indicator for identifying circularly polarized phonons (Bourgeois *et al.*, 2025; Ishito *et al.*, 2023a,b,c; Komiyama *et al.*, 2022; Yao *et al.*, 2008; Zhang and Niu, 2015; Zhang *et al.*, 2026a, 2023b; Zhang and Murakami, 2022; Zhang *et al.*, 2025c; Zhu *et al.*, 2018).

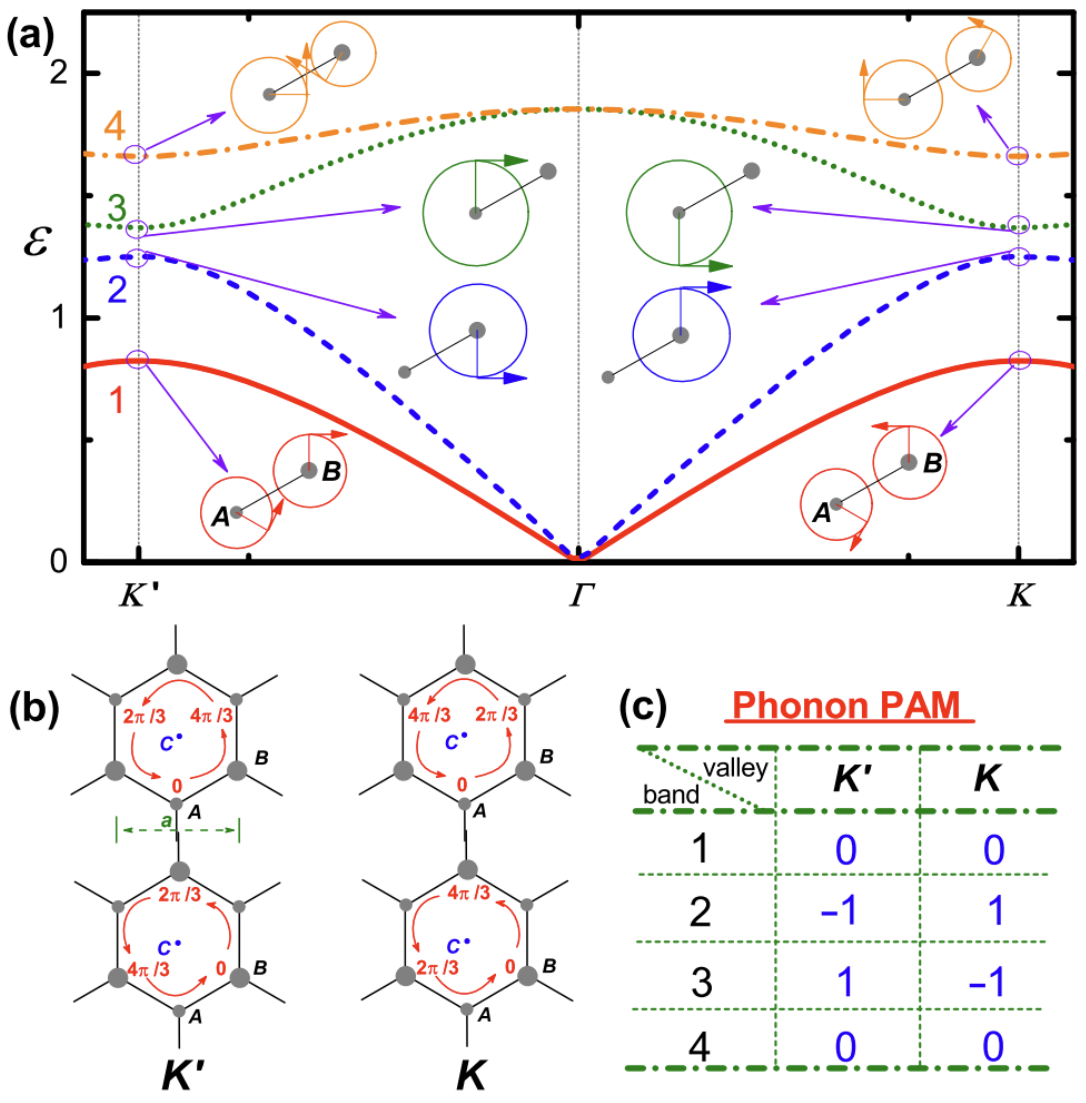

Figure 34 Valley phonons in a 2D honeycomb lattice. (a) Phonon dispersion of the honeycomb lattice. Insets show the phonon eigenmodes at K and K′ for sublattices A and B within one unit cell; circle radii represent vibration amplitudes, with phase and rotation direction indicated. (b) Phase correlation of the nonlocal phonon components between the two sublattices at the K and K′ valleys. (c) Phonon pseudoangular momentum (PAM) of the four bands at the two valleys. This figure is adapted from (Zhang and Niu, 2015).

Let's consider a phonon mode at momentum $\boldsymbol{q}$ which can be expressed as Fourier transformation of real-space atomic displacement $u_i^{\alpha}(\boldsymbol{q}) = \sum_{\ell} e^{-i\boldsymbol{q}\cdot\boldsymbol{r}_{\ell i}} u_{\ell i}^{\alpha}$. For any little space group symmetry operation $\mathcal{C}_n = \{R_n|\boldsymbol{t}\}$ ($R_n$ is a $n$-fold point group rotation operation) at $\boldsymbol{q}$, the representation matrix of $\mathcal{C}$ can be derived as follows:

$$\begin{aligned}\mathcal{C}u_i^{\alpha}(\boldsymbol{q}) &= \sum_{\ell} e^{-i\boldsymbol{q}\cdot(\boldsymbol{R}_\ell+\boldsymbol{d}_i)}\mathcal{C}u_{\ell i}^{\alpha} \\ &= \sum_{\ell} e^{-i\boldsymbol{q}\cdot(\boldsymbol{R}_\ell+\boldsymbol{d}_i)} P_{ij}\sum_{\beta} R_n^{\alpha\beta}u_{\ell' j}^{\beta} \\ &= \sum_{\ell'} e^{-i\boldsymbol{q}\cdot\boldsymbol{r}_{\ell' j}-i\boldsymbol{q}\cdot[(R_n-1)\boldsymbol{d}_i+\boldsymbol{t}]} P_{ij}\sum_{\beta} R_n^{\alpha\beta}u_{\ell' j}^{\beta} \\ &= e^{-i\boldsymbol{q}\cdot[(\mathcal{C}_n-1)\boldsymbol{d}_i+\boldsymbol{t}]} P_{ij}\sum_{\beta} R_n^{\alpha\beta}u_j^{\beta}(\boldsymbol{q}),\end{aligned} \tag{99}$$

$P_{ij} = \delta_{\mathcal{C}i,j}$ is the permutation matrix that interchanges sublattices $i$ and $j$ under symmetry operation $\mathcal{C}$. The inner product between $u_j^{\beta}(\boldsymbol{q})$ and $\mathcal{C}u_i^{\alpha}(\boldsymbol{q})$ yields the representation matrix for the rotational symmetry $\mathcal{C}_n$. Mathematically, this can be

expressed as:

$$D(\mathcal{C})_{ji}^{\beta\alpha} = e^{-i\boldsymbol{q}\cdot[(R_n-1)\boldsymbol{d}_i+\boldsymbol{t}]}P_{ij}R_n^{\alpha\beta} = e^{-i\boldsymbol{G}\cdot\boldsymbol{d}_i-i\boldsymbol{q}\cdot\boldsymbol{t}}P_{ij}R_n^{\alpha\beta}. \tag{100}$$

The idendities $\boldsymbol{q}\cdot(R_n-1)\boldsymbol{d}_i = \boldsymbol{q}\cdot R_n\boldsymbol{d}_i - \boldsymbol{q}\cdot\boldsymbol{d}_i = R_n^{-1}\boldsymbol{q}\cdot\boldsymbol{d}_i - \boldsymbol{q}\cdot\boldsymbol{d}_i$, and $R_n^{-1}\boldsymbol{q} = \boldsymbol{q}+\boldsymbol{G}$ have been used. $\boldsymbol{G}$ is a reciprocal lattice vector which should be determined by $R_n$ and $\boldsymbol{q}$. For $\boldsymbol{q}$ inside the first Brillouin zone (e.g. like $\Gamma$ point), $\boldsymbol{G}$ is always zero; for $\boldsymbol{q}$ at the Brillouin zone boundary (e.g. like K and K$'$ points of the honeycomb lattice), $\boldsymbol{G}$ can be nonzero.

The representation matrix $D(\mathcal{C}_n)$ shown in (99) can be diagonalized by phonon eigenvector $|\boldsymbol{\varepsilon}_{n\boldsymbol{q}}\rangle$ because the dynamical matrix $\mathsf{D}_{\boldsymbol{q}}$ commute with $D(\mathcal{C}_n)$. For purely rotational symmetry (i.e. $\boldsymbol{t}=0$), this can be expressed as:

$$D(\mathcal{C}_n)|\boldsymbol{\varepsilon}_{\nu\boldsymbol{q}}\rangle = e^{-i2\pi l_{\text{ph}}/n}|\boldsymbol{\varepsilon}_{\nu\boldsymbol{q}}\rangle. \tag{101}$$

$l_{\text{ph}}$ is the phonon PAM which reflects the eigenvalue of the representation matrix $D(\mathcal{C}_n)$ for the rotational symmetry $\mathcal{C}_n$, thus $l_{ph}$ can take arbitary integer values (modulo $n$).

In systems with screw rotational symmetry i.e. $\boldsymbol{t}$ is the $m/n$ fractional translation vector $\boldsymbol{\tau}_{m/n}$ along the rotational axis, the representation matrix for the screw rotation is modified to account for the fractional translation. Specifically, the representation matrix becomes $D(\mathcal{C}_n) = e^{-i\boldsymbol{q}\cdot\boldsymbol{\tau}_{m/n}}\cdot D(R_n)$. Thus we have:

$$D(\mathcal{C}_{n,\boldsymbol{\tau}_{m/n}})|\boldsymbol{\varepsilon}_{\nu\boldsymbol{q}}\rangle = e^{-i2\pi l_{\text{rot}}/n - i\boldsymbol{q}\cdot\boldsymbol{\tau}_{m/n}}|\boldsymbol{\varepsilon}_{\nu\boldsymbol{q}}\rangle. \tag{102}$$

In this case, PAM is expressed as:

$$l_{\text{ph}} = l_{\text{rot}} + l_{\boldsymbol{\tau}} = l_{\text{rot}} + \frac{\boldsymbol{q}\cdot\boldsymbol{\tau}_{m/n}}{2\pi/n}.$$

$l_{\text{rot}}$ represents the pure rotational part of the PAM, corresponding to the eigenvalue of the rotational symmetry $R_n$, thus it is an integer. $l_{\boldsymbol{\tau}} = \frac{\boldsymbol{q}\cdot\boldsymbol{\tau}_{m/n}}{2\pi/n}$ accounts for the contribution from the fractional translation $\boldsymbol{\tau}_{m/n}$ along the rotational axis. This decomposition reveals that the PAM $l_{ph}$ becomes $\boldsymbol{q}$-dependent in systems with screw rotational symmetries $\{R_n|\boldsymbol{\tau}_{m/n}\}$. As a result, $l_{\text{ph}}$ can take non-integer values, in contrast to systems with pure rotational symmetries where it is quantized to integers modulo $n$ (Zhang *et al.*, 2026a; Zhang and Murakami, 2022).

In phonon-mediated scattering processes, such as Raman scattering, both circularly polarized phonons and photons are eigenstates of the rotational operator. As a result, the PAM obeys a conservation law during the scattering process. These observations suggest that rotational eigenvalues, i.e., PAM, may serve as a unified descriptor for both topological and rotational chiralities. However, this correspondence holds only under specific symmetry constraints and is generally limited to systems with rotational symmetry, yet breaking $\mathcal{PT}$ symmetry and vertical mirror symmetry (e.g., $C_{nv}$). Although PAM can be used to selectively excite or detect phonon modes in an experiment, quantitative determination of the phonon AM, including its sign and associated rotational chirality, still requires detailed first-principles calculations.

## B. Angular Momentum *v.s.* Pseudo-angular Momentum

As discussed above, a number of recent works have employed PAM as a practical probe of phonon chirality. This widespread usage has, however, led to some ambiguity and potential confusion between the two concepts. Given that both quantities are of fundamental importance, it is therefore essential to clearly distinguish and systematically compare them. In this review, we aim to clarify their respective definitions, physical meanings, and applicability, thereby providing a unified perspective on their roles in characterizing topological chirality and rotational chirality.

The rotation operator in the $O(3)$ group with an arbitrary angle $\phi$ along the $\alpha$-th direction can be expressed as:

$$R(\alpha,\phi) = e^{-i\mathfrak{M}_\alpha\phi}, \tag{103}$$

Since phonon AM is the expectation value of $\mathfrak{M}_\alpha$, it can be defined in any systems, regardless of its symmetry. The definition in Eq. (97) demonstrates that AM solely encapsulates information about the phonon polarization and atomic motion within the primitive cell. It is independent of the relative phases between different atoms, reflecting the local property of the phonon mode, as discussed in (Streib, 2021; Zhang *et al.*, 2026a, 2025c).

Meanwhile, PAM reflects the eigenvalue of $\mathcal{C}_n$ operator, making it both an observable and a conserved physical quantity during phonon-related scattering processes in systems invariant under $\mathcal{C}_n$. When considering systems with translation symmetry, only discrete rotational symmetries $\mathcal{C}_n$ appear in crystals, where $n\in\{2,3,4,6\}$. Consequently, the rotational operator is expressed as:

$$\mathcal{C}_n = e^{-i\mathfrak{M}_\alpha 2\pi/n}. \tag{104}$$

Phonon PAM is defined by the eigenvalue of $\mathcal{C}_n$, as shown in Eq. (101), and thus can only be defined at $\mathcal{C}_n$-invariant momenta. Furthermore, PAM encodes information about the relative phase between $\mathcal{C}_n$-related atoms, potentially extending across multiple primitive cells. From this perspective, PAM can be regarded as a global property of a specific phonon mode (Streib, 2021; Zhang *et al.*, 2026a, 2025c).

As summarized in Tab. VI, phonon AM can be defined in any systems and reflects the local properties of a phonon mode. In contrast, PAM is defined only at (screw) rotation-invariant $\boldsymbol{q}$ points, as it reflects the eigenvalues of $\mathcal{C}_n$, revealing global information about the phonon mode. In specific systems such as Weyl phonons in Te (Zhang *et al.*, 2023b), or at high-symmetry points such as the K point discussed in (Zhang and Niu, 2015), symmetry constraints and band structure details allow for a correspondence between the sign of AM and PAM. However, there is in general no one-to-one correspondence between AM and PAM, and such relations require careful symmetry analysis and first-principles calculations to be established. Therefore, to observe circularly polarized phonons by PAM (or PAM conservation rule), first-principles calculations and symmetry analysis are essential, as demonstrated in studies on Te, $\alpha-$HgS and among others (Che *et al.*, 2024; Ishito *et al.*, 2023b; Yang *et al.*, 2025a; Zhang *et al.*, 2023b; Zhu *et al.*, 2018). The selection rules for PAM in the circularly polarized Raman scattering in systems with rotational symmetries are illustrated in Fig. 37.

Table VI The relationship between angular momentum and the pseudo-angular momentum.

| | Angular Momentum | Pseudo-Angular Momentum |
|---|---|---|
| Definition | $R(\alpha,\phi)=e^{i\mathfrak{M}_\alpha\phi}$<br>$l^\alpha_{\nu\boldsymbol{q}}=\sum_i^r l^{i\alpha}_{\nu\boldsymbol{q}}=\sum_i^r \hbar\langle\boldsymbol{\varepsilon}^i_{\nu\boldsymbol{q}}\vert\mathfrak{M}_\alpha\vert\boldsymbol{\varepsilon}^i_{\nu\boldsymbol{q}}\rangle$<br>$\mathfrak{M}_\alpha$ forms the Lie algebra of $O(3)$ | $\mathcal{C}_n=e^{i\mathfrak{M}_\alpha 2\pi/n}$<br>$D(\mathcal{C}_n)\vert\boldsymbol{\varepsilon}_{\nu\boldsymbol{q}}\rangle=e^{-i2\pi l_{\text{ph}}/n}\vert\boldsymbol{\varepsilon}_{\nu\boldsymbol{q}}\rangle$<br>$\mathfrak{M}_\alpha$ forms the Lie algebra of $O(3)$ |
| Difference | local property of phonon | global property of phonon |
| Correlation | There's no intrinsic relationship between PAM and AM, yet one can use PAM to identify topological information and chiral information in chiral systems combined with symmetry analysis and first-principle calculations. | |

## C. Topological chirality and rotational chirality

This subsection examines the interplay between topological chirality and rotational chirality in phonon systems (Fig. 35). We focus in particular on Weyl phonons, which simultaneously exhibit nontrivial topological property and well-defined rotational chirality, and analyze the key features that distinguish these modes. In many candidate materials, such phonon feature coexist with nontrivial electronic topology, such as Weyl fermions. An increasing number of studies suggest a correlation between electronic topology and phonon chirality; however, a comprehensive treatment of this connection lies beyond the scope of the present review (Che *et al.*, 2025c; Cui *et al.*, 2023; Luo *et al.*, 2023a; Yang *et al.*, 2025b; Zhang *et al.*, 2026b).

### 1. Weyl Phonons with Coexisting Topological and Rotational Chirality

Topological phonons are defined and classified by their topological invariants, leading to a rich variety of categories, including Weyl phonons, Dirac phonons, and nodal-line phonons, among others. Each classification is associated with distinct topological surface states and bulk properties, as governed by the principle of "bulk-surface correspondence." Weyl phonons, in particular, represent a unique class of quasiparticles that are both topological and chiral. They are characterized by a topological invariant, namely the Chern number. When the Chern number is nonzero, the phonon mode is topological, exhibiting helicoid surface states and unique pseudospin textures around the Weyl phonon degeneracy, as shown in Sec. IV.A.

The left panels of Fig. 35 illustrate an example of a twofold-degenerate quadruple Weyl phonon in the material BaPtGe (Li *et al.*, 2021a; Zhang *et al.*, 2025c, 2020b). This phonon mode features four parallel helicoid surface states due to its Chern number of 4, showcasing the interplay between topology and chirality in such systems.

The right panels show original definition of circularly polarized phonons relied on the concept of nonzero AM. Phonon mode exhibits circular vibrations in real space when AM is nonzero, providing a clear and unambiguous signature of rotational chirality. It also illustrate an example of the twofold-degenerate quadruple Weyl phonon in the material BaPtGe, which exhibits nonzero AM and circular atomic motions for the phonon mode with a Chern number of 4. This example highlights the connection between topological properties and chiral behavior in phonon systems.

We highlight two prototypical examples that decouple topological nature from rotational chirality: (i) The $A_1$ phonon in tellurium at general $k$-points (non-TRIM momenta) is non-topological yet circularly polarized, the nonzero AM arising from the chiral symmetry of the crystal structure. (ii) The nodal-line phonons in $MoB_2$ are topological, characterized by a Berry phase and $\mathcal{PT}$-symmetry protection, but are linearly polarized with vanishing AM, serving to decouple these concepts.

### 2. Diagnosing Topological Chirality by PAM

Topological classifications for Weyl quasiparticles with (screw) rotational symmetry $\mathcal{C}_n$ have been established (Fang *et al.*, 2012; Tsirkin *et al.*, 2017; Zhang *et al.*, 2026a). The monopole charge of a Weyl phonon on high-symmetry lines or at high-symmetry points depends on the symmetry of the bands forming the Weyl phonon, specifically the PAM in $\mathcal{C}_n$-invariant systems, as well as the little point group where the Weyl point is located (Zhang *et al.*, 2026a). Consequently, PAM can be used to diagnose Weyl phonons in $\mathcal{C}_n$-invariant systems.

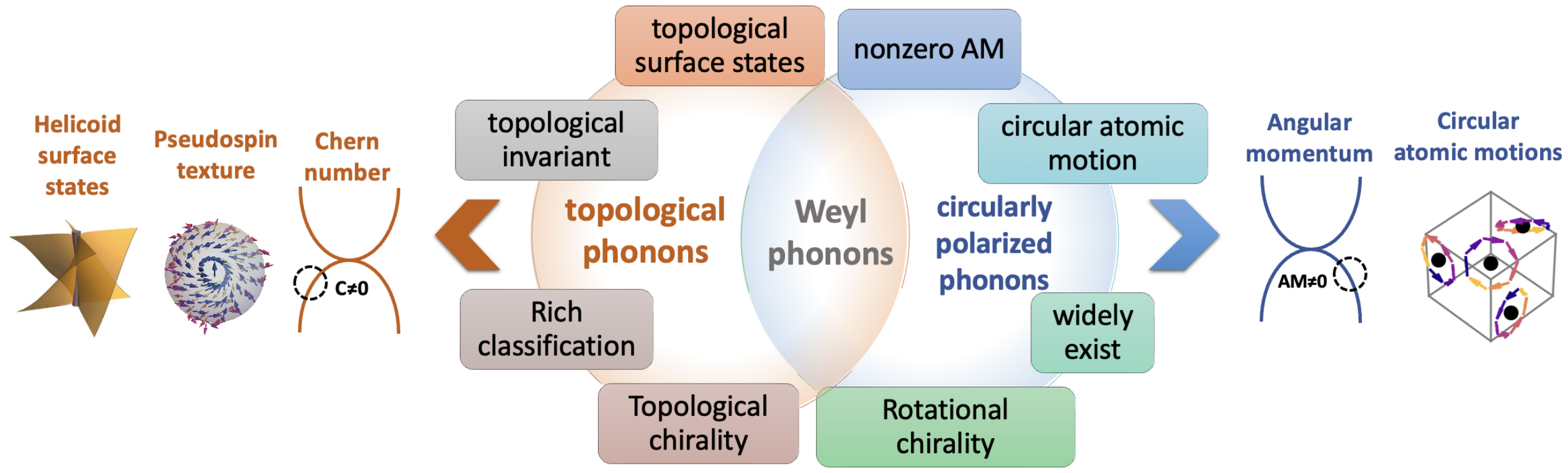

Figure 35 Weyl phonons provide a prototypical example of modes that simultaneously exhibit topological chirality and rotational chirality. Adapted from Ref. (Zhang *et al.*, 2025c).

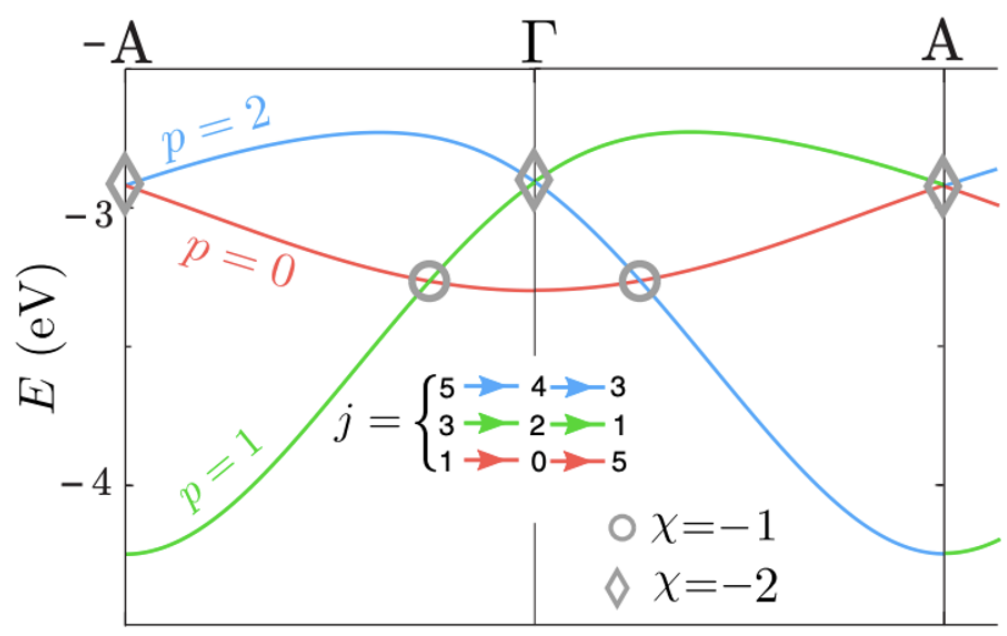

Figure 36 A group of band structures for a system along the $-A$-$\Gamma$-$A$ path, which preserves $\mathcal{C}_3$ rotational symmetry. Here, $p$ denotes the pseudo-angular momentum (PAM) value for each band, and $\chi$ represents the monopole charge of the Weyl phonons. It is important to note that the PAM for a specific phonon mode can be non-integer in systems with $\mathcal{C}_n$ screw rotational symmetry. Adapted from Ref. (Tsirkin *et al.*, 2017).

Figure 36 provides an example in $\mathcal{C}_3$-invariant systems, where $p$ represents the pure rotation part of PAM and $\chi$ denotes the Chern number (monopole charge) (Tsirkin *et al.*, 2017). The Weyl phonons at both $\Gamma$ and A, which are time-reversal invariant momenta, are double Weyl phonons with a monopole charge of $-2$, while those along the $\Gamma$-A path are conventional Weyl phonons with a monopole charge of $-1$. Thus, relationship between PAM and the Chern number for Weyl phonons can be established, providing a systematic framework for detecting topological chirality in the circularly polarized Raman scattering process. Moreover, the selection rules for circularly polarized Raman scattering in systems with rotational symmetries $\mathcal{C}_n$ are illustrated in Fig. 37 (Watanabe *et al.*, 2025; Zhang *et al.*, 2026a).

### 3. Diagnosing Rotational Chirality by PAM and DFT calculation

In several previous studies, the PAM has been employed to identify circularly polarized phonons. However, caution is required, since PAM is not intrinsically equivalent to the phonon AM, and therefore cannot, in general, be directly used to diagnose rotational chirality. In chiral systems, the phonon AM is generally nonzero at non-TRIM and can be probed experimentally through PAM-sensitive circularly polarized Raman scattering. Nevertheless, the sign of the phonon AM typically requires additional first-principles calculations for unambiguous determination.

Experimental observations of circularly polarized phonons in systems with $\mathcal{C}_3$ screw rotational symmetry have been reported in both $\alpha$-HgS (Ishito *et al.*, 2023b) and Te (Zhang *et al.*, 2023b), supported by first-principles calculations and symmetry analysis. These studies provide evidence for circularly polarized phonons in rotationally symmetric materials and highlight the interplay among symmetry, topology, and phonon circular polarization. In the following, we take Te as an example to illustrate that Weyl phonons can simultaneously exhibit both topological and rotational chirality, and further discuss in detail how such phonons can be experimentally probed through the PAM conservation rule.

Chiral crystals such as Te exist in two enantiomeric structures, as illustrated in Figs. 38 (a1) and (b1). These enantiomers exhibit threefold screw rotational symmetry along the atomic chain direction. Using the Weyl phonons in these two enantiomers as an example, we will show how to diagnose their topological and chiral properties through helicity-resolved Raman scattering, or circularly polarized Raman scattering.

Figures 38 (a2) and (b2) depict the atomic motions for the lower branch of the Weyl phonon along the $\Gamma$-A direction in right- and left-handed crystals, respectively. Figures 38 (a3) and (b3) are the PAM, (a4) and (b4) are the Chern number of the Weyl phonon in Te with right- and left-handed structures. If the chirality of the crystal structure changes,

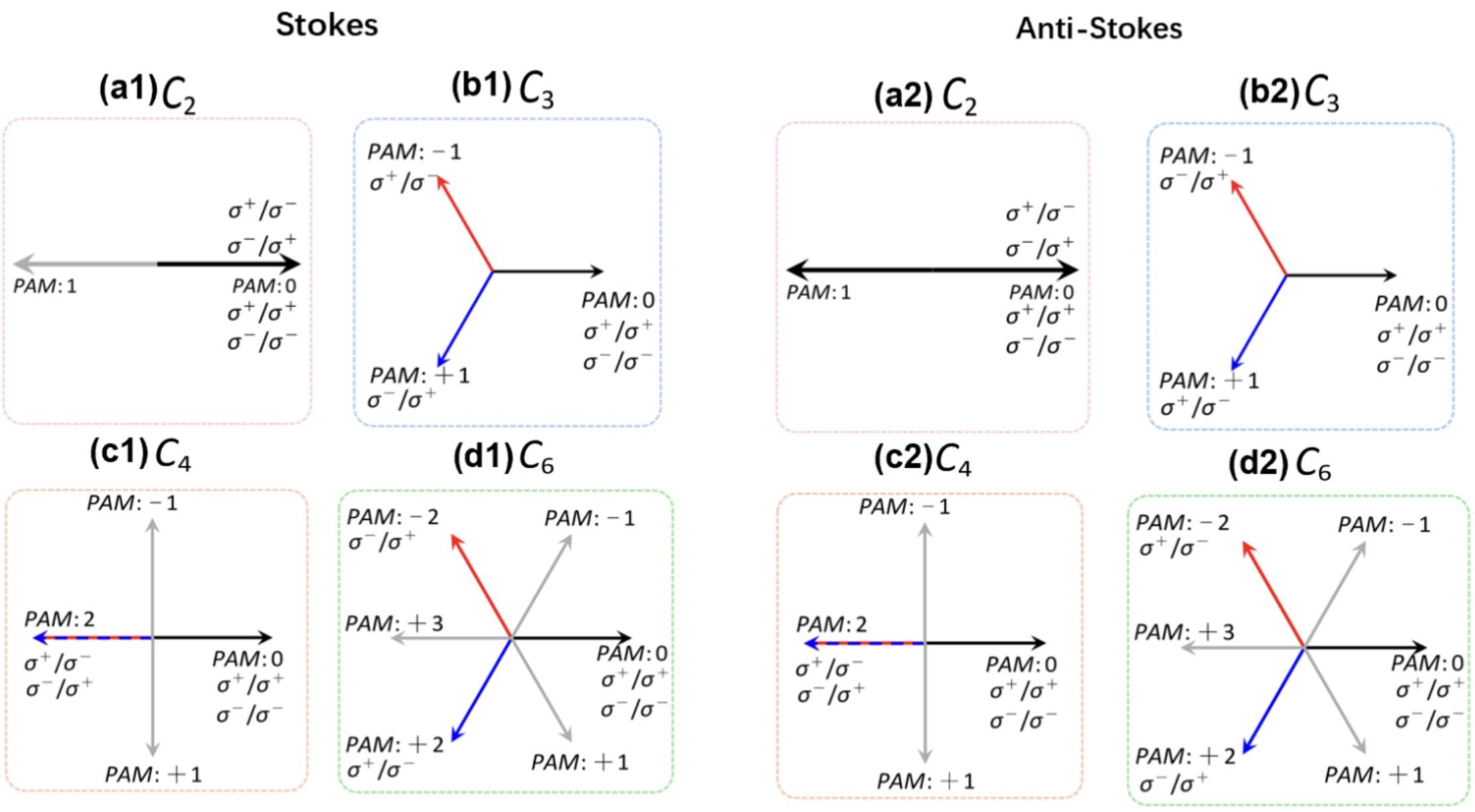

Figure 37 The selection rules for circularly polarized Raman scattering in the Stokes [(a1-d1)] and anti-Stokes [(a2-d2)] process for the systems with $\mathcal{C}_n = 2, 3, 4, 6$ rotation symmetries, where the incident and scattered light propagate along the rotation axis. Each arrow represents the eigenvalue of a phonon mode in the complex plane, corresponding to its PAM value. In each figure, phonon modes marked by the black arrow are Raman active in both the $\sigma^+/\sigma^+$ and $\sigma^-/\sigma^-$ processes, phonon modes marked by the blue (red) arrows are active in the $\sigma^+/\sigma^-$ ($\sigma^-/\sigma^+$) process. These selection rules highlight the dependence of Raman activity on the PAM values and the circular polarization of the incident and scattered light. Adapted from Ref. (Zhang *et al.*, 2026a).

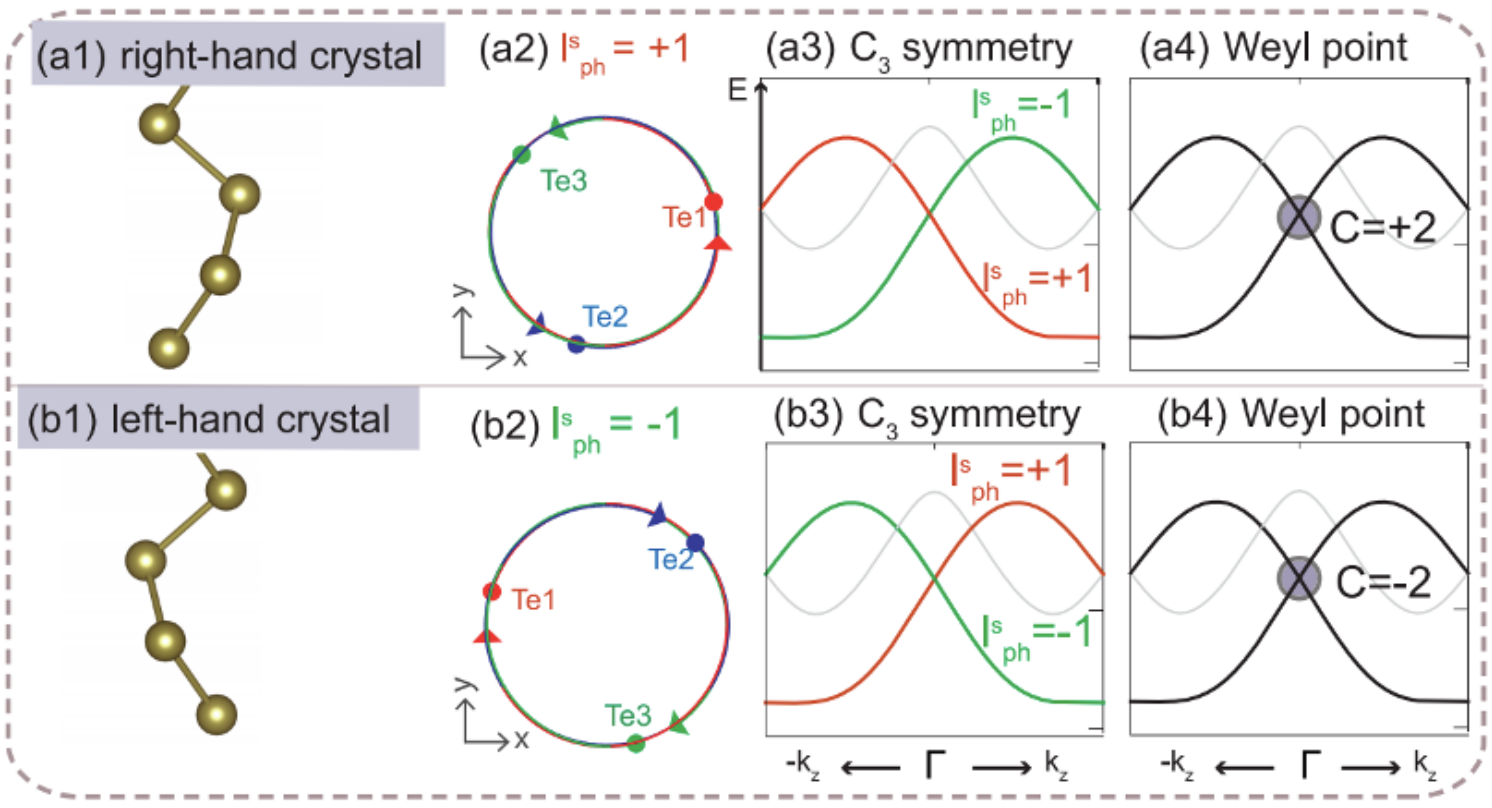

Figure 38 (a) and (b) are the different physical quantities for Te with the right-hand and left-hand structure, respectively. (a2) and (b2) shows the circular atomic motion for the same phonon modes but with different chiral crystal structures, showing opposite chiral motions in real space. (a3) and (b3) shows the pseudo-angular momentum for the same phonon modes with different chiral crystal structures. (a4) and (b4) shows the Chern number of the Weyl phonon in different chiral crystal structures. Thus, there is a close relationship between the chirality of crystal structure, the polarization of the atomic motion in the real space, the sign of pseudo-angular momentum, and the Chern number. Adapted from Ref. (Zhang *et al.*, 2026a).

the circularly polarization (obtained from the atomic motion), the sign of PAM and the Chern number will also be altered. Therefore, the chirality of the crystal structure is intrinsically linked to the sign of the AM (and thus the rotational chirality), the PAM, and the Chern number, demonstrating the profound connection between structural chirality and the topological and chiral properties of phonons (Bousquet *et al.*, 2024; Zhang *et al.*, 2026a, 2023b, 2025c).

In particular, we emphasize that the connection between PAM and the Chern number is rooted in topological band theory and is generally well-defined and robust within a given symmetry setting. By contrast, the relation between angu-lar momentum, which characterizes real-space circular motion (rotational chirality), and topological quantities such as the Chern number (topological chirality), is not universal and only holds in certain special cases. The experimental observation of topological phonons and circularly polarized phonons will be reviewed in Sec. VI.

## VI. EXPERIMENTAL PROGRESSES OF TOPOLOGICAL AND CIRCULARLY POLARIZED PHONONS

Theoretical advances have stimulated extensive experimental investigations of topological and circularly polarized phonons. In this section, we review the major experimental developments in this field.

### A. Topological Phonons

Similar to topological fermions, experimental studies of topological phonons have focused on both the phonon band dispersions, namely the eigenfrequencies $\omega_{n\boldsymbol{q}}$, and the associated wave functions characterized by the eigenvectors $|\varepsilon_{n\boldsymbol{q}}\rangle$. As discussed in Sec. I.C, these quantities can be directly probed through measurements of the phonon DSF using momentum- and energy-resolved scattering techniques.

#### 1. Topological Phonons Observed by IXS and INS

Soon after theoretical predictions of topological phonons in crystalline materials (Zhang *et al.*, 2018a), these states are experimentally observed in FeSi (Miao *et al.*, 2018). As described in Sec. IV, FeSi has B-20 structure that breaks inversion-symmetry. Its phonon excitations host spin-1 Weyl phonon at the $\Gamma$ point and charge-2 Dirac phonons at the $R$ point as shown in Fig. 25. The dispersion of these topological quasiparticles are experimentally determined using IXS. Figure 39 (a) shows the IXS determined $S(\mathbf{Q},\omega)$ of FeSi near the $R$ point. The second derivative plot of Fig. 39 (a) is shown in Fig. 39 (b). Green and red curves are DFT calculated phonon band dispersion. The overall excellent agreement between DFT calculation and IXS spectra supports the effective low-energy model for charge-2 Dirac phonon at the R point (Zhang *et al.*, 2018a). A detailed fitting of the IXS curve shown in Figs. 39 (c)-(f) further confirm the degenerate point and Dirac-like phonon band dispersion near the R-point.

Following the similar experimental protocol, $\mathcal{PT}$-symmetry–protected phononic nodal lines were observed in $MoB_2$ (Zhang *et al.*, 2019b). The IXS spectra revealed phonon band degeneracy points forming a nodal ring along the $L$-direction of the reciprocal space. Because these nodal rings are protected by $\mathcal{PT}$ symmetry, they can take arbitrary shapes and extend continuously across multiple Brillouin zones.

Topological quasiparticles, including Weyl fermions (Lv *et al.*, 2015b; Murakami, 2007; Pal, 2011; Wan *et al.*, 2011; Weng *et al.*, 2015; Xu *et al.*, 2015), phonons (Ding *et al.*, 2022; Li *et al.*, 2021a, 2018; Liu *et al.*, 2019b, 2020a, 2021e; Miao *et al.*, 2018; Xia *et al.*, 2019; Zhang *et al.*, 2023b, 2018a, 2020b), magnons (McClarty, 2022), and fractionalized excitations (Chen *et al.*, 2015; Ghaemi *et al.*, 2012; Levin and Stern, 2009, 2012; Maciejko and Fiete, 2015; Oshikawa and Senthil, 2006; Senthil and Fisher, 2001a,b; Seradjeh *et al.*, 2009; Shirley *et al.*, 2019; Song and Hermele, 2015; Stern, 2016; Sun and Mao, 2021; Vaezi, 2013; Wen, 1995), are defined on topological invariants that are derived from the bulk wave functions. This requires experimental measurements to go beyond the phonon band structure. As shown in Eq. (23), the phononic wave function enters $S(\boldsymbol{Q},\omega)$ through the term $\boldsymbol{Q}\cdot\boldsymbol{\epsilon}_{\mathbf{q}\sigma}(s)$. While a single measurement at a fixed $\boldsymbol{Q}$ measures phonon eigenvectors projected on $\boldsymbol{Q}$, the full phononic Hamiltonian can be extracted by fitting the experimentally determined $S(\boldsymbol{Q},\omega)$ in a large momentum space. This data driven procedure is essentially the same as determining the magnetic Hamiltonian using INS (Coldea *et al.*, 2001; Dahlbom *et al.*, 2025; Scheie *et al.*, 2022).Practically, since the DFT calculated $S(\boldsymbol{Q},\omega)$ is usually high accuracy, the topological phonons can be measured by comparing experimentally determined $S(\boldsymbol{Q},\omega)$ along high symmetry lines with first principles calculations.

For a topological chiral wave function, the three components of $\boldsymbol{\epsilon}_{\mathbf{q}\sigma}(s)$ are complex numbers with different phase factors, describing how atoms move out-of-phase with one another. Therefore, the circularly polarized phonon wave function will give rise to non-trivial interferences of the real and imaginary component of $\boldsymbol{\epsilon}_{\mathbf{q}\sigma}(s)$, which strongly modify the momentum dependent intensity distribution of $S(\boldsymbol{Q},\omega)$. One can visualize this effect by calculating $S^{\mathrm{DFT}}(\boldsymbol{Q},\omega)$, $R(\boldsymbol{Q},\omega)$ and $I(\boldsymbol{Q},\omega)$ (Li *et al.*, 2021a). Here $R(\boldsymbol{Q},\omega)$ and $I(\boldsymbol{Q},\omega)$ consider only the real and imaginary part of $\boldsymbol{\epsilon}_{\mathbf{q}\sigma}(s)$ in Eq. (23). The chiral phononic wave function can therefore be revealed by quantitatively comparing the experimental $S(\boldsymbol{Q},\omega)$ and $S^{\mathrm{DFT}}(\boldsymbol{Q},\omega)$. This approach has been applied in the IXS study of BaPtGe that hosts twofold quadruple Weyl (TQW) phonons described in Sec. IV (Li *et al.*, 2021a). It has been shown that the wave functions near the twofold quadruple Weyl are primarily real along the $X_1-\Gamma-X_2$ and $M_1-\Gamma-M_2$ high-symmetry directions. In contrast, the phononic wave function along the $R_1-\Gamma-R_2$ direction is mainly imaginary (Li *et al.*, 2021a). To understand the direction-dependent wave function near the twofold quadruple Weyl point at the $\Gamma$ point, (Li *et al.*, 2021a) developed a simplified model that considers only four Pt atoms in the unit cell with locations $\boldsymbol{r}_{\mathrm{Pt1}}=(c,c,c)$, $\boldsymbol{r}_{\mathrm{Pt2}}=(-c+\frac{1}{2},-c,c+\frac{1}{2})$, $\boldsymbol{r}_{\mathrm{Pt3}}=(-c,c+\frac{1}{2},-c+\frac{1}{2})$, $\boldsymbol{r}_{\mathrm{Pt4}}=(c+\frac{1}{2},-c+\frac{1}{2},-c)$, where $c$ is an arbitrary value satisfying $c\leq 1$. Based on the group theory, the atomic motions at the $\Gamma$ point can be described as three non-degenerate basis states with irreducible representations of $\Gamma_1^{(1)}$, $\Gamma_2^{(1)}$, $\Gamma_3^{(1)}$, and three threefold degenerate basis states with irreducible representations of $\Gamma_4^{(3)}$. The basis states of $\Gamma_2^{(1)}$ and $\Gamma_3^{(1)}$ that form the TQW can be derived by imposing chiral cubic crystal symmetry and time-reversal

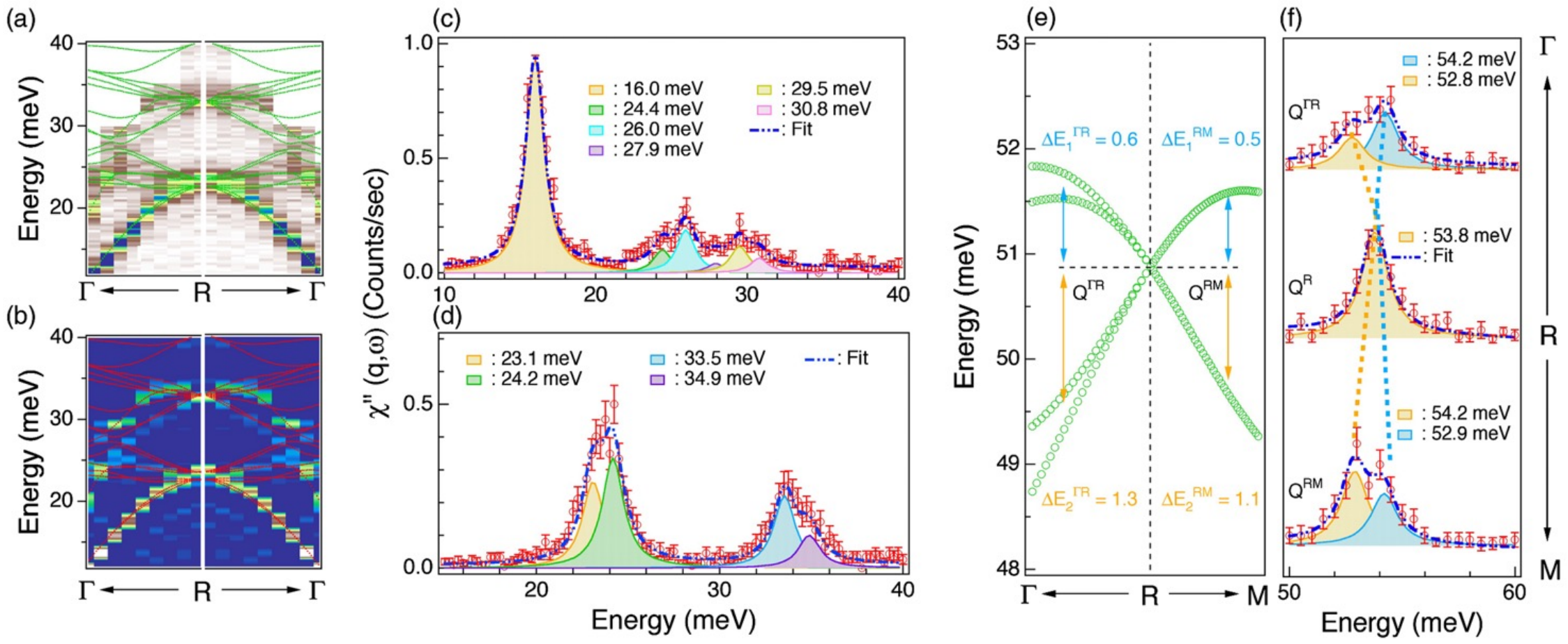

Figure 39 Evidence of charge-2 Dirac phonon in FeSi. (a), (b) Color plot and second derivative plot of phonon dynamical structure factor near the R point. (c) and (d) show high-statistic IXS spectra at $\boldsymbol{Q}$=(4.26, 0.26, 0.26) and (4.5,0.5,0.5), respectively. The dashed blue curves are the fitted result. (e) The charge-2 Dirac phonon in the 50 meV energy range. (f) shows the high statistic IXS spectra at $\boldsymbol{Q}^{\Gamma R}$, $\boldsymbol{Q}^{R}$, and $\boldsymbol{Q}^{RM}$. Adapted from Ref. (Miao *et al.*, 2018).

symmetry:

$$\phi_{\Gamma_2} = (e^{\Gamma_2}_{\mathrm{Pt1}}, e^{\Gamma_2}_{\mathrm{Pt2}}, e^{\Gamma_2}_{\mathrm{Pt3}}, e^{\Gamma_2}_{\mathrm{Pt4}}) = \frac{(1, \omega, \omega^2, -1, -\omega, \omega^2, -1, \omega, -\omega^2, 1, -\omega, -\omega^2)}{\sqrt{12}}, \tag{105}$$

$$\phi_{\Gamma_3} = (e^{\Gamma_3}_{\mathrm{Pt1}}, e^{\Gamma_3}_{\mathrm{Pt2}}, e^{\Gamma_3}_{\mathrm{Pt3}}, e^{\Gamma_3}_{\mathrm{Pt4}}) = \frac{(1, \omega^2, \omega, -1, -\omega^2, \omega, -1, \omega^2, -\omega, 1, -\omega^2, -\omega)}{\sqrt{12}} \tag{106}$$

where $\omega = e^{i2\pi/3}$. $\phi_{\Gamma_2}$ and $\phi_{\Gamma_3}$ have opposite chirality following $\phi_{\Gamma_2} = \phi^*_{\Gamma_3}$. In the vicinity of the $\Gamma$ point, low-energy effective Hamiltonian for the twofold quadruple Weyl:

$$H(\mathbf{q}) = \begin{pmatrix} A q_x q_y q_z & B^*(q_x^2 + \omega^2 q_y^2 + \omega q_z^2) \\ B(q_x^2 + \omega^2 q_y^2 + \omega q_z^2) & -A q_x q_y q_z \end{pmatrix}, \tag{107}$$

where $\omega = e^{2\pi i/3}$, $A$ is a real constant, and $B$ is a complex constant. The $q$-dependent wave functions can be obtained by diagonalizing Eq. (107) in the $\phi_{\Gamma_2}$ and $\phi_{\Gamma_3}$ basis states. Along the [100] and [011] directions, the diagonal components of Eq. (107) are zero, imposing purely real wave functions, $\psi_1 = (e^{i\theta/2}\phi_{\Gamma_2} + e^{-i\theta/2}\phi_{\Gamma_3})/\sqrt{2}$ and $\psi_2 = -i(e^{i\theta/2}\phi_{\Gamma_2} - e^{-i\theta/2}\phi_{\Gamma_3})/\sqrt{2}$, where $\theta = \arg(B)$. In contrast, along the [111] direction, the off-diagonal components are zero yielding $\psi_1 = \phi_{\Gamma_2}$ and $\psi_2 = \phi_{\Gamma_3}$, which have large imaginary components. This simplified model analysis thus explains the directional wave function reported in (Li *et al.*, 2021a).

Beside the general theory-experiment-comparison approach to determine topological and circularly polarized phonons, (Jin *et al.*, 2022) showed that IXS and INS can be used to directly extract the Chern number of a topological Weyl point. As an example, (Jin *et al.*, 2022) considered a twofold Weyl Hamiltonian:

$$H_{2\times 2}(\boldsymbol{q}) = \sum_{i=x,y,z} f_i(\boldsymbol{q}) \cdot \sigma_i + f_0(\boldsymbol{q})\sigma_0 \tag{108}$$

where $\sigma_i$ are the Pauli matrix. They then define a pseudospin:

$$\boldsymbol{S}(\boldsymbol{q})_{1,2} = \frac{\boldsymbol{f}(\boldsymbol{q})}{|\boldsymbol{f}(\boldsymbol{q})|} = \langle \boldsymbol{\phi}_{1,2} | \boldsymbol{\sigma} | \boldsymbol{\phi}_{1,2} \rangle \tag{109}$$

where $\boldsymbol{\phi}_{1,2}$ is the eigenvector of Eq. (108). The subscript 1 and 2 denote the upper and lower bands that form the Weyl node. The coherent scattering contribution for the inelastic neutron scattering intensity (Eq. (32)) can then be written in the pseudospin quantities:

$$\mathcal{S}_{coh} \propto |\mathbf{V}|(1 + \cos\langle \mathbf{S}(\mathbf{q}) \cdot \mathbf{V}(\mathbf{G}) \rangle) \tag{110}$$

where $\mathbf{V}(\mathbf{G})$ is a constant vector that does not sensitively depend on $\mathbf{q}$ but varies by changing $\mathbf{G}$. Equation (110) shows that the INS intensity is sensitive to the projection of $\mathbf{S}(\mathbf{q})$ on $\mathbf{V}(\mathbf{G})$. This effect has been confirmed by an INS study of MnSi (Jin *et al.*, 2022). As shown in Fig. 41, the averaged INS intensity over solid $\mathbf{q}$ sphere show maximum and minimum intensity in agreement with Eq. (110). Similar approach has also been applied to topological magnons (McClarty, 2022).

### 2. Topological Phonons Observed by EELS

Early experimental studies of topological phonons primarily focused on three-dimensional single crystals, while direct experimental confirmation in two-dimensional materials remained elusive until the work of Ref. (Li *et al.*, 2023a).

Using surface-sensitive high-resolution electron energy-loss spectroscopy (HR-MEELS), (Li *et al.*, 2023a) mapped the phonon spectra of atomically thin graphene over the entire two-dimensional Brillouin zone, as shown in Fig. 40. By comparing the measured phonon dispersions with first-principles calculations, the authors identified two nodal-ring phonons and four Dirac phonons [Fig. 40(c)]. Figure 40(d) further presents a three-dimensional visualization of the phonon spectra combining the two-dimensional momentum space and energy dimensions. Using the same experimental methodology, the same group subsequently demonstrated experimental evidence for topological phonons in monolayer hexagonal boron nitride (Tao *et al.*, 2025).

While topological phonons in the 3D and 2D bulk state have been extensively studied, the bulk-edge correspondence of topological phonons remain to be experimentally established. For this purpose, surface and local probes of phonon excitations are required. The ability to reveal the phonon spectrum of graphene in large momentum space making HR-MEELS a promising tool to uncover phononic surface state in 3D systems (Miao *et al.*, 2018; Zhang *et al.*, 2018a, 2019b). The phononic surface mode may also be encoded in electronic spectral functions through quasiparticle interactions. For instance, the drumhead surface phonon mode has been predicted in $MoB_2$ and high-$T_c$ conventional superconductor $MgB_2$. Due to the strong electron-phonon coupling, the topological phonon edge mode can couple with the conducting electrons on the surface, giving rise to electronic kinks that can be revealed using angle-resolved photoemission spectroscopy.

## B. Circularly Polarized Phonons

As we discussed in previous sections (see Fig. 35), topological and circularly polarized phonons are deeply connected in systems with Weyl phonons (Zhang *et al.*, 2025c). In this context, experimental studies of Weyl phonons and phononic wave functions (Li *et al.*, 2021a; Miao *et al.*, 2018) have provided important insights into phonon band topology and polarization properties. A definitive identification of circularly polarized phonon modes, however, typically requires additional input from symmetry analysis and first-principles calculations. On the other hand, since circularly polarized phonons are defined on AM, they are naturally connected to magnetism (Baydin *et al.*, 2022b; Che *et al.*, 2025a; Cheng *et al.*, 2020c; Fransson, 2023; Hernandez *et al.*, 2023; Hu *et al.*, 2021; Juraschek *et al.*, 2022b; Kim *et al.*, 2023; Li *et al.*, 2024b; Mankovsky *et al.*, 2022; Nova *et al.*, 2017b; Peters *et al.*, 2022; Ren *et al.*, 2021b; Saparov *et al.*, 2022; Wu *et al.*, 2023b; Xue *et al.*, 2025b; Yang *et al.*, 2025a), thermal Hall effect (Chen *et al.*, 2020, 2022b; Grissonnanche *et al.*, 2020, 2019; Kasahara *et al.*, 2018; Li *et al.*, 2020d, 2023b; Ohe *et al.*, 2024; Qin *et al.*, 2012; Zhang *et al.*, 2019c) and light-matter scattering processes (Bourgeois *et al.*, 2025; Che *et al.*, 2025a; Ishito *et al.*, 2023a,b; Li *et al.*, 2024a; Ueda *et al.*, 2023; Yang *et al.*, 2025a; Zhang and Niu, 2015; Zhang *et al.*, 2026a, 2023b; Zhang and Murakami, 2022; Zhang *et al.*, 2025c; Zhu *et al.*, 2018). For this reason, circularly polarized phonons have attracted significant interests in the community. In this section, we highlight some recent experimental progresses that are specifically related to circularly polarized phonons.

### 1. Experimental Signature of Phonon PAM and AM

Circular dichroism (CD) in light–matter scattering has been widely used to probe circularly polarized phonons. Owing to the conservation of pseudoangular momentum (PAM), microscopic light–matter scattering processes involving circularly polarized phonons can exhibit pronounced CD signals. We emphasize that PAM and AM are distinct physical quantities. While symmetry analysis based on PAM can help identify and selectively probe circularly polarized phonons in experiments, first-principles calculations are still required to quantitatively determine the sign of phonon AM and confirm its rotational chirality. Interestingly, it has been found that (Bourgeois *et al.*, 2025; Ishito *et al.*, 2023a,b; Zhang and Niu, 2015; Zhang *et al.*, 2026a, 2023b; Zhang and Murakami, 2022; Zhang *et al.*, 2025c; Zhu *et al.*, 2018) the light-matter scattering can display new selection rules when the incident photon momentum, $\boldsymbol{k_i}$, and scattered photon momenutm, $\boldsymbol{k_f}$, are along the rotational axis.

Figure 42 (a) shows Stokes and anti-Stokes Raman spectra of the circularly polarized phonon system $\alpha$-HgS. The red and blue curves correspond to the $\Gamma_3^{(2)}$ doublet phonon modes and are obtained in ($\varepsilon_i$, $\varepsilon_f$)=(L, R) and (R, L) polarization conditions, respectively. Here $L$ and $R$ stand for circular left and circular right polaritons with photon PAM $\sigma_L$=$-1$ and $\sigma_R$=$+1$. Notably, the PAM conservation for a $C_3$-symmetric direction enforces a selection rule: $\sigma_f - \sigma_i = -l_q^{ph}$ $mod$ 3, in agreement with experimental observations. The circularly polarized-phonon-driven CD has also been observed in a resonant inelastic x-ray scattering (RIXS) study of chiral quartz crystals as shown in Fig. 42 (b). It is important to note that the CD in RIXS measured at finite $\boldsymbol{Q}$ is not a direct consequence of PAM conservation. This is because unlike Raman scattering that measures phonon near $\boldsymbol{Q}$=0, for RIXS measures phonon at finite $\boldsymbol{Q}$. The Raman selection rules derived from the symmetry representations and pseudo-angular momentum (PAM) conservation under crystalline rotational symmetry are no longer valid for a typical RIXS measurement. Instead, the CD in RIXS arises from the coupling between circularly polarized phonons and the RIXS intermediate states (Devereaux *et al.*, 2016), which shows different scattering cross-section under $L$ and $R$ polarizations.

Circularly polarized phonons have also been observed in 2D using ultrafast pump-probe techniques (Britt and Siwick, 2023; Zhang *et al.*, 2025d; Zhu *et al.*, 2018). Figure 42 (c) shows the indirect optical transitions in a transition metal dichalcogenides $WSe_2$. The absence of inversion symmetry in $WSe_2$ lifts the degeneracy of circularly polarized phonon modes at the $K$ and $K$' valleys. The photon excited hole, $h_A(K)$, at the valence band of the $K$-valley are scattered, through its polarization dependent interactions with circularly polarized phonons, to the spin-split conduction band, $h_B(K')$, of the $K'$-valley, yielding CD in the time domain [Fig. 42 (b2)] and a characteristic circularly polarized phonon energy, 29±8 meV, at delay time $t = 0.8$ ps [Fig. 42 (b3)].

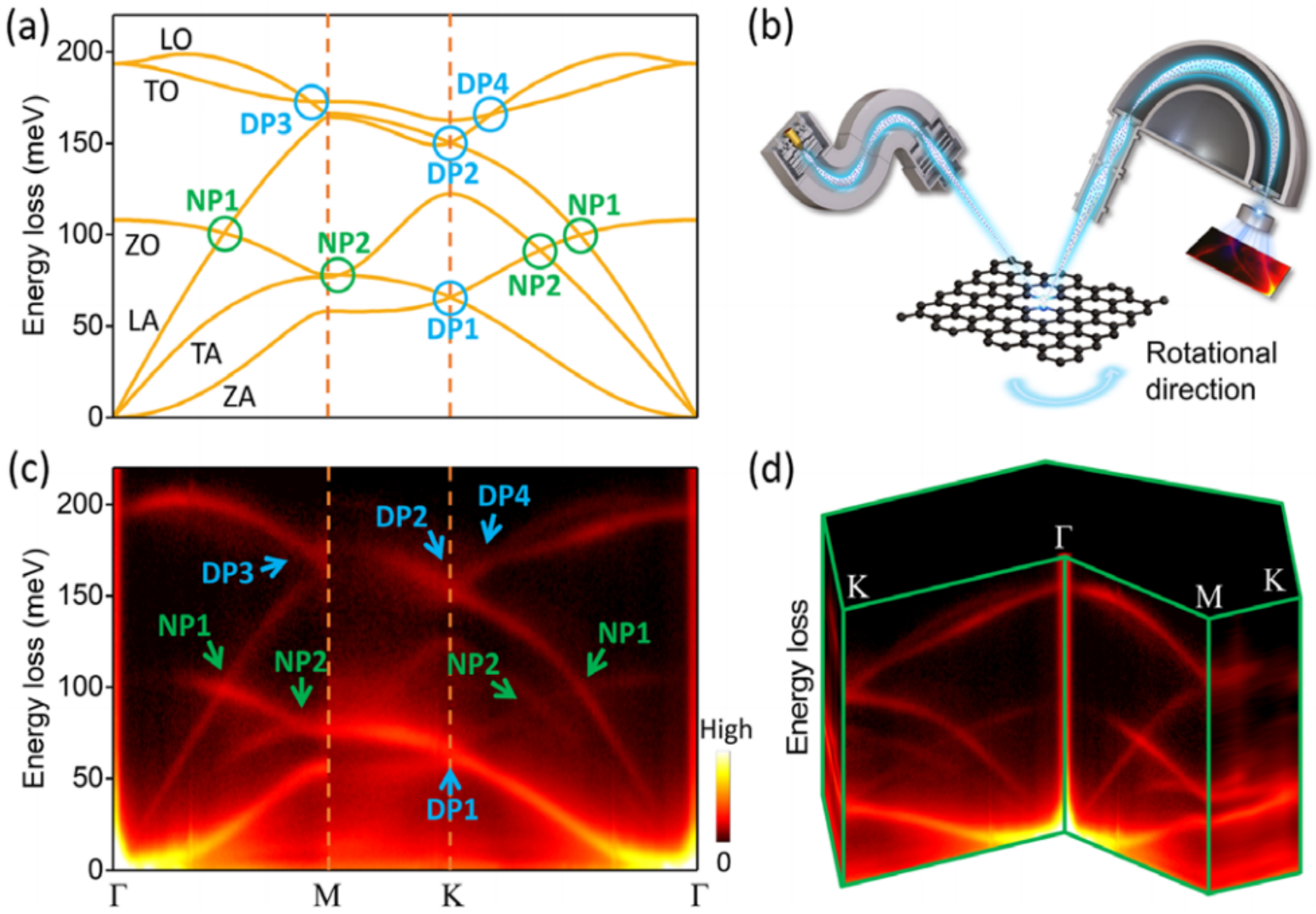

Figure 40 Phonon spectra and experimental setup for graphene. (a) Phonon bands obtained by first-principle calculations. Topological nodal-ring phonons (NPs, green circles) and Dirac phonons (DPs, blue circles) are highlighted. (b) Schematic illustration of the high-resolution electron energy loss spectroscopy (HR-EELS) experimental setup. (c) Experimental phonon spectra of graphene along the same high-symmetry paths as (a), with NPs and DPs marked by green and blue arrows, respectively. (d) 3D HR-EELS mapping of graphene phonon spectra, integrating momentum-space and energy-resolved measurements. Adapted from Ref. (Li *et al.*, 2023a).

Although CD provides important signature of circularly polarized phonons, it should be emphasized that CD is sensitve to mirror-symmetry breaking rather than the AM. It thus calls for experimental protocals to directly reveal the AM of phonons. Historically, the AM of microscopic quantum objects, such as the AM of electron spin, has been determined through the Einstein-de Hass (EdH) effect (Einstein, 1915; Frenkel, 1979; Gans, 1916). Due to total AM conservation, the suspended ferromagnet displays a macroscopic mechanical rotation under external magnetic field to compensate the total AM of microscopic spins. Motivated by this concept, it is proposed that the AM of circularly polarized phonon can induce mechanical rotations under a thermal gradient that breaks the time-reversal symmetry (Hamada *et al.*, 2018). The phonon-AM can be expressed as:

$$\boldsymbol{J}_{\text{ph}} = \sum_{n\boldsymbol{q}} \boldsymbol{l}_{n\boldsymbol{q}} \left( f_{n\boldsymbol{q}} + \frac{1}{2} \right), \tag{111}$$

where $f_{n\boldsymbol{q}}$ is the phonon distribution function. At thermodynamic equilibrium condition the occupation number is just the Bose distribution function $f_0 = \left[e^{\hbar\omega_{n\boldsymbol{q}}/(k_B T)} - 1\right]^{-1}$. In the presence of small thermal gradient, the phonon distribution function, $f_{n\boldsymbol{q}}$, can be approximately described by the Boltzmann equation (under relaxation-time approximation) as:

$$\begin{aligned} f_{n\boldsymbol{q}} &= f_0 - \tau_0(\boldsymbol{v}_{n\boldsymbol{q}} \cdot \boldsymbol{\nabla} T)\frac{\partial f_0}{\partial T} + O(\tau_0^2) \\ &= f_0 + \tau_0(\boldsymbol{v}_{n\boldsymbol{q}} \cdot \boldsymbol{\nabla} T)\frac{\omega_{n\boldsymbol{q}}}{T}\frac{\partial f_0}{\partial \omega_{n\boldsymbol{q}}} + O(\tau_0^2) \end{aligned} \tag{112}$$

where $v_{n\boldsymbol{q}}$ is the group velocity of the phonon mode, and $\tau_0$ is the average relaxation time. Following Eq. (111), the thermal gradient induced total phonon AM is:

$$\delta \boldsymbol{J}_{\text{ph}} = \frac{\tau_0}{T} \sum_n \int \frac{\mathrm{d}^3\boldsymbol{q}}{(2\pi)^3} \boldsymbol{l}_{n\boldsymbol{q}} (\boldsymbol{v}_{n\boldsymbol{q}} \cdot \boldsymbol{\nabla} T) \omega_{n\boldsymbol{q}} \frac{\partial f_0}{\partial \omega_{n\boldsymbol{q}}}. \tag{113}$$

Fundamentally, Eqs. (111)-(113) are similar to the Edelstein effect in electronic systems (Einstein, 1915; Frenkel, 1979; Gans, 1916). The presence of thermal gradient will induce mechanical torque, $\tau \sim \frac{\delta J_{\text{ph}}}{\tau_0}$. Recently, using a cantilever-based method [Fig. 43], the phonon AM induced mechanical torque has been observed in chiral single crystal Te (Zhang *et al.*, 2025a). The experimentally determined $\tau \sim 5 \times 10^{-11} N \cdot m$ is remarkably consistent with theoretical calculations, supporting phonon AM in chiral crystals.

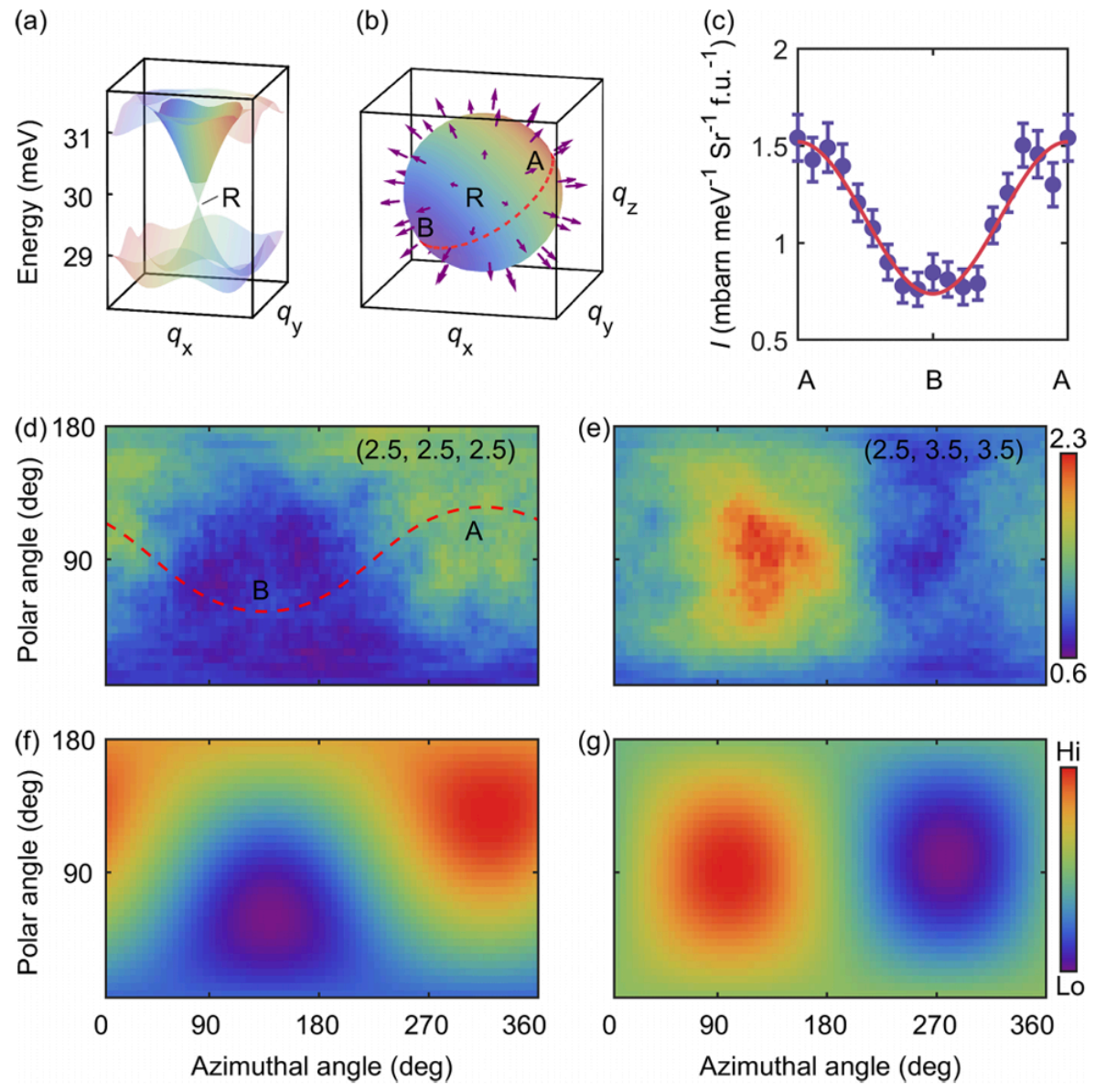

Figure 41 (a) Phonon dispersions near the charge-2 Dirac point in MnSi. Solid color indicates energy integration range (30–32 meV). (b) Pseudospin texture enclosing the R point. (c) INS intensity along the red circle in (b). (d) and (e) INS intensities over $\mathbf{q}$ spheres enclosing the momentum of (2.5, 2.5, 2.5) and (2.5, 3.5, 3.5), respectively. (e) and (f) Fitted-force-constant model calculations according to Eq. (110), for the same $\mathbf{q}$ spheres as in (d) and (e), respectively. Adapted from Ref. (Jin *et al.*, 2022)

## 2. Experimental Evidence of Circularly polarized Phonons Coupled with Magnetism

In real space, circularly polarized phonons correspond to collective circular motions of charged ions. Therefore, circularly polarized phonons can carry finite magnetic moment on the order of $\mu_{ph} \sim e\hbar/(2m_{ion})$, where $m_{ion}$ is the ionic mass that is much larger than the electron mass, $m_e$, yielding $\mu_{ph}/\mu_B \sim 10^{-3} - 10^{-5}$. Surprisingly, recent experimental studies have shown giant phonon magnetic moments, $g_{ph}\mu_{ph} \sim g\mu_B$, in various semimetals and semiconductors, including $CeCl_3$, $Cd_3As_2$, PbTe, and $Fe_2Mo_3O_8$, $Co_3Sn_2S_2$, etc. (Basini *et al.*, 2024; Baydin *et al.*, 2022b; Che *et al.*, 2025a; Cheng *et al.*, 2020c; Davies *et al.*, 2024; Lujan *et al.*, 2024b; Luo *et al.*, 2023c; Mustafa *et al.*, 2025; Wu *et al.*, 2023b; Yang *et al.*, 2025a). To understand these observations, Ren et al. pictured the phonon magnetic moment as electronic orbital magnetization in an adiabatic response to the underlying ionic circular motions (Ren *et al.*, 2021b). They reformulated the phonon magnetism by introducing a momentum $k$-dependent Born effective charge and discovered a topological magnetization term, which arises from the circularly polarized phonon-modified electronic energy with non-trivial Berry curvature. In topological semimetals and insulators, integration of the Berry curvature near the Yang's monopole gives rise to giant phonon magnetic moment, $g_{ph}\mu_{ph} \sim g\mu_B$.

Interestingly, a recent time-domain spectroscopy study of $Pb_{1-x}Sn_xTe$ films revealed that crossing the phase boundary from the topological-insulator regime to the trivial-insulator regime leads to a three-orders-of-magnitude suppression of $g_{\rm ph}\mu_{\rm ph}$, in agreement with theoretical predictions. Moreover, two independent theoretical studies successfully reproduced the circularly polarized phonon splitting observed in $Co_3Sn_2S_2$ (Che *et al.*, 2025a; Yang *et al.*, 2025a) using first-principles approaches. (Wang *et al.*, 2025b) attributed the splitting to a Raman term arising from spin-dependent phonon self-energies, whereas (Zhang *et al.*, 2026b) explained the effect within the framework of molecular Berry curvature. Despite the different theoretical formalisms, both studies obtained consistent results, suggesting that the observed phonon splitting originates from symmetry breaking induced by intrinsic magnetic order. Further theoretical developments on phonon magnetic moments can be found in Refs. (Bao *et al.*, 2023; Chaudhary *et al.*, 2024b, 2025; Chen *et al.*, 2026a; Geilhufe and Hergert, 2023; Juraschek *et al.*, 2022b; Wang *et al.*, 2025b; Wu *et al.*, 2025; Zhang *et al.*, 2026b, 2023d).

The coupling between circularly polarized phonon and magnetism has also been suggested in ultrafast optical studies of nonequilibrium states. Figure 42 (c) shows an experimental observation of a chiral-phonon-activated spin Seebeck effect (CPASS). Shining an ultrafast laser pulse builds a transient temperature gradient across the heterostructure and excite circularly polarized phonons from the non-magnetic 2D hybrid organic-inorganic perovskite (HOIP). It has been argued that the AM of circularly polarized phonons is transferred to electrons in the Cu, resulting in spin currents without external magnetic fields. The generated spin current is then probed by the time-resolved Kerr rotation angle $\Delta\theta_{kerr}$. By switching the handedness of HOIP from left (S, red) to right (R, blue), the oscillating Kerr signal in the time domain shows a $\pi$-phase shift, consistent with the chirality-dependence of the spin current. Intriguingly, Kim et al. estimated the CPASS coefficient in their setup and obtained a value that is 1$\sim$3 orders greater than the conventional spin Seeback coefficient based on magnetic materials (Kim *et al.*, 2023). (Tauchert *et al.*, 2022) performed ultra-fast diffraction study of magnetic Ni film. The authors observed that the Ni film loses its magnetic order almost completely within femtosecond (fs) timescales. In contrast, an almost instantaneous, long-lasting, non-equilibrium population of anisotropic high-frequency phonons were observed within 150–750 fs. The authors interpreted their observations as a rapid AM transfer process between electrons and circularly polarized phonons that occurs before the macroscopic rotation happens. Despite these interesting observations, it should keep in mind that the AM is not conserved in the reported ultrafast experimental setups. In addition, the time-scale of spin-pin and spin-phonon interactions is also unknown in these system. Theoretically, in the presence of spin-orbital coupling, circularly polarized phonons can indirectly induce spin magnetization via dynamical modulation of electronic states (Hamada and Murakami, 2020a; Yao and Murakami, 2024). Further controlled experiments maybe re-

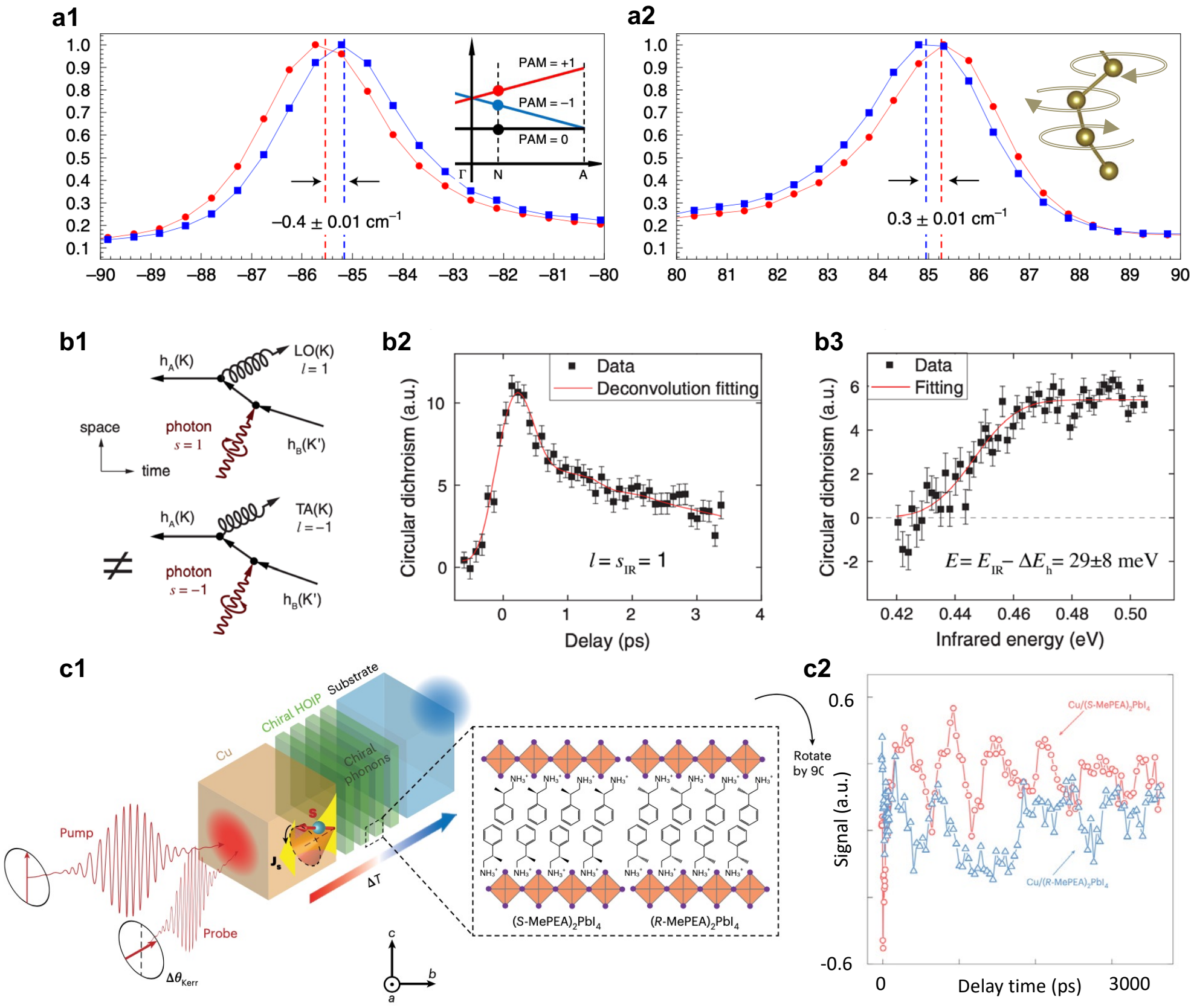

Figure 42 Anti-Stokes (a1) and Stokes (a2) spectra of a doubly degenerate circularly polarized phonon modes in $\alpha$-HgS. Chiral optical phonons slightly away from the $\Gamma$ point with PAM=+1 (red) and -1 (blue) are selectively probed under $(\hat{\varepsilon}_i, \hat{\varepsilon}_f)$=$(L, R)$ and $(R, L)$ polarization conditions. Here, $L$ and $R$ stand for left and right circular polarization, respectively (Ishito *et al.*, 2023a). (b1)-(b3) show circularly polarized phonon driven circular dichroism (CD) in the indirect optical transitions in $WSe_2$ (Zhu *et al.*, 2018). $h_A(K)$ and $h_B(K')$ represent holes at the valence band of the $K$ valley and the spin-split band of the $K'$ valley, respectively. The experimentally determined CD at 82 K and transient CD acquired at delay time, $\tau$=0.8 ps [(b2)], are used to estimate a circularly polarized phonon energy 29±8 meV [(b3)], in agreement with first principles calculations. (c1)-(c2) show experimental signature of a chiral-phonon-activated spin Seebeck effect (Kim *et al.*, 2023). Applying transient thermal gradient on left- and right- handed materials [(c1)] yields spin currents with a $\pi$-phase difference [(c2)]. Adapted from Ref. (Ishito *et al.*, 2023b; Kim *et al.*, 2023; Zhu *et al.*, 2018)

quired to firmly establish the AM-transfer between phonons and spin.

Circularly polarized phonons have shown application potentials in quantum information science (Chen *et al.*, 2019). A recent study from (Kong *et al.*, 2022) showed that circularly polarized phonons can selectively light up dark excitons from different valleys of $MoSi_2N_4$ in the photoexcitation process, yielding a circularly polarized phonon-triggered photoluminescence. Circularly polarized phonons have also been considered for non-reciprocal thermal transport (Chen *et al.*, 2020, 2022b; Grissonnanche *et al.*, 2020, 2019; Kasahara *et al.*, 2018; Li *et al.*, 2020d, 2023b; Ohe *et al.*, 2024; Qin *et al.*, 2012; Zhang *et al.*, 2019c). More recently, circularly polarized phonons have been proposed to detect dark matter through its non-gravitational interactions with standard-model matter (Romao *et al.*, 2023). Furthermore, quasiparticle interactions between topological magnon, polaron, and phonon were predicted to give rise to chiral edge phonon modes for future heat management of microelectronic devices (Debnath *et al.*, 2025; Ding *et al.*, 2025; Ghader and Jabakhanji, 2026; Husain and Ramesh, 2025; Li *et al.*, 2026; Liu *et al.*, 2025; Park *et al.*, 2025; Sukhanov *et al.*, 2025; Weißenhofer *et al.*, 2025).

## VII. CONCLUSIONS AND PERSPECTIVES

We end this review by summarizing main theoretical and experimental achievements and discussing open questions in topological phonons and circularly polarized phonons.

*Topological Phonons.* The theory of topological phonons is derived from the topological band theory (Bansil *et al.*, 2016; Hasan and Kane, 2010; Qin *et al.*, 2011). Theoretical predictions motivated experimental studies to directly measure Dirac

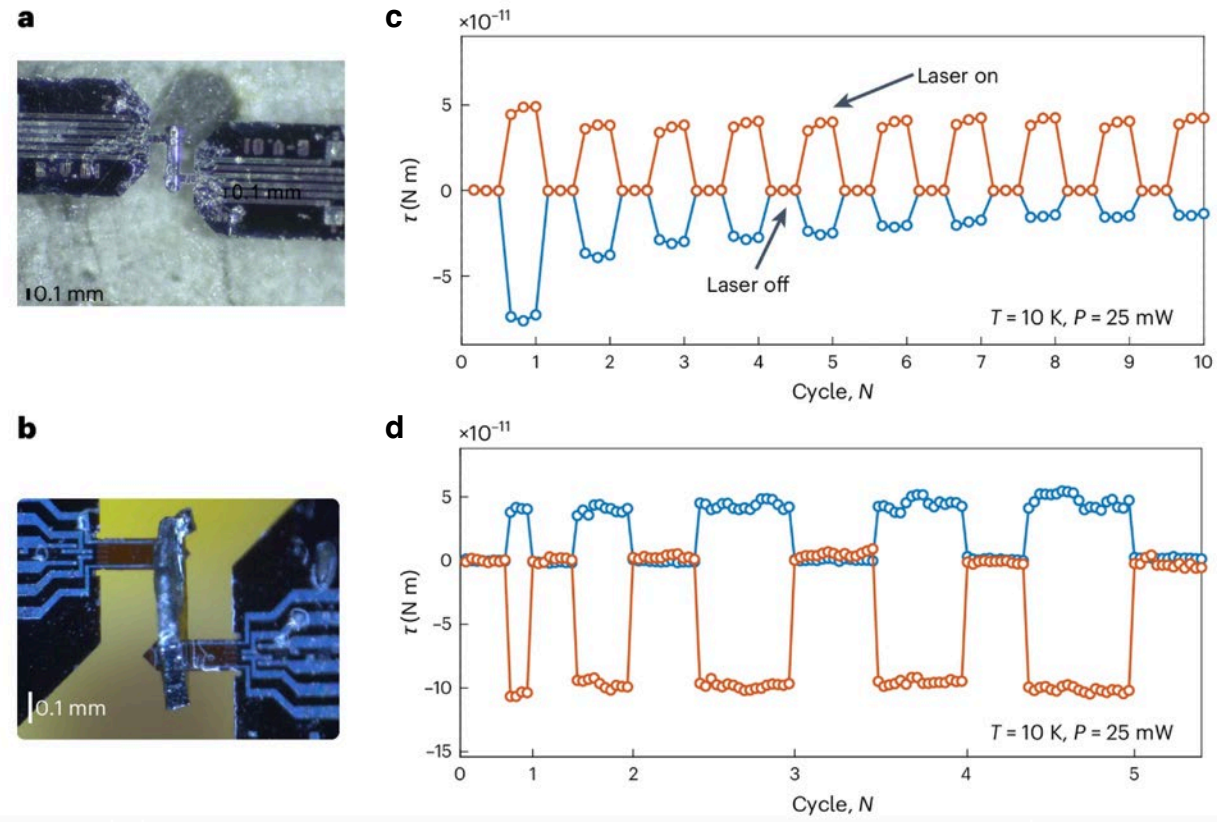

Figure 43 (a) and (b), Photos of experimental set-ups for the left- and right-handed Te single crystals. (c) Torque responses for the left-handed Te single crystal. Turning on the heating laser drives opposite sign of torque $\tau$ from the two cantilevers over multiple cycles. (d), The same measurement for the right-handed Te single crystal. Adapted from Ref. (Zhang *et al.*, 2025a) and Weyl phonons and chiral phononic wave functions in the bulk states. Like electronic systems, topological phonons can give rise to phononic edge mode. Double-helicoid surface phonon modes have been predicted to occur in transition-metal monosilicides, including FeSi and CoSi (Jin *et al.*, 2022; Miao *et al.*, 2018; Zhang *et al.*, 2018a), whereas flat surface phonon modes are expected in $MoB_2$ (Zhang *et al.*, 2019b) and the conventional high-$T_c$ superconductor $MgB_2$ (Li *et al.*, 2020c). In $MgB_2$, angle-resolved polarized Raman spectroscopy has further provided valuable structural information for elucidating the interplay between topological phonons and the material's superconducting properties under external perturbations (Cheng *et al.*, 2024). Direct experimental observation of these unconventional topological surface phonon modes would represent an important step toward establishing their existence and clarifying the physical consequences of topological bosonic excitations. Surface sensitive probes, such as HR-MEELS, will be a powerful technique to resolve this issue. RIXS and IXS with grazing incident condition may also provide key insights. With new experimental capabilities enabling selective excitation and detection of phonon modes, further topological phonon-driven phenomena in solids are likely to emerge. These advances also motivate applications in metamaterials, exemplified by topological phonon polaritons (Guddala *et al.*, 2021) and robust phononic devices such as disorder-immune delay lines and one-way waveguides (Cha *et al.*, 2018; Xi *et al.*, 2025; Zhang *et al.*, 2018c).

An important and largely unexplored direction concerns the interplay between topological phonons and topological electronic states. Such coupled excitations are expected to be particularly relevant in quantum materials hosting both phononic and electronic band topology, such as Weyl semimetals (e.g., CoSi and BaPtGe), where Weyl phonons and Weyl fermions may jointly affect transport behavior. More broadly, it remains an open question whether electron–phonon coupling in these topological materials can generate emergent transport phenomena beyond those arising from the electronic sector alone. In particular, it is of considerable interest to investigate whether the geometric and topological properties of phonon wavefunctions, including Berry-curvature-related effects, can contribute to or modify nonlinear transport responses that have thus far been primarily studied in topological electronic systems (Atencia *et al.*, 2025; Cao *et al.*, 2025; Desai *et al.*, 2026; Gao *et al.*, 2020; He *et al.*, 2019; Hossain *et al.*, 2025). The topological quasiparticle-interactions can also be important for spontaneous symmetry-breaking orders, such as charge/spin density waves (Luo *et al.*, 2018; Miao *et al.*, 2024; Yang *et al.*, 2024a; Zhang *et al.*, 2024), quantum Hall effects (Luo *et al.*, 2018), and superconductivity (Gao *et al.*, 2023; Grissonnanche *et al.*, 2020). The localized surface/edge states induced by nontrivial bulk band topology can help to enhance electron-phonon couplings and stabilize superconductivity at the interfaces (Di Miceli *et al.*, 2022). Understanding of these topological quasiparticle-interactions may also pave the way for practical applications, *e.g.*, nonreciprocal transport and enhanced thermoelectric conversion efficiency (Xu *et al.*, 2014).

Another interesting open question is the quantum geometry of phonons (Huang and Fang, 2026; Wu *et al.*, 2026). Given that the imaginary part of quantum geometry corresponds to the Berry curvature, phonon systems are expected to exhibit nontrivial quantum metrics (Albert *et al.*, 2016; Berry, 1989, 1984b; Bleu *et al.*, 2018; Gianfrate *et al.*, 2020; Kang *et al.*, 2025; Kim *et al.*, 2025; Neupert *et al.*, 2013; Provost and Vallee, 1980; Tan *et al.*, 2019; Yu *et al.*, 2024, 2020), i.e., the real part of quantum geometry, which may also play an important role in phonon-mediated processes. Recently, it has been shown that under the Gaussian approximation, the lower bound of electron-phonon coupling is determined by the quantum geometry of the electronic structure (Yu *et al.*, 2024). The role of quantum geometry of phonons in the electron-phonon coupling is yet to be understood.

*Circularly polarized Phonons.* Unlike topological phonons, the concept and study of circularly polarized phonons have a much longer history. A recent conceptual development has been the introduction of pseudo-angular momentum (PAM), which can be directly probed using polarization-resolved optical techniques (Che *et al.*, 2025a; Ishito *et al.*, 2023a; Watanabe *et al.*, 2025; Yang *et al.*, 2025a; Zhang *et al.*, 2026a, 2023b; Zhu *et al.*, 2018). As discussed in Sec. V.B, the PAM is fundamentally distinct from the angular momentum (AM) of phonons. While PAM can be directly accessed through Raman and infrared spectroscopies, experimental detection of the phonon AM itself was achieved only very recently using a mechanical torque measurement setup (Zhang *et al.*, 2025a).

While AM can be transferred between phonons and electronic excitations through quasiparticle-interactions, the magnitudes of these scattering channels remain unresolved, which may complicate the interpretations of experimental data. For instance, it has been known that chiral structure can induce spin polarized current (Göhler *et al.*, 2011; Naaman *et al.*, 2019; Naaman and Waldeck, 2012; Ray *et al.*, 1999). A recent theoretical study showed that this chirality-induced spin selectivity arises from spin-orbital coupling, which polarizes electron spin rather than filters spin through AM transfer from circularly polarized phonons (Wolf *et al.*, 2022).

As one of the main heat carriers, phonons are critical for thermal conductivity. For a long time, the phonon contribution to the thermal Hall conductivity, $\kappa_{xy}$, has been overlooked due to the small phonon magnetic moment. Recently, giant $\kappa_{xy}$ was observed in many strongly correlated insulators, including the cuprate high-$T_c$ superconductors (Grissonnanche *et al.*, 2020), Kitaev spin liquid candidate $\alpha$-$RuCl_3$ (Kasahara *et al.*, 2018), quantum paraelectric $SrTiO_3$ (Li *et al.*, 2020d). Empirically, the thermal Hall effect in these insulators displays two ubiquitous features: (1) longitudinal and transverse thermal conductivity peak at the same temperature, and (2) $\kappa_{xy}/\kappa_{xx} \sim 10^{-3}$ around B$\sim$10 T. These observations suggest that the unusually large $\kappa_{xy}$ has a phonon origin. Phenomenologically, the transverse conductivity can be written as $\kappa_{xy} = \frac{1}{\nu}\sum_\gamma C^\gamma(v^\gamma_{xy}l^\gamma + l^\gamma_{xy}v^\gamma)$, where $C$, $v$, and $l$ are the specific heat, phonon velocity, and mean-free path, respectively. $\gamma$ is the index of phonon modes, and $\nu$ is a dimension-dependent constant. Like electrons, the phonon thermal Hall effect also has intrinsic contribution from phonon Berry curvature (Qin *et al.*, 2012; Zhang *et al.*, 2010) and extrinsic contribution from skew scattering and side-jump (Chen *et al.*, 2020). However, it has been found that the intrinsic effect is usually orders of magnitude smaller than experimental observations. These experimental observations call for further theoretical study of the coupling between phonons and their embedded environments (Chen *et al.*, 2022b; Grissonnanche *et al.*, 2019; Jin *et al.*, 2025; Li *et al.*, 2023b; Zhang *et al.*, 2019c).

*Phonons with Both Topological Chirality and Rotational Chirality.* Breaking $\mathcal{PT}$ symmetry is the essential requirement for realizing nonzero Chern numbers and phonon angular momentum, thereby enabling the coexistence of topological chirality and rotational chirality (Zhang *et al.*, 2025c, 2018a). Under open boundary conditions or at interfaces between distinct topological domains, topological surface phonons can emerge and remain robust against impurities and structural imperfections. For example, thermal transport studies in chiral $\alpha$-quartz have shown that phonons can carry chirality-dependent angular momentum (Ohe *et al.*, 2024), which may be converted into measurable spin signals. This observation suggests that topological surface phonons could likewise possess chirality-dependent angular momentum, potentially enabling robust spin and heat transport. Furthermore, the introduction of magnetic order or other $\mathcal{T}$-symmetry-breaking perturbations can generate both topological and rotational chirality even in centrosymmetric materials (Che *et al.*, 2025a; Yang *et al.*, 2025a; Zhang *et al.*, 2026b), thereby giving rise to nonreciprocal phonon transport (Hirokane *et al.*, 2020). In such systems, topological protection suppresses dissipative scattering of surface phonons, while rotational chirality facilitates angular-momentum transfer. As a consequence, the longitudinal thermal conductivity can depend on whether the heat current is parallel or antiparallel to the vector product of electric polarization and magnetization, opening a possible route toward controllable phononic thermal diodes.

In summary, topological and circularly polarized phonons constitute a rapidly emerging frontier in condensed matter physics. These unconventional lattice excitations exhibit nontrivial topology and quantum geometry and can carry nonzero angular momentum, thereby playing an important role in light–matter scattering, thermal transport, and emergent symmetry-breaking and topological phases. Together, these concepts substantially broaden our understanding of symmetry, topology, and transport in crystalline solids, with promising implications for both fundamental physics and future quantum technologies.

## VIII. ACKNOWLEDGEMENTS

T.Z. acknowledges the support from the National Key R&D Project (Grant Nos. 2023YFA1407400 and 2024YFA1409200) and the National Natural Science Foundation of China (Grant Nos. 12374165 and 12047503). Y.L. is supported by the Innovation Program for Quantum Science and Technology (Grant No. 2023ZD0300500), the National Natural Science Foundation of China (Grant No. 12404279), and Shanghai Pujiang Program (Grant No. 23PJ1413000). H.M. is supported by the US Department of Energy, Office of Science, Basic Energy Sciences, Materials Sciences and Engineering Division. S.M. is supported by JSPS KAKENHI Grant Nos. JP22H00108 and JP24H02231.